\documentclass[12pt]{article}
\usepackage{amsmath,amsfonts, amssymb,verbatim}
\newcommand{\Pp}{{\rm P}}
\newcommand{\Qq}{{\rm Q}}
\newcommand{\A}{{\cal A}}
\newcommand{\Aa}{\mathrm{A}}
\newcommand{\EE}{\mathrm{E}}

\newcommand{\Q}{{\mathbb Q}}
\newcommand{\C}{{\mathbb C}}
\newcommand{\N}{{\mathbb N}}
\newcommand{\Z}{{\mathbb Z}}
\newcommand{\R}{{\mathbb R}}
\newcommand{\HH}{{\rm H}}

\newcommand{\F}{{\mathbb F}}

\newcommand{\V}{\mathcal{V}}
\newcommand{\bV}{{\bf V}}
\newcommand{\VV}{{\rm V}}

\newcommand{\uu}{{\bf u}}
\newcommand{\vv}{{\bf v}}
\newcommand{\hh}{\mathrm{h}}

\newcommand{\e}{{\rm e}}
\newcommand{\s}{{\bf s}}
\newcommand{\rr}{{\bf r}}
\newcommand{\dd}{\mathfrak{c}}

\newcommand{\cc }{\mathfrak{a}}

\newcommand{\ga}{K_{\rm fr}}
\newcommand{\gau}{\mathbf{G}}
\newcommand{\M}{{\bf M}} 
\newcommand{\Lim}{\mathsf{lim}\,}  
\newcommand{\la}{\langle}
\newcommand{\ra}{\rangle}
\newcommand{\acl}{\mathrm{acl}}
\newtheorem{pkt}{}[section]  
\newcommand{\bpk}{\begin{pkt}\rm }  
\newcommand{\epk}{\end{pkt}} 
\newcommand{\inv}{^{-1}}   

\newcommand{\be}{\begin{equation}} 
\newcommand{\ee}{\end{equation}}  
 \newcommand{\ev}{{\rm ev}}
\newcommand{\spec}{{\rm Sp}}

\newcommand{\subs}{\subseteq}
\newcommand{\ZZ}{{^*\Z}}
\newcommand{\QQ}{{^*\Q}}

\newcommand{\CC}{{^*\C}}
\newcommand{\EEE}{{^*\EE}(\mathbf{1})}
\newcommand{\Ee}{{^*\EE}}
\newcommand{\pk}{\epk \bpk}
\newcommand{\x}{\mathbf{x}}
\newcommand{\Nn}{\mathrm{N}}
\title{The semantics of the canonical commutation relation }
\author{B. Zilber}

\begin{document}
\maketitle
\begin{abstract}
We treat the canonical commutation relations and the conventional calculus based on it as an algebraic syntax of quantum mechanics and establish a geometric semantics of this syntax.  
This leads us to a geometric model, the space of states with the action of time evolution operators, which is a limit of finite models. The finitary nature of the space allows to give a precise meaning and calculate various classical quantum mechanical quantities. 
\end{abstract}

\section{Introduction}
\bpk This paper is a part of a broader project which aims to establish,  for a typical 'co-ordinate algebra' $\A$ in the sense of  non-commutative geometry, or a scheme, a geometric counterpart $\bV_\A.$ 
Here ``geometric'' should also be understood in some broad but well-defined sense. We believe this is possible in the setting of model theory, where 
a crucial new experience in generalisation of semantics of algebraic and analytic geometry, and applications of this, has been accumulated in the last 50 years.  

In fact, we aim to establish a rigorous mathematical duality 
\be\label{A_M} \A_\bV\ \longleftrightarrow \ \bV_\A\ee
between algebras $\A_\bV$ and structures $\bV_\A$ of geometric flavour.

Let us recall that a project with similar aims, but mainly concentrated around classical structures of mathematical physics, has been suggested by C.Isham and J.Butterfield and is being developed by C.Isham, A.D\"oring and others, see e.g. \cite{ID} and later publications. In that approach $\bV_\A$ is supposed to be a topos. 

Our approach is different in many details: $\bV_\A$ is a {\em multi-sorted} structure, each sort of which is a {\em Zariski geometry} in the sense of \cite{Zbook}. An essential component of the multi-sorted structure $\bV_\A$ is the set of morphisms between sorts functorially agreeing with embeddings between certain sublagebras of $\A,$ which makes  the left-to right  arrow a functor between a category of those subalgebras $A$ and sorts $\bV_A$ of $\bV_\A.$  This functor, in fact, defines a quite rich sheaf over the category of sublagebras. This makes an interesting point of contact of our approach with the topos-theoretic approach to foundations of physics.  
\epk
\bpk At this stage of our project we decided to keep the general context to minimum and instead concentrate on developing and illustrating the approach in a very special case of Heisenberg-Weyl algebras and the {\em canonical commutation} relation. And even there we quite quickly drop the generality of the $n$-th Heisenberg-Weyl  algebras for the sake of developing rather sophisticated calculus with one position and one momentum operators. 
\epk
\bpk The $n$-th Heisenberg-Weyl  algebra $\A^{(n)}$
 has generators $P_1,\ldots, P_n, Q_1,\ldots, Q_n,$ the ``co-ordinates''  of $n$-dimensional quantum mechanics, 
with  commutation relations
\be \label{ccr} Q_lP_k-P_kQ_l=2\pi i\delta_{lk}I,\ee
where the $P_k$ and $Q_l$ are seen as self-adjoint operators,
$I$ is the ``identity operator'' and  $\delta_{kl}$  the Kronecker symbol.
(We have chosen the multiplier $2\pi i$ 
for convenience of further
notations and to mean that $\exp 2\pi i I=I.$)

The infinite-dimensional Heisenberg algebra is
$$\A^{(\omega)}=\cup_n \A^{(n)}.$$

The representation theory of $\A^{(n)}$
is quite complicated, even for $n=1,$
 due to the fact that 
the algebra can not be represented as an algebra of bounded operators on a Hilbert space, not to speak about finite-dimensional representations. 

However, as suggested by Herman Weyl, we may instead consider the representation theory of algebras generated by the {\em Weyl
operators} which can be formally defined as 
\be \label{df}U_k^{a_k}=\exp  ia_kQ_k,\ \ V_l^{b_l}=\exp ib_lP_l\ee 
 for  $a_k ,b_l \in \R.$ These  are unitary (and so bounded!) operators if the $P_k $ and $Q_l $ are self-adjoint, and the following commutation relation holds: 
\be \label{ccr1} \begin{array}{ll}U_k^{a_k}V_l^{b_l}=q_{kl}V_l^{b_l}U_k^{a_k}\\
\mbox{where } q_{kl}=\exp { 2\pi i a_kb_l \delta_{kl}}\end{array}
\ee

Given a choice of real $a_1,\ldots,a_n, b_1,\ldots,b_n$ the complex $C^*$-algebra generated by Weyl operators $V_1^{\pm a_1},\ldots, V_n^{\pm a_n}, U_1^{\pm b_1},\ldots, U_n^{\pm b_n},$
$$A(a_1,\ldots,a_n, b_1,\ldots,b_n)=\C[V_1^{\pm a_1},\ldots, V_n^{\pm a_n}, U_1^{\pm b_1},\ldots, U_n^{\pm b_n}]$$
is called (in this paper) a Weyl algebra.

 The all-important Stone -- von Neumann theorem states that the representation theory of all the Weyl algebras
$A(a_1,\ldots,a_n, b_1,\ldots,b_n)$ together is equivalent to the Heisenberg commutation relation (\ref{ccr}). 
One can say that $\A^{(n)}$ can be fully replaced by the entirety
of its Weyl subalgebras  $A(a_1,\ldots,a_n, b_1,\ldots,b_n)$ which have a good  Hilbert space representation theory.

Especially nice finite-dimensional representations have Weyl algebras for which the multipliers $\e^{ 2\pi i a_k b_l }$ are roots of unity. This is the case when all the $a_k, b_l$ are rational. We call such algebras {\bf rational Weyl algebras}.


 In a certain sense, 
made precise in the paper, the  rational  Weyl algebras approximate the full Heisenberg-Weyl algebra just as rational points approximate points of the $n$-dimensional Euclidean space.
In fact, the paper uses the correspondence between rational Weyl algebras $A$ and certain  geometric objects $\bV_A$. The structure $\bV_A$  encapsulates the
representation theory of $A$ and is at the same time a well-understood object of model-theoretic studies, {\em a quantum Zariski geometry}, see \cite{QZG}. 

The quoted paper  established the duality (\ref{A_M})
between quantum algebras $A$ at roots of unity (rational Weyl algebras in our case) and corresponding quantum Zariski geometries $\bV_A,$ which extends the classical duality between commutative affine algebras and affine algebraic varieties.  The new step implemented in the present work is to extend the duality on algebras $A$ approximated by rational Weyl algebras. In order to achieve this we need to develop a notion of approximation on the side of structures $\bV_A.$
Such a notion, of {\em structural approximation}, has essentially been developed
in \cite{Appr} (quite similar to the context of {\em positive model theory} developed by I. Ben-Yaacov \cite{BY} and others). 
\epk 
\bpk \label{str_appr} A few introductory words about structural approximation:

We work with the category whose objects are rational Weyl algebras   $A(a,b),$ $a,b\in \Q,$ and morphism are embeddings. It is useful to have in mind that the embedding $A(a,b)\subset A(c,d)$  can be determined in terms of divisibility relations between denumerators of the fractions of integers representing $a,b,c$ and $d.$
 
We  consider an ultrafilter $\mathcal{D}$  on $\Q^2$ to be defined more specifically below. Consider the ultraproduct of algebras $\{ A(a,b): \la a,b\ra\in \Q^2\}$ modulo $\mathcal{D},$  call it  ${\tilde{\mathrm{A}}}  .$ This can be seen as a {\em non-standard} algebra ``generated'' by  operators $U^aV^b,$ for $a,b\in \QQ,$ the field of { \em non-standard rational numbers.}

We consider then the corresponding ultraproduct modulo $\mathcal{D}$ of Zariski geometries $\bV_{A(a,b)}$, denote it
$\bV_{\tilde{\mathrm{A}}}  .$

Now we are ready to define a limit object, {\em the space of states}
$$\mathbb{S}=\Lim_{\mathcal{D}}\bV_{A(a,b)}$$ and the procedure of taking the limit. 
By definition in \cite{Appr} the limit structure is 
  a {\em homomorphic} image of $\bV_{\tilde{\mathrm{A}}} $ and $\Lim$
is a  surjective {\em homomorphism} 
$$ \Lim: \bV_{\tilde{\mathrm{A}}}  \twoheadrightarrow \mathbb{S},$$
which means a map preserving basic relations of the language.

The canonical equivalence on $\bV_{\tilde{\mathrm{A}}}$
associated with $\Lim,$
$$x_1\approx x_2 \Leftrightarrow \Lim x_1=\Lim x_2$$ 
should be read as ``$x_1$ is infinitesimally close to $x_2$''.

This definition is very sensitive to the choice of the basic relations  (primitives) of the language and is not preserved under the transition to an inter-definable language. Effectively, the choice of the ``right'' primitives amounts to the choice of what physicists would consider as {\bf observables}. 

The fact that we construct our structures as Zariski geometries plays a key role here: the notion of homomorphism in the above definition agrees with  the notion of morphisms in the category of respective Zariski geometries, structures with (a generalised) Zariski topology on them.
 
Further choices of primitives of the language lead to the introduction of  probability (equal to the absolute value of the inner product) as an observable as opposed to the inner product, which is well-defined for each $\bV_{A(a,b)}$ but can not be defined uniformly  without breaking the symmetries required by Zariski structure.    
\epk
\bpk \label{DD}
The  ultrafilter $\mathcal{D}$ of \ref{str_appr} is chosen     
to be a Fr\'echet
ultrafilter with respect to the partial ordering between the $A(a,b),$ that is:

 for
each $A(a,b)$  the set
$\{ \la c,d\ra\in \Q^2: A(a,b)\subset A(c,d)\}$ is in 
$\mathcal{D}.$

We also add a condition on $\mathcal{D}$ which depends on a choice of a rational number 
$\hh$ and which results in the fact
that ${\tilde{\mathrm{A}}}$  is
``generated'' by a pair of operators $U^\frac{1}{\mu}$ and $V^\frac{1}{\nu},$ for
non-standard integers $\mu$  and $\nu$ such that  $$\hh=\frac{\mu}{\nu}.$$

The fact that  $\mathcal{D}$  was  Fr\'echet has an important consequence:
both $\mu$ and $\nu$ are {\em divisible by all the standard integers.}

Consequently we will get that  $A(a,b)\subset {\tilde{\mathrm{A}}}  $ for any
(standard) rational $a,b$ which is one of the desirable properties;  $\tilde{\mathrm{A}}$ is universal for the family of all rational algebras.

\epk

\bpk Let us turn now to more detailed description of
the objects $\bV_A$ and their limit, the space of states $\mathbb{S}.$ 

For a rational Weyl algebra $A$ the object $\bV_A$ 
is defined, following \cite{QZG},
as a bundle (not locally trivial!) of irreducible $A$-modules over an algebraically closed field, say $\C.$
In the current paper we identify a more interesting structure of 
what we call ``algebraic-Hilbert'' modules on $\bV_A.$
 ``Algebraicity'' of  the modules first of all reflects the fact that the modules are of finite dimension $n_A.$ More serious algebraic feature of $\bV_A$ is in the fact that the submodule over a much smaller subfield,
$\R_0[\sqrt{-1}]\subset \C,$ where $\R_0$ is the field of algebraic totally real numbers,  has an inner product structure with values in  $\R_0[\sqrt{-1}].$ To express the sesquilinearity of the inner product we use the formal complex conjugation on $\R_0[\sqrt{-1}],$ which does not require definability of the reals. This is crucial since we want to keep the structure $\bV_A$ a Zariski structure (hence $\omega$-stable). In fact, for each particular $A$ we need just a finitely generated subfield $\Q[q]$ of $\R_0[\sqrt{-1}],$ where $q$ is a root of unity introduced in (\ref{ccr1}). 

With the inner product and a formal complex conjugation defined we can also speak of adjoint operators, unitarity and the related notions.
\epk

\bpk Each algebraic-Hilbert space $\bV_{A(a,b)}$ is acted upon by respective Weyl operators $U^a$ and $V^b.$ In the limit procedure 
the rational parameters $a$ and $b$  become infinitesimals and for this reason can not be observed in the limit structure. The same can be said about  $U^a$ and $V^b.$  

Instead of $U^a$ and $V^b$   we introduce in the language of $\bV_{A(a,b)}$
 the new names $\Pp$ and $\Qq$ which can be used for all values of $a,b\in \Q:$
$$\Qq:=\frac{U^a-U^{-a}}{2ia},\ \ \Pp:=\frac{V^b-V^{-b}}{2ib}$$
(compare with (\ref{df})). 
Note that in each $A(a,b)$ the operators $\Qq$ and $\Pp$ are inter-definable with $U^a$ and $V^b$ respectively (but by different formulas!) The effect of redefining the basic operators is that the new operators $\Qq$ and $\Pp$ have a meaning throughout the whole construction and in the limit of it. So 

{\em the operators $\Qq$ and $\Pp$ are  observables.}
 \medskip

Crucially, we prove: 
\medskip

{\bf Theorem} (see \ref{QP}) {\em The canonical commutation relation (\ref{ccr}) holds in $\mathbb{S}.$}

\medskip

This can be seen as another version of Stone - von-Neumann Theorem.

\epk
\bpk Other observable operators can be constructed as e.g. linear combinations $c\Qq+d\Pp,$ $c,d\in \R.$
To each such operator in the conventional quantum mechanical setting one associates a Lagrangian subspace of the phase space. A symplectomorphism between two such Lagrangian subspaces can be identified as a (generalised) Fourier transform (see e.g. \cite{LV}). 

In our finitary setting this picture translates into considering commutative subalgebras of $\tilde{A}$ of the form $A(c,d)$ where $c,d$ are rationals (eventually approximating real values).  To each such commutative subalgebra one can associate the space of eigenvectors, a subset of $\bV_{\tilde{A}},$ which is an equivalent of the lagrangian, or rather the lagrangian subspace with a line bundle with connection. 
 In place of the Fourier transforms we define a class of Zariski morphisms with special properties which we call
 {\em  regular unitary transformations} since they preserve the orthonormality. 

\medskip
 
 {\bf Theorem.} {\em The   regular unitary transformations survive to the limit and form a group which has
 $\mathrm{SL}(2,\R)$ as its homomorphic image.}

In $n$-dimensional case   $\mathrm{SL}(2,\R)$ should become the symplectic group $\mathrm{Sp}(2n,\R),$ but we do not do this case in the current version of the paper.  
  
\epk
\bpk The fact that the construction of the limit structure involves non-standard (infinite) integers $\mu$ and $\nu$ (as in \ref{DD})  is one of the causes of emergence of non-observable values. More concretely, the absolute value of the inner product
of norm 1  eigenvectors $|x\ra$ and $|p\ra$ of operators $\Qq$ and $\Pp$ is 
    $$|\la x\, | p\ra| =\frac{1}{\sqrt{\mu\nu}}$$
    and so is an infinitesimal. This can be used to define the {\em probability density} but for more subtle calculations one needs a {\em renormalisation}, a systematic procedure which produces a finite value. We introduce such a universal procedure in terms of our construction. This is
    called in the paper the {\em Dirac rescaling}.  As an intermediate stage the Dirac rescaling introduces a ``discrete'' Dirac delta-function on each lagrangian 
    and in the above example
    gives us 
       $$\la x\, | p\ra_{Dir} =\frac{1}{ \sqrt{2\pi\hbar}}\e^{ixp}, $$
       where the right-hand side is defined up to a phase factor.
\epk

\bpk It is well-known that some (not all) of the generalised Fourier transforms can be identified as {\em time evolution operators} for certain particles. The same is true for regular unitary transformations. We consider such transformations $K_\mathrm{free}^t$ and $K_\mathrm{QHO}^t$ corresponding to time evolution operators for the free particle and the quantum harmonic oscillator and calculate the Feynman propagator, which coincides with the well-known results: 

$$ \la x_1| K_\mathrm{free}^t x_2\ra_{Dir}=\frac{1}{\sqrt{2\pi i\hbar t}}\e^\frac{i(x_1-x_2)^2}{2t\hbar}$$ and $$\la x_2|K^t_{\mathrm{QHO}}|x_1\ra_{Dir}= \sqrt{\frac{1}{2\pi i \hbar\sin   t}}\,
\exp i\frac{(x_1^2+x_2^2)\cos   t-2x_1x_2}{2\hbar\sin   t},$$
where $t$ are arbitrary positive real numbers such that  $\hbar=2\pi \hh >0$ and $\sin t\neq 0.$     

\epk
\bpk The most interesting example we calculated is the formula
$$\mathrm{Tr}(K^t_{\mathrm{QHO}})=\frac{1}{ i|\sin \frac{t}{2}|}$$
for the trace of the evolution operator for the quantum harmonic oscillator. A naive approach to calculating the trace would be to consider the sum of eigenvalues of $K^t_{\mathrm{QHO}}$ which are well known
$$ \mathrm{Tr}(K^t_{\mathrm{QHO}})=^?\sum_{n=1}^\infty \e^{i (n-\frac{1}{2})t}.$$
But this does not make sense since  $|\e^{i (n-\frac{1}{2})t}|=1.$
Our calculation takes place for arbitrary rational values of the parameter $\arcsin t$  in big enough finite-dimensional algebraic-Hilbert spaces $\bV_A.$ In each such space the trace can be calculated and the result has the same form as above (for the given value of the parameter).  Note that the easy of obtaining the formula is compensated by the difficulty of calculating eigenvalues of $K^t_{\mathrm{QHO}}$ in the finite-dimensional space. In fact, we have not calculated these eigenvalues.
\epk

\bpk Of course, discrete models have been in permanent use throughout the history of quantum physics as a heuristic tool, see the reference \cite{Zeidler I}  or more advanced treatment in similarly defined spaces in \cite{GN}.
 The analysis of such a discrete model usually concludes with the phrase ``now we pass to continuous limit'', which actually is an ill-defined notion and a source of main troubles of quantum physics. 
 
 Perhaps the main achievement of this paper is the
 approximation procedure $\Lim$ suggesting a rigorous interpretation of the above.

Moreover, the mathematical model developed in this paper may have a much stronger physical interpretation. The pseudo-finite-dimensional Hilbert space which is associated with our main object
$\bV_{\tilde{A}}$ corresponds to a pseudo-finite space of states. Here ``pseudo'' refers to the non-standard finite (that is infinite) size of the space. This is a good mathematical substitute for the notion of a {\em ``huge finite'' universe}. The limit $\mathbb{S}$ of this space should then be considered a ``continuous approximation'' to the huge finite universe.       The claim of the paper is that by assuming this as a model for reality we get the correct mathematical theory of quantum mechanics. 
\epk
\bpk {\bf The role of logic and model theory.} We see canonical commutation relations and more general formulas of algebraic quantum mechanics as rules of a syntax that speaks correctly about quantum mechanics since the time of Heisenberg and Dirac. However, unlike the case of the syntax  of classical Hamiltonian mechanics  (the calculus of differential manifolds),  no geometric semantic interpretation of this syntax has been established. And this is arguably the main cause of  troubles in quantum physics, see \cite{ID0}. From our point of view this is also  a true challenge to logicians and, more concretely, model-theorists.
  
The quoted paper \cite{ID0} suggest toposes as possible semantical interpretations for similar syntactic construction (general von Neumann algebras), and there have been a few other constructions of category-theoretic origin. 
 
 The construction we use in \cite{QZG}, in the present paper and in other ongoing joint works has many common features with the above mentioned approaches, but the role of category theory is taken by model theory. The advantage is that the semantical interpretation is supposed to take shape of a concrete structure with crucial model-theoretic characteristics. These characteristics supposed also to justify the adjective {\em geometric} in the definition {\em geometric semantics}. Model theory has tools for distinguishing structures of this desired type,  tools developed in {\em stability theory}. In particular, the notion of Zariski geometry (and its generalisations) proved to be a very convenient generalisation of objects of algebraic geometry, including non-commutative one. The insistence that the model of quantum mechanics must be a Zariski geometry severely restricts our choices but also brings in strong parallels with classical geometric structures. Zariski geometries come with an internally defined notion of {\em dimension}. Another key property  is {\em homogeneity}: any two subobjects which are indiscernible by the language are conjugated by an automorphism of the structure. In the current setting this property explains the inherent symmetries of quantum mechanics and exhibits their common nature with Galois automorphisms of algebraic geometry. 
 
 In the end the approach works, and this paper is here to prove the statement.
 
\epk
\bpk {\bf Acknowledgements}. I would like to thank Alex Cruz Morales for his interest to this work and his many questions which essentially effected the form and the content of this paper and led to our joint paper \cite{withA} surveying an early version of the present work. I would also like to thank Andreas D\"oring who explained to me the idea of the topos-theoretic approach to foundations of physics and who introduced me to  people interested in foundations of physics. I am also grateful to Bruce Bartlett who draw my attention to some recent works in mathematical foundations of quantum mechanics.   
\epk

\section{Preliminaries}\label{prelim}
\bpk {\bf The Heisenberg group and its rationally generated subgroups.}
We start with the Heisenberg real Lie algebra $L^{(n)}$ with 
generators  $iP_1,\ldots,iP_n,$ $ iQ_1,\ldots, iQ_n, 2i\pi $ and defining relations (\ref{ccr}).  This is a $2n+1$-dimensional nilpotent Lie algebra with the 1-dimensional centre $i\R.$

Applying Lie exponentiation one obtains a real Heisenberg group $\mathrm{Heis}^{(n)},$ 
$$L^{(n)}\to  \mathrm{Heis}^{(n)},$$
which is generated by the $2n$ 1-parameter subgroups
$$U^\R_k:=\{ U^a_k: a\in \R\},\ \    V^\R_k:=\{ V^b_k: b\in \R\},\ k=1,\ldots n,$$
where $U^a_k$ and $V^b_k$ are the images of $aiQ_k$ and $biP_k$ respectively, as defined in   (\ref{df}).
This is a $2n+1$-dimensional nilpotent Lie group of nilpotency class 2 and the centre $S^1:=\exp i\R$
equal to the commutator subgroup $[ \mathrm{Heis}^{(n)}, \mathrm{Heis}^{(n)}].$

We denote $\HH(a_1,\ldots,a_n,b_1,\ldots,b_n)$ the subgroup of  $\mathrm{Heis}^{(n)}$
generated by $U^{a_1}_1,\ldots, U^{a_n}_n, V^{b_1}_1,\ldots, V^{b_n}_n.$ 

Our particular interest is with the groups with rational values of parameters $a_1,\ldots,a_n,b_1,\ldots,b_n.$
The centre of such a group $\HH,$ equal to the commutator $[\HH,\HH]$ is a finitely generated periodic abelian group with generators $q_k:= \exp 2\pi i a_kb_k,$ and so is finite. In group-theoretic terms 
  $\HH(a_1,\ldots,a_n,b_1,\ldots,b_n)$ is a group of rank $2n.$ 
  
  We will also consider  subgroups of $\mathrm{Heis}^{(n)}$  of the more general form, which we call {\bf Heisenberg groups of rank $2k.$} These are subgroups of some  $\HH(a_1,\ldots,a_n,b_1,\ldots,b_n)$ with rational parameters of rank exactly $2k.$

Most of the time will be interested in the case 
$n=1$ and the Heisenberg group $\HH(a,b),$ which is a group of rank $2.$
 More generally a Heisenberg group of rank 2 is a  subgroup $\HH$ of $\HH(a,b)$ such that the quotient $\HH/[\HH,\HH]$ is non-cyclic. 
\epk
\bpk \label{H1} {\bf Lemma.} {\em A Heisenberg group of rank $2$ is isomorphic to $\HH(c,d)$ for some $c,d.$ If  $\HH\subs \HH(a,b)$ for rational $a,b,$ then $c,d$ are rational as well.}
 
 {\em A Heisenberg group of rank $2n$ is 
of the form $$\HH_1\times\ldots\times \HH_n$$
where the $\HH_i$ are  Heisenberg groups of rank $2.$ }

{\em A Heisenberg group $\HH$ of rank 2 which is a subgroup of some $\HH(a,b)$
 has generators of the form
$U^{g_{11}}V^{g_{12}}$ and $U^{g_{21}}V^{g_{22}},$ for rational $g_{11},g_{12},g_{21},g_{22}$ such that $\det \left( g_{ij}\right)\neq 0.$ }

We write, correspondingly,
$$\HH=\la U^{g_{11}}V^{g_{12}}, U^{g_{21}}V^{g_{22}}\ra.$$

\medskip

{\bf Proof.} Easy. $\Box$ 

\epk

\bpk 
We also will consider an algebraically closed field $\F$ of characteristic $0$ (e.g. the field of complex numbers $\C$) and the Weyl algebras,  which are defined as a group algebras $\F \HH$ of the respective Heisenberg groups. In particular,
$$A(a_1,\ldots,a_n,b_1,\ldots, b_n):=\F \HH(a_1,\ldots,a_n,b_1,\ldots, b_n),$$
where $$\HH(a_1,\ldots,a_n,b_1,\ldots, b_n)=\la U^{a_1},\ldots,U^{a_n}, V^{b_1},\ldots,V^{b_n}\ra.$$
\epk
\bpk \label{H2}
By \ref{H1} any Heisenberg group of rank $2n$ is isomorphic to one of the form $\HH(a_1,\ldots,a_n,b_1,\ldots, b_n)$ and every Weyl algebra is isomorphic to one of the form
$A(a_1,\ldots,a_n,b_1,\ldots, b_n).$
\epk
\bpk \label{symH} {\bf A symplectic structure associated with the Heisenberg group.} 

Recall that $\HH$ is a nilpotent group of class $2$ and hence
the commutator operation $[h,g]:= hgh\inv g\inv$
determines an alternating $\Z$-bilinear form on $\HH/[H,H],$
$$[\cdot,\ \cdot]:\ \HH/[H,H]\times \HH/[H,H]\to [H,H]\cong\Z.$$ 
For the case of real Heisenberg group this is called the symplectic structure on $\HH.$ The same terminology can be applied to $\HH(a_1,\ldots,a_n,b_1,\ldots, b_n)$
without the risk of ambiguity.
 
\epk

\bpk A {\bf rational Weyl $\F$-algebra of rank $2n$} is a Weyl algebra isomorphic to one of the form $A(a_1,\ldots,a_n,b_1,\ldots, b_n)$ with rational non-zero  $a_1,\ldots,a_n,$ $b_1,\ldots, b_n.$

In this case the $q_{kk}$ in (\ref{ccr1}) are roots of unity.
 Such algebras, and more generally {\em quantum algebras at roots of unity}, can be seen as ``coordinate algebras'' of corresponding Zariski geometries. This has been studied in \cite{QZG}. 
 
\epk

\bpk\label{T1}{\bf Theorem.} {\em Let $A=A(a_1,\ldots,a_n,b_1,\ldots, b_n)$ be a rational Weyl $\F$-algebra. Then 
\begin{enumerate}
\item[(i)] $A$ and the centre $Z(A)$ are prime noetherian PI rings. Let $n_A$ be the PI degree of $A.$
\item[(ii)] Every maximal ideal of $A$ is of the form $\tilde \alpha=A\alpha,$ for $\alpha\subseteq Z(A),$ $\alpha \in \mathrm{Spec}(Z(A)),$ a maximal ideal of the centre.
\item[(iii)] The unique irreducible $A$-module $\VV_A(\alpha)$ with annihilator ${\rm Ann}(\VV)=\tilde{\alpha}$ is of dimension $n_A$ over $\F.$
The irreducible representations of $A$ are in bijective correspondence with points $\alpha\in \mathrm{Spec}({Z(A)}),$ the maximal spectrum of $Z(A),$
which can be identified with ($\F$-points of) an affine irreducible
variety.
\item[(iv)] There exists an isomorphism $A/\tilde{\alpha}\cong {\rm M}(n_A,\F),$ the full $n_A\times n_A$-matrix algebra.  
\item[(v)] The irreducible $A$-module $\VV_A(\alpha)$ is isomorphic to a tensor product of corresponding irreducible $A_k$-modules
$\VV_{A_k}(\alpha_k),$
$$\VV_A(\alpha) \cong \bigotimes_{k=1}^n \VV_{A_k}(\alpha_k)$$
for $A_k=A(a_k,b_k)$ and some maximal ideals $\alpha_k$ of $Z(A_k).$
\end{enumerate}}

{\bf Proof.} For (i)-(iv) see   \cite{BG}, III.1.1 and III.1.6.

(v) follows from the fact that \be \label{prod} \HH(a_1,\ldots,a_n,b_1,\ldots, b_n)=\prod_{k=1}^n \HH(a_k, b_k).\ee
$\Box$

\epk
\bpk
Note that the  isomorphism in (iv) is given by a representation $j_a:A\to {\rm M}(n_A,\F),$  which  in its own right is
 determined by a choice of a maximal ideal $a\subset Z(A)$ and a basis ${\bf e}$ in the module $\VV_A(a),$ so $j=j_{a,e}.$
\epk  
\bpk \label{Aut} {\bf Definition.} Given the Heisenberg group $H(a,b)$ with rational $a,b,$ it is easy to describe its automorphism group $\mathrm{Aut}\,H(a,b).$ It consists
of automorphisms $\xi_{g,n,m},$ for  $g=(g_{ij})\in \mathrm{SL}(2,\Z),$ $n,m\in 0,1,\ldots,N-1$ where $N$ is the order of 
defined on the generators of $H(a,b)$ the root of unity $q=e^{ 2\pi i ab}:$

$$\xi_{g,n,m}: \begin{array}{ll} U^a\mapsto q^nU^{g_{11}a}V^{g_{12}b}\\

V^b\mapsto q^mU^{g_{21}a}V^{g_{22}b}
\end{array}$$

Note that by isomorphism the commutator of the image of $U^a$ and $V^b$ is equal to the same number $q,$ hence $\det (g_{ij})=1.$

Set
$\Gamma_{H(a,b)}$ to be the subgroup of   $\mathrm{Aut}(H(a,b))$ which fixes $H(a,b)/Z(H(a,b)),$
the quotient of the Heisenberg group modulo its centre.

Given a rational Weyl algebra  $A=A(a_1,\ldots,a_n,b_1,\ldots, b_n)$ we set
$$\Gamma_A=\Gamma_{A_1}\times \ldots \times\Gamma_{A_n}$$
 for $ A_k=A(a_k,b_k),\ k=1.\ldots,n,$
with the action on $A=A_1\oplus \ldots \oplus A_n$ coordinate-wise. 
 
\epk

\bpk\label{STI2} {\bf Lemma.} {\em For $A=A(a,b)$ the group
$\Gamma_{A}$ is generated by the two automorphisms $\mu$ and $\nu$ defined on the generators of $A$ as follows

\be \label{munu}\mu: \begin{array}{ll}U^a\mapsto q U^a\\
V^b\mapsto V^b
\end{array} \ \ \ \ \ \ 
\nu: \begin{array}{ll}U^a\mapsto U^a\\
V^b\mapsto qV^b
\end{array}\ee

$\Gamma_A$ and $\Sigma_A$ fix the centre $Z(A)$ point-wise.

$$\Gamma_A\cong (\Z/N\Z)^2.$$}
  
 {\bf Proof.} By definition any $\gamma\in \Gamma_A$ acts by $U^a\mapsto q^k U^a$ and $V^b\mapsto q^mV^b,$ for some integers $k$ and $m.$ Then $\gamma: 
 U^aV^bU^{-a}V^{-b}\mapsto U^aV^bU^{-a}V^{-b},$ that is $\gamma(qI)=qI$ for all $\gamma\in \Gamma_A.$ 

Clearly, $\mu,\nu\in \Gamma_A.$ Also $\gamma\inv \mu^k\nu^m$ acts as  identity on $U^a,V^b,$ hence $  \gamma= \mu^k\nu^m.$ The transformation $\phi$ fixes $qI$ by (\ref{ccr1}).

Finally, it is immediate from (\ref{munu}) that the generators $\mu,\nu$ of $\Gamma_A$ commute and are of order $N.$ 
  $\Box$
\epk

\bpk
In what follows we will extensively use $(\ref{prod})$ and reduce the study of $n$-th Weyl algebras to the study of 1-st Weyl algebras. 

\epk

\bpk \label{6.6} {\bf Lemma.} {\em The centre $Z(A)$ of  rational $A=A(a_1,\ldots,a_n,b_1,\ldots, b_n)$ is of the form  $A(N_1a_1,\ldots,N_na_n,N_1b_1,\ldots,N_n b_n),$
where $N_k$ is the order of the root of unity $\exp 2\pi i a_kb_k,$ that is equal to the denominator of the reduced rational number $a_kb_k.$}

{\bf Proof.} Immediate. $\Box$
\epk
\bpk \label{maxH} {\bf Lemma.} {\em Suppose $$A(a'_1,\ldots,a'_n,b'_1,\ldots, b'_n)\subs A(a_1,\ldots,a_n,b_1,\ldots, b_n)$$
and $A(a'_1,\ldots,a'_n,b'_1,\ldots, b'_n)$ is commutative and maximal among commutative.

Then 

(i)  $a'_k=M_ka_k,\ b'_k=L_kb_k,$ for some $M_k,L_k\in \Z$ such that $M_kL_k$ is equal to the denominator of the reduced fraction $a_kb_k,$ $k=1,\ldots,n.$
 
(ii)  In particular, $A(N_1a_1,\ldots,N_na_n, b_1,\ldots, b_n)$ 
and $A(a_1,\ldots,a_n, N_1b_1,\ldots, N_nb_n)$
are maximal commutative subalgebras of $A(a_1,\ldots,a_n,b_1,\ldots, b_n).$

(iii) Maximal commutative subalgebras $A(N_ka_k, b_k)$  and $A(a_k, N_kb_k)$ of $A(a_k,b_k)$ 
are 1-generated by $V^{b_k}$ and $U^{a_k},$ respectively, over the centre of the algebra.  
 }

{\bf Proof.}  (i)  We may assume that $n=1$ and we consider $A(a',b')\subs A(a,b)$ maximal commutative.
 Then  $a'=Ma,\ b'=Lb,$ for some $M,L\in \Z$ by the embedding assumption.
Commutativity requires that $MLab\in \Z.$ Maximality implies the last of the conditions. 

(ii) Immediate from (i).

(iii) Just note that  $U^{N_ka_k}$ and $V^{N_kb_k}$ are in the centre of the algebra. 
$\Box$
\epk

\bpk \label{6.5} {\bf Lemma.} {\em Given a rational commutative $n$-th algebra $A=A(a_1,\ldots,a_n,b_1,\ldots, b_n),$  there is finitely many commutative   rational $n$-th Weyl algebras $B$ extending $A.$}

{\bf Proof.} Any element of $B$ is of a product of some
$U_k^{a'_k}$ and $V_k^{b'_k},$ $k=1,\ldots, n.$
Commutativity implies that $a'_kb_k, a_kb'_k\in \Z$ for all $k.$ This puts bounds on the denominators of $a'_k$ and $b'_k.$ On the other hand, since $U_k^{a_k},V_k^{b_k}\in B,$ we can reduce every product in $B$ modulo $A$ so that $|a'_k|\le |a_k|$ and  $|b'_k|\le |b_k|.$ There are finitely many such $a'_k$ and $b'_k$ with the bound on denominators.
 $\Box$

\epk
\bpk \label{Zup} {\bf Lemma.} {\em Given a commutative  rational $n$-th Weyl algebra  

$C=A(c_1,\ldots,c_n,d_1,\ldots, d_n),$ let $C^\uparrow$ be the algebra generated by all the commutative   rational $n$-th Weyl algebras $B$ extending $C.$ Then 
\begin{itemize}
\item[(i)] $C^\uparrow=A(\frac{c_1}{N_1},\ldots,\frac{c_n}{N_n},
\frac{d_1}{N_1},\ldots,\frac{ d_n}{N_n}),$
where $N_k=c_kd_k,$ $k=1,\ldots,n.$

\item[(ii)] $C^\uparrow$ is generated by two maximal commutative subalgebras

$A(\frac{c_1}{N_1},\ldots,\frac{c_n}{N_n},{d_1},\ldots,{ d_n})$ and $A({c_1},\ldots,{c_n},\frac{d_1}{N_1},\ldots,\frac{ d_n}{N_n}).$

\item[(iii)]
 If $C$ is the centre of $A=A(a_1,\ldots,a_n,b_1,\ldots, b_n),$  then 
$C^\uparrow=A.$
\end{itemize}
}

{\bf Proof.} (i) It is clear that $C^\uparrow$ contains $A(\frac{c_1}{N_1},\ldots,\frac{c_n}{N_n},{d_1},\ldots,{ d_n})$ and $A({c_1},\ldots,{c_n},\frac{d_1}{N_1},\ldots,\frac{ d_n}{N_n})$ as the two subalgebras are commutative and contain $C.$

(ii) It is easy to see that both are maximal commutative subalgebras of $C^\uparrow$ as presented in (i). It is obvious from the explicitly given parameters that the two sublagebras generate $C^\uparrow.$ 

 (iii) By  \ref{6.6} $C=A(N_1a_1,\ldots,N_na_n,N_1b_1,\ldots,N_n b_n),$ where the $N_k$ are denominators of the reduced rational numbers $a_kb_k.$ Hence, by setting $c_k=N_ka_k$ and $d_k=N_kb_k$ we get 
by (i) $C^\uparrow=A.$ $\Box$
\epk
\bpk \label{ZupCor}{\bf Corollary.} {\em The functors $A\mapsto Z(A)$ and $C\mapsto C^\uparrow$ between categories of  rational $n$-th Weyl algebras
and commutative  rational $n$-th Weyl algebras are inverse to each other.}
\epk

\bpk \label{Cgamma}{\bf Lemma.}

 {\em The family of all  the commutative Weyl subalgebras $C$ of a rational Weyl algebra $A$
containing the centre $Z(A)$ of $A$ is finite.}

{\bf Proof.} It is enough to note that the generators $U_k^{a_k},V_k^{b_k}$ of $A$ are
periodic modulo the centre. Hence 
 the images of canonical generators $U_k^{c_k},V_k^{d_k}$
of $C$  modulo $Z(A)$ take only finitely many values. $\Box$
 
\epk
\bpk \label{ass2}{\bf Definition.} 
Given a rational Weyl algebra $A$ 
 $O(A)$ be the set of  maximal commutative Weyl subalgebras $C\subseteq A.$
 
 \medskip


 \epk
\bpk \label{O-lagr} {\bf Proposition.} {\em There is a natural bijection between $O(A)$ and the set of Lagrangian subgroups of the Heisenberg group $\HH$ such that $A=A(\HH).$}

{\bf Proof.} Immediate by definition. $\Box$
 \epk
\bpk \label{Cgamma2}{\bf Lemma.} {\em Suppose now $A=A(a,b)$ and $C\in O(A).$ Then $C=A(a,Nb)^\gamma$ for some $\gamma\in \mathrm{Aut}\HH(a,b).$ }

{\bf Proof.} Note that $A(a,Nb)$ is maximal commutative in $A.$
By \ref{maxH}(iii) $C$ is generated by one element $W\in A$ over $Z(A).$ Then
$W=U^{g_{11}a}V^{g_{12}b},$ $g_{11},g_{12}\in \Z,$ co-prime modulo $N.$ Choose $g_{21},g_{22}\in \Z$ such that $g_{11}g_{22}-g_{12}g_{21}=1$ and consider $W':=U^{g_{12}a}V^{g_{21}b}.$ We see that $$\gamma: \begin{array}{ll} U^a\mapsto W\ V^b\mapsto W'\end{array}$$
is an automorphism of $H(a,b).$ 
$\Box$

\epk
\bpk\label{O-gen}{\bf Lemma.} {\em For a rational $n$-th Weyl algebra $A=A(a_1,\ldots,a_n,b_1,\ldots,b_n),$
 $$O(A):=\{ C_1\times\ldots\times C_n: C_k\in O(A(a_k,b_k)),\ k=1,\ldots,n\}.$$ }
 
 {\bf Proof.} We may consider maximal commutative subgroups $C\subset \HH$ of a rank-$2n$ Heisenberg group instead. Recall that $$\HH=\prod_{i=1}^n \HH_i$$
where $\HH_i$ are Heisenberg groups of rank 2. The projection $C_i$ of $C$ to the $i$-th subgroup must be commutative. It also must be maximal, since otherwise there is an element $h_i$ with $h_i\notin C_i,$ $[h_i,C_i]=1,$ on the $i$-th co-ordinate and equal to $1$ on other co-ordinates, and such an $h$ commutes with $C,$ contradicting maximality.

Since $C\subseteq \prod_{i=1}^n C_i$ and the product is commutative we have the equality. $\Box$ 
\epk
\bpk {\bf Corollary.} {\em There is a canonical bijection between the set $O(A)$ of maximal commutative subalgebras  of $A$ and the set of all Lagrangian subspaces on the symplectic space associated with the respective  Heisenberg group $\HH$ by \ref{symH}.}
\epk
\section{Categories $\A_{fin}$ and $\mathcal{C}_{fin}.$} \label{GWA}
 
\bpk \label{def1} {\bf Definition.} We fix  $\F$ to be an algebraically close field of characteristic zero.

$\A_{fin}$ is the category of all the rational Weyl $\F$-algebras  with  canonical embeddings  as morphisms. 

$\A^{(n)}_{fin}$ is the subcategory of $n$-th Weyl algebras of $\A_{fin}$ 

$\mathcal{C}_{fin}$ and $\mathcal{C}^{(n)}_{fin}$ are the subcategories of commutative algebras of   $\A_{fin}$ and  $\A^{(n)}_{fin}$ respectively.

\epk 
 
\bpk \label{1.1old} {\bf Properties.}

It is easy to check:

\begin{itemize}
 \item[(i)] 
$\A_{fin}$ and $\mathcal{C}_{fin}$  are  small categories of unital $\F$-algebras with 
at most one morphism between two objects.

\item[(ii)]
The functor $A\mapsto Z(A)$ is an isomorphism between $\A_{fin}^{op}$ and $\mathcal{C}_{fin}.$ We will describe    
the inverse functor $Z(A)\mapsto A:\ Z\mapsto Z^\uparrow$ below.

\item[(iii)] $$\bigcap_{A\in \A_{fin}}Z(A)=\F.$$

\end{itemize}

{\bf Proof.} (i) and (iii) are immediate from definitions.  (ii) is just \ref{ZupCor}.$\Box$

\epk
\bpk \label{F0} {\bf Subfield $\F_0$ with complex conjugation.}

For most of our purposes it is suficient to work within a subfield $\F_0$ of $\F.$
 
 We set $\R_0\subset \F$ to the subfield of the {\bf totally real } algebraic numbers and $\F_0=\R_0[i],$ the extension of $R_0$ by $i=\sqrt{-1}.$
In particular, $\F_0\subs \Q^{alg},$ the algebraic closure of $\Q.$

Note that $\mu,$ be the  group of all roots of unity in $\F,$ is a subset of 
$\F_0.$    
Note also that $q+q\inv$ is a totally real number for $q\in \mu.$

We will think
of $\mu$ as the group of ``elements of modulus 1'' and introduce an {\bf formal (complex) conjugation}  
$x\mapsto x^*,$  an automorphism of the field $\F_0,$
  defined as $$^*:\ r+is\mapsto r-is,\mbox{ for }r,s\in \R_0.$$
In particular,  
  $u^*=u\inv,$ for $u\in \mu$ and $r^*=r$ for $r\in \R_0.$
\medskip

In particular, given $n\times n$ matrix $X$ over $\F_0,$ we define an adjoint $X^*$ as the matrix obtained by transposition and formal conjugation.  We call $X$ {\bf $\F_0$-unitary} if $X^*X=1.$    
   
\epk
\bpk
{\bf Remark.} Suppose $\F$ is of cardinality continuum (which may be assumed without loss of generality). Then $\F$ is abstractly isomorphic to the field of complex numbers. So, in this case $\F_0\subset \C$ and by construction the involution $x\mapsto x^*$ coincides on $\F_0$ with the complex conjugation. This is immediate from the definition. 

\epk

\bpk \label{ass1}{\bf Proposition.} {\em An $A\in \A_{fin}$ is an affine prime algebra over $\F,$ finitely generated 
over its centre $Z(A)$ as a module. $A$ is finitely generated as an $\F$-algebra with the field of definition $\Q[q_1,\ldots,q_n],$ for $q_k= \exp 2\pi i a_kb_k$ $k=1,\ldots, n.$ 

Every maximal ideal of $A$ is regular.}

{\bf Proof.} Immediate.
\medskip

Such algebras are considered in \cite{BG} which contains in particular the following statement.

\epk

\bpk \label{ass3}
{\bf Proposition.} { Let $A,B\in \A_{fin}.$ Then  
 $Z(A)\cap Z(B)\in \mathcal{C}_{fin}$ and so $(Z(A)\cap Z(A))^\uparrow=\la A, B\ra\in \A_{fin}.$
}

{\bf Proof.} Immediate by \ref{Zup}. $\Box$
\epk
\bpk {\bf Notation.} Given $A\in \A_{fin}$ we write $\spec_A$ for $\mathrm{SpecMax}(Z(A)).$
\epk
\bpk \label{piBA} { \bf Lemma.} {\em Suppose $A,B\in \A_{fin},$ $B\subs A,$ so $Z(A)\subs Z(B).$ Given $\beta\in \spec_B,$  a maximal ideal of $Z(B),$ we get a maximal ideal $\beta\cap Z(A)$ of $Z(A).$ The  map
$$b\mapsto \beta\cap Z(A), \ \ \pi_{BA}:\spec_B\to \spec_A$$
is surjective. }

{\bf Proof.} Follows from \ref{T1}(ii). $\Box$

\epk

\bpk \label{1.1(vi)}{\bf Lemma.} {\em Let $A\in \A_{fin}$ and $C\in O(A).$ Let
$B=C^\uparrow.$
Then $$\spec_C=\spec_B\mbox{ and }\pi_{BA}:\spec_B\to \spec_A$$
is unramified of order $n_A.$}

{\bf Proof.}  Since $C=Z(B)$, we have by definition $\spec_C=\spec_B.$

 Let $n=n_A.$
Recall that by \ref{1.1old}  the image of $Z(A)$ in $A/\tilde{\alpha}$ is $\F\cdot \mathbf{1},$ the subalgebra of scalar matrices. By \ref{ass2}(ii) the image of $C$ in the algebra of matrices  is the diagonal subalgebra, that is of the form $\F\cdot \mathbf{i}_1+\ldots+\F\cdot \mathbf{i}_n,$  where $\mathbf{i}_1,\ldots, \mathbf{i}_n$ are orthogonal idempotents. 

Note that every maximal ideal of $C/\tilde{\alpha}\cap C$ is of the form $\sum_{j\neq k}\F\cdot \mathbf{i}_j,$ 
for some $k=1,\ldots,n.$ It follows that every maximal ideal of $C$ is of the form
$$\xi_k=\sum_{j\neq k}\F\cdot {i}_k +\tilde{\alpha}\cap C, \mbox{ where }i_k\in C,\ \mathbf{i}_k=i_k+\tilde{\alpha}\cap C.$$ 

On the other hand, $\xi_k\cap Z(A)=\alpha.$ Indeed, 
$(\xi_k\cap Z(A))/(\tilde{\alpha}\cap C\cap Z(A))=0$ since $\F\cdot 1\cap \sum_{j\neq k}\F\cdot \mathbf{i}_k=0.$ Hence $\xi_k\cap Z(A)\supseteq \tilde{\alpha}\cap C\cap Z(A)=\alpha.$ The inverse inclusion is obvious by maximality of $\alpha.$

Hence $\pi_{BA}: c\mapsto \alpha=c\cap Z(A)$ is an $n-$to$-1$ map $\spec_B\to \spec_A.$ 
$\Box$ 

\epk
\bpk \label{TS} {\bf Corollary.} {\em Suppose $B\subs A,$ $A,B\in \A_{fin}.$  Then
\begin{itemize}
                                  \item[(i)] $$\pi_{BA}:\spec_B\to \spec_A$$
is an unramified projection of order $n_A:n_B.$

\item[(ii)] for a generic $\alpha\in \spec_A$ and $\beta\in \pi_{BA}\inv(\alpha)$ the extension $(\Q(\beta): \Q(\alpha))$ is Galois, and the
 Galois group of the extension acts transitively on $\pi_{BA}\inv(\alpha).$
 
                                 \end{itemize}
}

{\bf Proof.} (i) is immediate from \ref{1.1(vi)}.

(ii)  follows from the fact that $\spec_B$ is irreducible (see \ref{1.1old}(ii)) and so
$\pi_{BA}\inv(a)$ is irreducible over $\Q(\alpha).$ 
$\Box$

\epk

\bpk \label{1.5} 
 {\bf Remark.} The  projection $\pi_{BA}$
 can be seen in a  graphical way if we take into account that $Z(B)$ is an extension of $Z(A)$
by a finite set  of operators $X.$
Correspondingly, any element of  $\spec_B$ can be determined by $\alpha^{\smallfrown}\xi$ (concatenation) where $\alpha\in \spec_A$ and $\xi$ correspond to eigenvalues of operators $X\in Z(B).$ Now
\be \label{pr}\pi_{BA}: \alpha^{\smallfrown}\xi \mapsto \alpha.\ee  
\epk
\bpk \label{Cbasis}
Now suppose $C\in O(A),$ $\alpha\in \spec_A.$
 Then $\alpha=Z(A)\cap \tilde{\alpha}\subs C\cap \tilde \alpha.$ 
Consider the irreducible $A$-module $\VV_A(\alpha)$ introduced in  \ref{T1}(iii). Let $n=n_A=\dim_\F \VV_A(\alpha).$ So
 there is a basis $\{ e_1(\alpha),\ldots, e_n(\alpha)\}$ of 
$\VV_A(\alpha)$ consisting of common eigenvectors of all the operators in $C.$

 On the other hand, given such a basis $\{ e_1(\alpha),\ldots, e_n(\alpha)\},$ $\alpha\in \spec_A,$ we will have
$$C=\{ X\in  A:  \forall \alpha\in \spec_A\ \exists x_i\in \F\ \bigwedge_{i=1}^{n_A} X_\alpha e_i(\alpha)=x_i e_i(\alpha)\},$$   
where $X_\alpha$  denotes the image of $X$ in the representation $A/\tilde{\alpha}.$ 

We call such a basis a $C$-basis. 


\epk

\bpk \label{uv}
Now suppose $A$ is the  Weyl algebra $A(a,b)$ and  $C=A(a,Nb),$ a maximal commutative Weyl subalgebra as defined in \ref{maxH}(ii). Recall that $A(a,Nb)$ is generated over the centre by $U^a.$
Maximal ideals $\alpha$ of the centre $A(Na,Nb)$ of $A(a,b)$ are generated by two elements 
$U^{aN}-\hat{u}I$ and $V^{bN}- \hat{v}I$ of the centre, where 
$\hat{u},\hat{v}\in \F^\times$ and $I$ is the unit of $A(a,b).$ 

Choose two elements of $\F^\times,$ suggestively denoted $u^a$ and $v^b,$ such that   
$\hat{u}=(u^{a})^N$ and  $\hat{v}= (v^{b})^N.$  Let $q=\exp 2\pi i ab,$ and let $N$ be the order of  the root of unity $q.$

We define a {\bf canonical $A(a,Nb)$-basis} $\{ \uu\}_{\alpha}$ for $\VV_A(\alpha),$ \\
$\alpha=\la U^{aN}-u^{aN}I, V^{bN}-v^{bN}I\ra,$
$$\uu^{a,b}_{\VV_A(\alpha)}=\{  \uu^{a,b}(u^a,v^b),\uu^{a,b}(q u^a,v^b),...,\uu^{a,b}(q^{N-1}u^a,v^b)\}$$  to satisfy 
\be \label{UV1}\begin{array}{ll}
U^a: \uu^{a,b}(q^k u^a,v^b) \mapsto q^{k} u^a\uu^{a,b}(q^k u^a,v^b)\\

 V^b: \uu^{a,b}(q^k u^a,v^b) \mapsto  v^b\uu^{a,b}(q^{k-1} u^a, v^b)
\end{array}
\ee
We often abbreviate the terminology and call such a basis a $U^{a}$-basis.
\medskip

We also introduce symmetrically a canonical $A(Na,b)$-basis, or $V^b$-basis, for the same module,
 
$$\vv^{a,b}_{\VV_A(\alpha)}=\{  \vv^{a,b}(v^b,u^a),\vv^{a,b}(q v^b,u^a),...,\vv^{a,b}(q^{N-1}v^b),u^a\}$$  to satisfy 
\be \label{VU1}\begin{array}{ll}
V^b: \vv^{a,b}(q^k v^b,u^a) \mapsto q^{k} v^b\vv^{a,b}(q^k v^b,u^a)\\

 U^a: \vv^{a,b}(q^k v^b, u^a) \mapsto  u^a\vv^{a,b}(q^{k+1}  v^b,u^a)
\end{array}
\ee

\epk

\bpk \label{w} Suppose now $A=A(a,b)$ and $C\in O(A).$ Then by 
\ref{Cgamma2} there is an automorphism $\gamma$ of $H(a,b)$
such that $U^{a\gamma}=S,$ $V^{b\gamma}=T$ and $S$ generates $C$
 over $Z(A).$ A canonical $C$-basis $\{ \s\}_\alpha$ of $\VV_A(\alpha)$
 is defined following (\ref{uv}) as
 \be \label{st}\begin{array}{ll}
S: \s(q^k s,t) \mapsto q^{k} s\,\s(q^k s,t)\\

 T: \s(q^k s,t) \mapsto  t\,\s(q^{k-1} s, t)
\end{array}
\ee
\epk
\bpk \label{uuvv}
In the general case of $A=A(a_1,\ldots,a_n,b_1,\ldots,b_n)$ we will also consider
maximal commutative subalgebras of the form $C=A(a_1,\ldots,a_n,N_1b_1,\ldots,N_nb_n),$ generated
by $U^{a_1},\ldots,U^{a_n}$ over the centre. 
Using the fact that an irreducible $A$-module $\VV_A$ can be represented as a tensor product of
modules $\VV_{A_k(\alpha_k)}$ for $A_k=A(a_k,b_k)$ and the corresponding $\alpha_k.$

{\bf A canonical 
$A(a_1,\ldots,a_n,N_1b_1,\ldots,N_nb_n)$-basis} (or $\la U^{a_1},\ldots,U^{a_n}\ra$-{\bf basis})  is defined correspondingly as the tensor product of the $A(a_k,N_kb_k)$-bases.

\epk
\bpk \label{check} Below, unless  indicated otherwise, we assume $A=A(a,b),$ $\check{U}=U^a,$ $\check{V}=V^b$ and the $A$ module $\VV_A(\alpha)$ is determined by parameters
$u=u^a,\ v=v^b.$ We abbreviate the notation for  canonical $\check{U}$-bases as $\uu=\uu^{a,b}.$
\epk
\bpk \label{twobases}
{\bf Lemma.} {\em (i) Any two canonical bases 
$ \{ \uu(u,v),\uu(q u,v),...\uu(q^{N-1}u,v)\}$ and $\{ \uu'(u,v),\uu'(q u,v),...\uu'(q^{N-1}u,v)\}$ given by the condition (\ref{UV1}) can only differ  by a scalar multiplier, that is for some $c\neq 0,$ 
$$\uu'(q^k u,v)=c\,\uu(q^k u,v),\ \ \ k=0,\ldots,N-1.$$

(ii) Given also a canonical base $ \{ \uu(u,vq^m),\uu(q u,vq^m),...\uu(q^{N-1}u,vq^m)\},$ there is a $c\neq 0$ such that
$$\uu(q^k u,vq^m)=cq^{km}\uu(q^k u,v),\ \ \ k=0,\ldots,N-1.$$

(iii) Let $\hat{V}\in \HH(a,b)$ be such that 
$$\check{U}\hat{V}=q\hat{V}\check{U}$$
and $\hat{v}$ an $\hat{V}$-eigenvalue   in the module $\VV_A(\alpha)$
and let $ \{ \hat{\uu}(u,\hat{v}),\hat{\uu}(q u,\hat{v}),...,\hat{\uu}(q^{N-1}u,\hat{v})\}$ be a canonical $\check{U}$-basis in $\VV_A(\alpha)$ with regards
to $\check{U}$ and $\hat{V}.$ Then there is a $c\neq 0$ and an integers $n$  such that
$$\hat{\uu}(q^k u,\hat{v})=cq^{-n\frac{k(k+1)}{2}}\uu(q^k u,v),\ \ \ k=0,\ldots,N-1.$$
}

Proof. (i) We will have $\uu'( u,v)=c\,\uu( u,v)$ for some $c$ since the space of $\check U$-eigenvectors with a given eigenvalue $u$ in an irreducible 
$(\check U,\check V)$-module is one-dimenional. The rest follows from
the definition (\ref{UV1}) of the action by $\check V$ on the bases. 

(ii) We have $\uu(q^ku,vq^m)=c\,\uu(q^k u,v)$ and so by induction on $k,$ applying $\check V\inv$ and (\ref{UV1}) to the both parts we get the desired formula.

(iii) First note that from assumptions $\hat{V}=q^l\check{V}\check{U}^n$ for some
integers $n$ and $l.$ In particular, it follows $\hat{v}=q^{r+l}u^nv$ for some integer $r.$ 

Assume the equality holds for a given $k.$ Apply $\hat{V}\inv$ to both sides:
$$\hat{V}\inv:\hat{\uu}(q^k u,\hat{v})\mapsto \hat{v}\inv\hat{\uu}(q^{k+1} u,\hat{v})$$
$$\hat{V}\inv: cq^{-n\frac{k(k+1)}{2}}\uu(q^k u,v)\mapsto cq^{-n\frac{k(k+1)}{2}}q^{-n(k+1)}\hat{v}\inv \uu(q^{k+1} u,v).$$
Comparing we get the required.
$\Box$
\epk
\bpk \label{defv}{\bf Lemma - definition.} {\em Set
 $$\vv(vq^{m},u):= \frac{1}{\sqrt{N}}\sum_{k=0}^{N-1} 
q^{-mk}\uu(uq^{k},v), \ m=0,\ldots, N-1.$$
The system   $$\{ \vv(v,u),\ldots, \vv(vq^{N-1},u)\}$$ is a canonical $\check{V}$-basis of $\VV_{A(a,b)}$
satisfying (\ref{VU1})
}

{\bf Proof.} One checks directly that the system satisfies (\ref{VU1}). $\Box$
\epk

\bpk \label{group}{\bf Groups  $\Gamma_A(\alpha).$ }

For  $A\in  \A_{fin},$ $\alpha\in \spec_A,$ we define  
the  $\F$-linear transformations of $\VV_A(\alpha)$ 
   given in a canonical $\check{U}$-basis as:

 $$\mu_\alpha: \uu(q^m u,v)\mapsto \uu(q^{m-1} u,v),$$
and 
 $$\nu_\alpha :\uu(uq^{m},v)\mapsto q^{-m} \uu(uq^{m},v).$$ 

We define 
$\Gamma_A(\alpha)$ to be the group 
 generated by  $\mu_\alpha$ and $\nu_\alpha.$
 
 

\epk
\bpk \label{Gam-invar} {\bf Lemma.} {\em 
The definition of $\mu_\alpha$ and $\nu_\alpha$ depend on the choice of an $\hat{U}$-basis $ \{ \uu(q^m u,v)\}$ in the following way:

(i) any other basis is of the form
$$ \uu(q^m u',v'):=c\cdot\gamma\,  \uu(q^m u,v), \ \ m=0,1\ldots, N-1$$
for $\gamma\in \Gamma_A(\alpha)$ and $c\in \F;$

(ii) for  $\mu'_\alpha$ and $\nu'_\alpha$ corresponding to the basis $ \{ \uu(q^m u',v')\}$ we will have

$$\mu'_\alpha=\mu_\alpha^\gamma, \ \ \nu'_\alpha=\nu_\alpha^\gamma;$$

(iii) the group $\Gamma_A(\alpha)$ does not depend on the choice
of the canonical $\hat{U}$-basis, that is the group generated by   $\mu'_\alpha,\nu'_\alpha$ is $\Gamma_A(\alpha);$

(iv) moreover, 
 $\Gamma_A(\alpha)\cong \Gamma_A(\beta)$ for any $\beta\in \spec_A.$
}

{\bf Proof.} (i) is by \ref{twobases}. 
(ii) is immediate and (iii) follows from the fact that the change of basis corresponds to the inner automorphism of the groups.

(iv) For any  $\beta\in \spec_A$  a vector space isomorphism  $\VV_A(\beta)\to   \VV_A(\alpha)$ sending an $\hat{U}$-basis to $\hat{U}$-basis
induces, by (ii), a group isomorphism  
$\Gamma_A(\alpha)\to \Gamma_A(\beta).$ 
$\Box$
\epk

Given $\xi\in \Gamma_A(\alpha)$ and $X\in A$ let $X_\alpha$ be the image of $X$ in $A(\alpha),$ an operator acting on $\VV_A(\alpha),$ and $$X_\alpha^\xi:=\xi X_\alpha \xi\inv.$$

\bpk \label{mst}{\bf Lemma.} {\em We have

$$\begin{array}{ll}  \check{U}_\alpha^{\mu_\alpha}=q \check{U}_\alpha\\
 \check{V}_\alpha^{\mu_\alpha}=\check{V}_\alpha
\end{array} \ \ \ \ \
\begin{array}{ll}  \check{U}_\alpha^{\nu_\alpha}=\check{U}_\alpha\\
 \check{V}_\alpha^{\nu_\alpha}=q\check{V}_\alpha
\end{array}
$$
}

\epk
\bpk {\bf Corollary.} {\em The map $h_\alpha: \Gamma_A\to \Gamma_A(\alpha)$ defined on the generators
as $$h_\alpha:\begin{array}{ll} \mu\mapsto \mu_{\alpha}\\ 
\nu\mapsto \nu_\alpha
\end{array}$$
extends to a surjective homomorphism of groups.}

\epk
\bpk \label{rem1.9(iii)} \label{ass4}{\bf Remarks.} (i) For $A$ commutative  we have $\Gamma_A(\alpha)=\mathbf{1},$
for all $\alpha\in \spec_A.$ Indeed, $n_A=1$ in this case.

(ii) In a canonical $\check{U}$-basis the group $\Gamma_A(\alpha)$ is represented by a subgroup ${\rm SU}(n_A,\F_0)$ of $\F_0$-unitary matrices. This is immediate by definition.
\epk

\bpk \label{Lemmag}
{\bf Lemma.} (i) {\em If $\xi\in \Sigma_A(\alpha)$ and $X_\alpha^\gamma=X_\alpha$ for all $X\in A,$ then $\xi=gI,$ for some $g\in \F[n_A],$ root of unity of order $n_A.$ }

(ii) {\em $\Sigma_A(\alpha)$ is finite.}

{\bf Proof.} (i) Suppose $X_a\xi=\xi X_a,$ for all $X\in A.$ Let $e\in \VV_A(a)$ be an eigenvector of $\xi$ with eigenvalue $g.$ Then $\xi':=\xi-gI$ is a linear transformation of $\VV_A(a)$ which commutes with all the $X_a$ and 
$$\xi'\VV_A(a)=W\subset \VV_A(a),\ W\neq \VV_A(a).$$ 
But $X_aW=X_a\xi'\VV_A(a)=\xi'X_a\VV_A(a)=\xi'\VV_A(a)=W,$ that is $W$ is an $A$-submodule of $\VV_A(a).$ By irreducibility of the latter, $W=0$ and so $\xi=gI.$ But since $\det \xi=1,$ $g^{n_A}=1.$

(ii) By \ref{Cgamma} $\Sigma_A$ contains a subgroup $\Sigma^0_A$ of finite index which fixes setwise $O(A).$  

By definition elements of $\Sigma^0_A(\alpha)$ send a $C$-basis  to a $C$-basis, for any $C\in O(A).$ So, a possibly smaller subgroup of finite index preserves
$C$-eigenvalues, for all $C\in O(A).$ We may assume $\Sigma^0_A(\alpha)$ has this property.

Hence any $\xi\in  \Sigma^0_A(\alpha)$ commutes with any $X_\alpha,$ for $X\in C.$ By \ref{T1}(ii) $\xi$ commutes with $A/\tilde{\alpha}.$ By (i)  $\xi=gI,$ $g\in \F[n_A].$ $\Box$

\epk

\bpk \label{EA}{\bf $U$-frames $\EE^U_A(\alpha)$.}
Given $\alpha\in \spec_A,$ choose a $\check{U}$-eigenvector  $e\in \VV_A(\alpha)$ and
set $$\EE^U_A(\alpha):= \Gamma_A(\alpha)\cdot e.$$

We call $\EE^U_A(\alpha)$ an $U$-frame of $\VV_A(\alpha).$

Clearly, $e$ can be included in a canonical basis \\
$\{ \uu(uq^m,v): m=0,\ldots,N-1\}$ as, say, $e=\uu(uq^k,v).$ Then  $$\{ \uu(uq^m,v): m=0,\ldots,N-1\}=\{ \mu^m_\alpha e: m=0,\ldots,N-1\}\subseteq
\EE^U_A(\alpha).$$
By \ref{twobases} and the fact that $\nu_\alpha\in \Gamma_A(\alpha),$ any other canonical $\check{U}$-basis
is a subset of  $\EE^U_A(\alpha)$ of the form $\Delta(\alpha) e$ for a cyclic subgroup $\Delta(\alpha)\subset 
\Gamma_A(\alpha).$
\epk
\bpk \label{plustwobases} {\bf Corollary.} Assuming that the bases in \ref{twobases} are all in  $\EE^U_A(\alpha),$ the scalar $c$ is of the form $q^l$ for some $l\in \{ 0,1,\ldots, N-1\}.$
\epk
\bpk \label{Inprod} {\bf Inner product}. 
Let $\F_0[\EE^u_A(\alpha)]$ stand for the $\F_0$-linear span of $\EE^U_A(\alpha),$ that is the  $\F_0$-vector subspace of $\VV_A(\alpha)$  span over $\EE^U_A(\alpha).$ 
 By above $\{ \uu(uq^m,v): m=0,\ldots,N-1\}$ is a basis of the vector space 
$\F_0[\EE^U_A(\alpha)].$

  We introduce an {\bf inner product}  on $\F_0[\EE^U_A(\alpha)]$ by declaring the 
basis  {\bf orthonormal}. By \ref{rem1.9(iii)}(ii)
 the action of $\Gamma_A(\alpha)$ on $\F_0[\EE^U_A(\alpha)],$ and so on $\VV_A(\alpha),$ is given by {\bf unitary} $\F_0$-matrices. 
Since any other canonical basis can be obtained by a transformation in $\Gamma_A(\alpha),$ the inner product structure does not depend on the choice of the initial basis. We denote the inner product between two elements $f,g\in \F_0[\EE^U_A(\alpha)]$ by
$$\la f|g\ra_\alpha.$$
We usually omit the subscript.

Following physics tradition we write the property of $\F_0$-sesquilinearity
as
$$\la f|\tau g\ra= \tau \la f|g\ra= \la \tau^* f|g\ra $$
for $\tau\in \F_0.$
\epk

\bpk \label{exm_uv} {\bf Example.} The definition of a $V$ basis in terms of $U$-basis in \ref{defv} gives
$$\la \uu(uq^k,v)|\vv(vq^m, u)\ra=\frac{1}{\sqrt{N}}q^{km}.$$ 
\epk
 \bpk
 We call each instance of $\VV_A(\alpha)$ endowed with an $U$-frame $\EE^U_A(\alpha)$  {\bf an algebraic-Hilbert space} and denote it $\bV_A(\alpha).$ Note $\EE^U_A(\alpha)$ gives rise to the unique inner product on  $\F_0[\EE^U_A(\alpha)].$

\epk
\bpk \label{FC}
{\bf Remark.} As in the remark in \ref{F0} assume $\F=\C.$ Then we may use the same canonical basis to introduce an inner product on
$\C[\EE^U_A(\alpha)]=\VV_A(\alpha).$  This gives $\VV_A(\alpha)$
 a structure of a Hilbert space which extends the algebraic-Hilbert space structure on $\F_0[\EE^U_A(\alpha)].$
 
\epk

\bpk We call $S\in  A(a_1,\ldots,a_n,b_1,\ldots,b_n)$ a {\bf pseudo-unitary operator} if $S\in H(a_1,\ldots,a_n,b_1,\ldots,b_n). $ We set a {\bf formal adjoint of $S$} to be $S\inv.$ 

\epk
\bpk We call $P\in  A(a_1,\ldots,a_n,b_1,\ldots,b_n)$ a {\bf pseudo-selfadjoint operator} if $$P=\sum_{i=1}^n a_iS_i,\ \mbox{ some pseudo-unitary }S_i\mbox{ and } 
a_i\in \F_0$$
such that  $$P=P^*:=\sum_{i=1}^n a_i^*S_i^*,$$
for formal conjugates $a_i^*$ and adjoints $S_i^*$ of $a_i$ and $S_i$ respectively.

 $a_1,\ldots,a_n\in k.$
\epk

\bpk \label{W-inner_pr}{\em Suppose $S\in A=A(a_1,\ldots,a_n,b_1,\ldots,b_n)$ is pseudo-unitary, $\alpha\in \spec_A$ and $s$ is an eigenvalue of $S$ acting on $\VV_A(\alpha).$
Then inner product in $\F_0[\EE^U_A(\alpha)]$  is invariant under the action of $s\inv S,$ that is for any $f,g\in \F_0[\EE^U_A(\alpha)]$
$$\la s\inv Sf| s\inv Sg\ra_\alpha=\la f|g\ra_\alpha.$$

In particular, the matrix of $s\inv S$ in a canonical $U$-basis is $\F_0$-orthogonal. }

{\bf Proof.} We may assume without loss of generality that $n=1.$
By skew-linearity we just need to check the identity for $f,g$ belonging to a given canonical $U$-basis, so we may assume $f=\uu(uq^m,v)$ and $g=\uu(uq^k,v).$ Clearly, if the statement is true for $S=\check{U}$ and for $S=\check{V},$ then it is true for any group word of these, so is true for $S.$ The lemma follows. $\Box$
 \epk
 \bpk \label{principal} For each rational Weyl algebra $A$ one particular module
corresponding to the the ideal $\alpha=\la 1,1\ra=:\mathbf{1}$ (i.e. $u^N=1=v^N$) will play at the end a special role. We call the module $\VV_A(\mathbf{1})$ and the respective algebraic-Hilbert space  $\bV_A(\mathbf{1})$  {\bf the principal $A$-module } and {\bf the principal  algebraic-Hilbert space.}
  The main property of this module is that the eigenvalues of $U^a$ and $V^b$ are roots of unity.

\medskip

Again, the notions introduced above have natural extension to rational Weyl $n$-algebras.
\epk

 \bpk {\bf Proposition.} {\em Let $\bV_A(\mathbf{1})$ be the principal algebraic-Hibert space, $S$ a pseudo-unitary and $P$ a pseudo-selfadjoint operators.
Then \begin{itemize}
\item[(i)] there is an orthonormal basis of $S$-eigenvectors of the space;  
\item[(ii)] eigenvalues of $S$ are in $\F_0[N]$ for some $N$ (i.e. roots of unity);
\item[(iii)] there is an orthonormal basis of $P$-eigenvectors of the space;
\item[(iv)] eigenvalues of $P$ are totally real algebraic numbers.
\end{itemize}
 }

{\bf Proof.}  (i) and (ii) is immediate by \ref{W-inner_pr}. 

Note also that by the argument in \ref{W-inner_pr} $$\la Pf| g\ra= \la f| Pg\ra$$
for any $f,g\in \F_0[\EE_A(\mathbf{1})].$ Thus $P$ is a self-adjoint operator
in the (formal) Hilber space over field $\F_0\subset \C$ with complex conjugation. 
(iii) and (iv) follow. $\Box$ 
\epk
\bpk \label{W-basis} {\em Suppose $A=A(a,b),$
$C\in O(A)$ and $\mathbf{e}=\{ e_0,\ldots,e_{N-1}\}$ is a canonical $C$-basis of $\VV_A(\alpha)$ such that $\la e_0|e_0\ra=1.$ Then
the basis is orthonormal and the transition matrix from a canonical $U$-basis to $\mathbf{e}$ is $\F_0$-unitary.}

{\bf Proof.} By \ref{w} 
there are pseudo-unitary generators $S,T$ of $A$ such that $C$ is generated by a pseudo-unitary $S$ and the basis $\mathbf{e}$ is of the form $\{ \s(sq^m,t): m=0,\ldots,N-1\}$ satisfying (\ref{st}). Hence it is orthonormal.

By definition $S=q^l\check{U}^k\check{V}^m$ for some integer $l,k$ and $m$ such that
$k$ and $m$ are co-prime modulo $N.$ We may also assume without loss of generality that $l=0.$ Then by (\ref{UV1}) we may set $s=u^kv^m.$

Denote $$w_{n,p}=\la \s(sq^n,t)| \uu(uq^p,v)\ra,\ p=0,1,\ldots, N-1.$$ 
Apply $s\inv S$ to both parts of the pairing. Then by \ref{W-basis} we get
$$w_{n,p}=\la \s(sq^n,t)| \uu(uq^p,v)\ra = \la q^n\s(sq^n,t)| q^{(p-m)k}\uu(uq^{p-m},v)\ra=$$ $$=q^{(p-m)k-n}\la \s(sq^n,t)| \uu(uq^{p-m},v)\ra
=q^{(p-m)k-n}\cdot w_{n,p-m}.$$
Using the fact that $\check{U}=q^lS^cT^d,$ for some integer $l,c$ and $d,$
and applying $u\inv \check{U}$ to the initial pairing one gets similarly
$$w_{n,p}= q^{r}w_{o,p}$$
for some integer $r$ depending on $l,c,d,n,p$ and $o.$

We thus get an $\F_0$-linear homogeneous system of equation for the transition matrix $W=\{ w_{np}\}.$ We may thus assume that $W$ is an $\F_0$-matrix,so conjugation is applicable, and since it takes an $\F_0$-orthonormal basis  to  an $\F_0$-orthonormal basis, $W$ is $\F_0$-orthogonal. $\Box$
\epk 
\bpk {\bf Remark.} It is not hard to write down the matrix $W$ in terms of  $\F_0.$ 
\epk
\bpk \label{E^u} {\bf Corollary.} {\em The binary relation $e\in \EE^U_A(\alpha)$ (between $e$ and $\alpha$) is definable if and only if the $N+1$-ary relation  ``$\{ e_0,\ldots,e_{N-1}\}$ is a canonical orthonormal $U$-basis of $\VV_A(\alpha)$'' is definable.}
\epk
\bpk {\bf Definitions.} Let $C\in O(A).$ Set $\EE^C(\alpha)$ for each $\alpha\in \spec_A$ to be the set of $C$-eigenvectors in $\F_0[\EE^U_A(\alpha)]$ of norm $1.$ 
Set
$$\EE_A(\alpha):=\bigcup_{C\in O(A)}\EE^C_A(\alpha).$$

\epk
\bpk \label{E^C} {\bf Lemma.} {\em Let $C\in O(A)$ and suppose that the binary relation $e\in \EE^U_A(\alpha)$ is definable. 

Then the binary relation  $e\in \EE^C_A(\alpha)$ (between $e$ and $\alpha$) is definable. 
}

{\bf Proof.} 
By \ref{w} and \ref{W-basis} we can pick a canonical
$U$-basis $\uu(\alpha)$ in $\EE^U_A(\alpha),$ for each $\alpha\in \spec_A,$ and a $\F_0$ unitary matrix $R$ such that $\mathbf{e}(\alpha)=R\uu(\alpha)$ is a canonical
orthonormal $C$-basis. So, $\mathbf{e}(\alpha)\subseteq \EE^C_A(\alpha)$ for each $\alpha.$ By \ref{twobases} there is a finite group of $\F_0$-unitary transformations (independent on $\alpha$) such that any element in $\EE^C_A(\alpha)$ can be obtained by applying these transformations to  $\mathbf{e}(\alpha).$ $\Box$
\epk

\bpk \label{EUC}
{\bf Corollary.} {\em `` $e\in \EE_A(\alpha)$'' is definable if and only if `` $e\in \EE^U_A(\alpha)$'' is definable.}

\epk
\bpk \label{ass4+} {\bf Lemma-definition} {\em
For all $\alpha\in\spec_{A},$  the algebraic-Hilbert spaces $\bV_A(\alpha)=(\VV_A(\alpha), \EE_A(\alpha))$ are of the same isomorphism type.  We denote it  $$\bV_A=(\VV_A, \EE_A).$$    }

{\bf Proof.} Immediate by \ref{Gam-invar} and definitions. $\Box$

\epk

\bpk \label{selfadj} {\bf Lemma.} {\em Suppose $\F=\C$ and consider a Hilbert space structure on $\VV_A(\alpha)$  extending the algebraic-Hilbert space structure on $\F_0[\EE_A(\alpha)]$ (see \ref{FC}).

Let $C\in O(A),$ $C_\alpha$ be the image of $C$ in the representation  $A/\tilde{\alpha}$ of $A$ on $\VV_A(\alpha)$ and let $X\in C_\alpha.$
Then the adjoint $X^*_\alpha\in \C_\alpha.$ 
}

{\bf Proof.} By \ref{T1}(iii) in any canonical $C$-basis the algebra $C_\alpha$ is represented as the algebra of the diagonal complex matrices. This is closed under taking adjoints. $\Box$
\epk
\bpk \label{semisimple} {\bf Lemma.} {\em Let $B\subs A,$ $B\in \A_{fin}$ and let $B_\alpha\subs A/\tilde{\alpha}$ be the image of $B$ in $A/\tilde{\alpha}.$
Then $B_\alpha$ is a semi-sipmle algebra.}

{\bf Proof.} Without loss of generality we may assume that $\F=\C.$ Then each $C\in O(B)\subs O(A),$  $C_\alpha$ is closed under taking adjoints. Since  $B_\alpha$ is generated by all such $C_\alpha$ (see \ref{T1}(ii)), $B_\alpha$ is closed under taking adjoints, in particular the involution
$X_\alpha\mapsto X^*_\alpha$ is an anti-automorphism of the ring $B_\alpha.$ This implies that the Jacobson radical of $B_\alpha$ is closed under the involution.

It remains to invoke a well-known argument that under the above condition the Jacobson radical $J$ of the matrix algebra $B_\alpha$ is trivial, that is $B_\alpha$ is semisimple. Indeed, suppose $Y\in J.$ Then $Y^*Y\in J$ and  is nilpotent, that is $(Y^*Y)^k=0$ for some $k.$  By positive-semidefiniteness of $Y^*Y$, this implies $Y^*Y=0.$ So the inner product $\la Yv|Yv\ra=0,$ for any vector $v.$ Hence $Y=0.$ 
$\Box$

\epk
\bpk \label{1.5-}{\bf The bundle of algebraic-Hilbert $A$-modules}

We treat the family $$\{ (\bV_A(\alpha): \alpha\in \spec_A\}$$ of $A$-modules  as
 a bundle  $$\VV_A:=\bigcup\{ \VV_A(\alpha):\ \alpha\in \spec_A\}$$
with the 
projection map $\ev: \VV_A\to \spec_A,$ $\ev(v)= \alpha \mbox{ iff } v\in \VV_A(\alpha),$
the algebraic-Hilbert space structure and the $A$-module structure on each fibre.
 
Note that as a  bundle of algebraic-Hilbert spaces it is trivial, but as a bundle of $A$-modules, in general it can be highly non-trivial. See \cite{QZG} for a statement and a proof of the fact that the bundle of $A$-modules is not definable in $\F$ even when 
$\EE_A$  is omitted.  
\epk

\bpk \label{1.7} Let $A,B\in \A_{fin},$ $B\subs A.$
Hence $\VV_A(\alpha)$ can also be considered a  
$B$-module, in general, reducible. By \ref{semisimple} $\VV_A(\alpha)$ splits into a direct sum of irreducible $B_\alpha$-modules, so
irreducible $B$-modules. The number of the irreducible components is determined by dimensions $n_A$ and $n_B,$ so is equal to
$n_A:n_B.$ Invoking \ref{TS}  we obtain,
 $$\VV_A(\alpha)=\bigoplus_{\pi_{BA}(\beta)=\alpha}\VV_{AB}(\beta),\ \ \VV_{AB}(\beta)\cong \VV_B(\beta),$$
for uniquely determined $B$-submodules $\VV_{AB}(\beta)\subseteq \VV_A(\alpha)$ for 
each maximal ideal $\beta\subseteq Z(B),$ $\alpha\subseteq \beta.$ 

\epk
\bpk \label{3.45} We will describe the embeddings of \ref{1.7} in terms of the canonical bases.

Let as before $A$ be generated by $\check{U}=U^a$ and $\check{V}=V^b,$ $n_A=N$ and
 $\{ \uu(uq^{m},v): m=0,\ldots,N-1\}$ be a canonical orthonormal $\check{U}$ basis in a $A$-module $\VV_A(\alpha),$ where $\alpha$ is determined by $u^N$ and $v^N.$

Then $B\subset A$ is the algebra $\la \check{U}^n, \check{V}^k\ra$ generated by $\check{U}^n$ and $\check{V}^k$ for some $k,n\in \N.$ In fact, we may assume that $B=\la \check{U}^n,\check{V}\ra$ or  
$B=\la \check{U},\check{V}^n\ra,$ since any $B$ can be reached by such two steps.

In case $B=\la \check{U}^n,\check{V}\ra$ a $B$-submodules $\VV_{AB}(\beta)$ can be constructed by
choosing an $\check{U}^n$-eigenvector $\uu^{n,1}(u^n,v)$ of modulus $1$ and then generating
a whole canonical $\check{U}^n$-basis by applying $\check{V}$ according to (\ref{UV1}). Since 
$\check{U}^N$ acts on the module as a scalar, we may assume that $n$ divides $N.$

It is easy to see that a canonical $\check{V}$-basis in $\VV_{AB}(\beta)$ can be identified as follows:
\be\label{vv1} \vv^{1,n}(vq^{pn},u^n)=\vv(vq^{pn},u),\ p=0,\ldots, \frac{N}{n}-1,\ee
where on the right we have elements of the $\check{V}$-basis in
$\VV_A(\alpha).$

Set, for $\ell=0,\ldots, n-1,$ $ m=0,\ldots, \frac{N}{n}-1,$ $\xi=\e^\frac{2\pi i}{n},$
\be \label{uu1} \uu^{n,1}(u^nq^{mn},vq^\ell):=\frac{1}{\sqrt{n}}\sum_{k=0}^{n-1}\uu(uq^{m}\xi^k,vq^\ell)=
\frac{q^{m\ell}}{\sqrt{n}}\sum_{k=0}^{n-1}\xi^{k\ell}\uu(uq^{m}\xi^k,v)
\ee
The second equality follows from \ref{twobases}(ii).

One checks immediately that for a fixed $\ell$ this is a canonical orthonormal $\check{U}^n$-basis for the  
$B$-submodule corresponding to the maximal ideal $\beta_\ell$ of $Z(B)=\la \check{U}^N, 
\check{V}^\frac{N}{n}\ra $ generated by
$\check{U}^N-u^NI$ and $\check{V}^\frac{N}{n}-v^\frac{N\ell}{n}I.$ There are $n$ distinct such $\beta$ and respective non-isomorphic  $B$-modules.

\medskip

In case $B=\la \check{U},\check{V}^n\ra$ set, for $\ell=0,\ldots, n-1,$
\be \label{uu2} 
\uu^{1,n}(uq^{nm+\ell},v^n):=\uu(uq^{nm+\ell},v),\ m=0,\ldots, \frac{N}{n}-1\ee
This is  a canonical orthonormal $\check{U}^n$-basis for the  
$B$-submodule corresponding to the maximal ideal $\beta_\ell$ of $Z(B)=\la \check{U}^\frac{N}{n}, 
\check{V}^N\ra $ generated by
$\check{U}^\frac{N}{n}-u^\frac{N\ell}{n}I$ and $\check{V}^N-v^NI.$
\epk
\bpk \label{nk}{\bf Lemma.} {\em Suppose as above $A=\la \check{U}, \check{V}\ra,$ $B=\la \check{U}^n, \check{V}^k\ra.$ 
Let  $\VV_{AB}(\beta)\subset \VV_A(\alpha)$ be a $B$-submodule and $\{ \uu^B(u'q_B^{m},v'): 0\le m<n_B\}$ and $\{ \vv^B(v'q_B^{p},u): 0\le p<n_B\}$ be canonical $\check{U}^n$- and $\check{V}^k$-bases in $\VV_{AB}(\beta).$ Then
$$\la \uu^B(u'q_B^{m},v')|\vv^B(v'q_B^{p},u)\ra_{\bV_A(\alpha)}=\la \uu^B(u'q_B^{m},v')|\vv^B(v'q_B^{p},u)\ra_{\bV_{AB}(\beta)}$$
  In particular, the inner product is preserved in the embedding $\bV_{AB}(\beta)\to \bV_A(\alpha).$ } 
  
  {\bf Proof.} As above it is enough to check the case $k=1.$
Then $q_B=q^n,$ where now $n=n_A:n_B.$ We can rewrite $$\uu^B((u'q_B^{m},v')=
\uu^{1,n}(u^nq^{mn},v) $$
and 
$$\vv^B((u'q_B^{m},v')=
\vv^{n,1}(vq^{mn},u^n).$$  By \ref{exm_uv}
 $$\la \uu^B(u'q_B^{m},v')|\vv^B(v'q_B^{p},u')\ra_{\bV_{AB}(\beta)}=\frac{1}{\sqrt{n_B}}q_B^{mp}=\frac{1}{\sqrt{n_B}}q^{nmp}.$$
By (\ref{uu1}) and (\ref{vv1}) we can calculate
$$\la \uu^{1,n}(u^nq^{nm},v)|\vv^{n,1}(vq^{np},u^n)\ra_{\bV_A(\alpha)}=\frac{1}{\sqrt{n}}\sum_{k=0}^{n-1}
\la \uu(uq^{m+\frac{n_A}{n}k},v)| \vv(vq^{np},u)\ra=$$
$$=
\frac{1}{\sqrt{n}}\cdot n \frac{1}{\sqrt{n_A}} q^{nmp}=\frac{\sqrt{n}}{\sqrt{n_A}}\cdot q^{nmp}$$
(taking into account that $q$ is of order $n_A.$)

It remains to note that 

\epk
\bpk \label{1.10} \label{bfg} Keeping the notation and assumptions of \ref{1.7} we will consider
 $B$-module isomorphisms ${\rm p}^\beta_{BA},$ abbreviated to ${\rm p}^\beta,$
$${\rm p}^\beta: \VV_B(\beta)\to \VV_{AB}(\beta)\subset \VV_A(\alpha)$$ such that
$${\rm p}^\beta(\EE_B(\beta))\subs  \EE_A(\alpha)\cap \VV_{AB}(\beta).$$

In case $B=\la \check{U}^n,\check{V}\ra$ we define the maps on the bases, according to (\ref{uu1}), as
\be \label{uu3} p^\beta: \uu^{n,1}(u^nq^{mn},vq^\ell)\mapsto \frac{1}{\sqrt{n}}\sum_{k=0}^{n-1}\uu(uq^{m+\frac{N}{n}k},vq^\ell),\ m=0,\ldots, \frac{N}{n}-1,  \ee
where $\uu^{n,1}(u^nq^{mn},vq^\ell)$ is now naming elements of a canonical basis of $\VV_B(\beta).$

In case 
$B=\la \check{U},\check{V}^n\ra$
\be \label{uu4} p^\beta:
\uu^{1,n}(uq^{nm+\ell},v^n)\mapsto \uu(uq^{nm+\ell},v),\ m=0,\ldots, \frac{N}{n}-1\ee

Note that in terms of the $\check{V}$ bases (\ref{uu3}) can be rewritten as 

\be \label{vv3} p^\beta:
\vv^{n,1}(vq^{nm+\ell},u^n)\mapsto \vv(vq^{nm+\ell},u),\ m=0,\ldots, \frac{N}{n}-1\ee
(just use the symmetry between $\check{V}$ and $\check{U}$).
\medskip

Finally, we would like to remark that the definition of $p^\beta$ does depend on the choice of the canonical orthonormal bases in $\VV_B(\beta)$ and $\VV_A(\alpha).$
\epk

\bpk \label{lemmapb}
{\bf Lemma.} Let ${\rm p}^\beta$ and ${{\rm p}'}^\beta,$ $\beta\in \spec_B$ be two isomorphisms  $\VV_B(\beta)\to \VV_{AB}(\beta).$ Then there is 
a $g_\beta\in \F^*[n_B]$   such that
${\rm p}^\beta= {{\rm p}'}^\beta g_\beta.$ In particular, there are exactly $n_B$ choices for ${\rm p}^\beta$ for a given $\beta\in \spec_B.$

 {\bf Proof.} 
 Consider $({\rm p'}^\beta)\inv{\rm p}^\beta.$  This is a linear transformation commuting with operators from
 $B/\tilde{\beta}$, since ${\rm p}^\beta$ and ${{\rm p}'}^\beta$
 are $B$-module isomorphisms.
By the argument in \ref{Lemmag}(i) $({\rm p'}^\beta)\inv{\rm p}^\beta=g_\beta I$ for some $g_\beta\in \F^*[n_B].$
 $\Box$
\epk

\bpk\label{dfp} {\bf Definition} A map ${\rm p}^\beta$ satisfying \ref{bfg} will be called a {\bf local morphism at $\beta$} of $\bV_B\to \bV_A.$ 

We call the collection  ${\bf p}^\beta$ of all the local morphism at $\beta$ a fibre of a morphism $\bV_B\to \bV_A$ at $\beta.$ 

We call the family of fibres $${\bf p}_{BA}=\{ {\bf p}^\beta:\ \beta \in \spec_B\}$$
{\bf the morphism} $\bV_B\to \bV_A.$
\epk

\bpk \label{riso}{\bf Remark.} Lemma~\ref{lemmapb} implies  that $\bV_A$ is determined by $A$ uniquely up to isomorphism, that is given 
$\bV_A$ and $\bV'_A$ both satisfying the definition in \ref{1.10}, there is a family $${\rm p}^\alpha:\bV_A(\alpha)\to \bV'_A(\alpha), \ \alpha\in \spec_A$$
of local isomorphisms establishing an isomorphism between the two structures. 
  
\epk

\bpk Continuing the notation of
 \ref{1.7}, set
$$\Gamma_{AB}(\beta)=\{ \gamma\in \Gamma_A(\alpha): \gamma \VV_{AB}(\beta)=\VV_{AB}(\beta)\}.$$

Let $$\gamma\mapsto \check \gamma,\ \Gamma_{AB}(\beta)\to {\rm GL}\left(\VV_{B}(\beta)\right)$$
be the group homomorphism defined as follows. For $\gamma\in \Gamma_{AB}(\beta)$ set $\check\gamma$ to be the unique
$\A_{fin}$-transformation $\check\gamma$ of $\VV_{B}(\beta)$ such that $$\gamma p^\beta v=p^\beta\check\gamma\,v,\ \mbox{ for all } v\in \VV_B(\beta)$$
(the transformation on $\VV_{B}(\beta)$ induced from $\VV_{AB}(\beta)$ by the isomorphism $p^\beta$).

In general, $\det \check\gamma$ does not have to be equal to $1$ (in fact, $(\det \check\gamma)^{\frac{n_A}{n_B}}=1$), so we define
$$\pi^*_{BA}(\beta):\gamma \mapsto  (\det \check\gamma)^{-\frac{1}{n_B}}\cdot \check\gamma,\ \ \check\gamma= (p^\beta)\inv\gamma p^\beta$$  
a $1$-$n_B$-correspondence, a multivalued map into $\Gamma_B(\beta),$ or a homomorphism 
 $$\pi^*_{BA}(\beta): \Gamma_A(\alpha)\to \Gamma_B(\beta)/\F^*[n_B]\subs {\rm PGL}\left(\VV_{B}(\beta)\right),$$
 with the domain $\Gamma_{AB}(b)\subset \Gamma_A(\alpha).$ Here  $\F^*[n_B]$ is the group of roots of unity of order $n_B,$ which is also
the centre of $\Gamma_B(\beta).$ We write ${\rm P}\Gamma_B(\beta)$ for $\Gamma_B(\beta)/\F^*[n_B]$ below for any $B\in \A_{fin}.$

Note that by definition, for $\alpha\in \F^*[n_A],$ $\check{\alpha\gamma}=\alpha\check\gamma,$ so in fact we have equivalently defined
  $$\pi^*_{BA}(\beta): {\rm P}\Gamma_A(\alpha)\to {\rm P}\Gamma_B(\beta),$$
 a homomorphism of the projective groups.

\epk

\bpk {\bf Remark.} Recall that the action of $\Gamma_A(\alpha)$ on the algebra $A/\tilde{\alpha}$ reduces to the action of ${\rm P}\Gamma_A(\alpha)$ on
the  algebra. So, in this regard we do not lose any information switching to the projective groups.
\epk
\bpk \label{bV} {\bf $\bV_A$ as  structures in the language $\mathcal{L}_A$.}

For each $A\in \A_{fin}$ the bundle $\bV_A$ can be described as a two sorted structure in a language  $\mathcal{L}_A$. 

{\bf Sorts.} First sort, named $F$ will stand for (the universe of)
the field $\F.$ 

The language  $\mathcal{L}_A$  will have names for all Zariski closed subsets of $\F^n,$ all $n,$ including the ternary
relations for graphs of $+$ and $\cdot,$ the field operations. One of the closed subsets of $\F^m,$ where $m$ is the number of generators
of the affine algebra $A,$ will correspond to the spectrum $\spec_A=\mathrm{Spec}(Z(A))$ of the centre $Z(A)$ of $A:$  for each of the 
generating operators $X_1,\ldots,X_m$
of $Z(A)$ and for each maximal ideal $\alpha$ of $Z(A)$ we put in correspondence the $m$-tuple $\la x_1,\ldots,x_m\ra\in \F^m$ such that
$X_1-x_1I,\ldots,X_m-x_mI$ generate the ideal $\alpha.$ We call this closed subset $\spec_A.$

The second sort named $\VV$ will stand for $\VV_A,$ that is as a universe (set of points) it is the set  
$\bigcup\{ \VV_A(\alpha):\ \alpha\in \spec_A\}$ described in \ref{1.5-}. 

There is in $\mathcal{L}_A$ a symbol $\ev$ for a map $\ev: \VV_A\to \spec_A\subs \F^m$ which is interpreted as described in 
\ref{1.5-}. In particular, for each $\alpha\in \spec_A,$ $\ev\inv(a)=\VV_A(a)$ is definable.

{\bf Algebraic structure.} The language  $\mathcal{L}_A$  contains a binary relation $R_X(v,w)$ for each $X\in A,$ which is interpreted on
sort $\VV_A$ as $X\cdot v=w$ for $v,w\in \VV_A(\alpha),$ for $\alpha\in \spec_A.$ 

There is also a ternary relation $S(\xi,v,w)$ between $\xi\in \F,$ 
$v,w\in \VV_A(a)$ allowing us to say that $\xi\cdot v=w.$ 

A ternary relation $v+_Au=w$ on $\VV_A$ will be  saying that there exists $\alpha$ such that $u,v,w\in \VV_A(\alpha),$ and that $v+u=w$ 
in the sense of the module structure on $\VV_A(\alpha)$.

{\bf Canonical bases.}
$\mathcal{L}_A$  contains a binary symbol $E^S,$ for each pseudo-unitary $S\in A,$
which distinguishes in  $\VV_A$ a relation 
 $$E^S(e,\alpha)\equiv \ \ e\in \EE^S_A(\alpha)\ \& \  \alpha\in \spec_{A}$$ 
 as well as the binary symbols $\mathrm{IP}_r,$ for non-zero $r\in \F_0,$ which are
interpreted on $\VV_A$
as $$\mathrm{IP}_r(e_1,e_2)\equiv\ \  \exists \alpha\in \spec_A\ e_1,e_2\in \EE_A(\alpha)\ \& \ \la e_1|e_2\ra_\alpha=r,$$ the value of the {\bf inner product} (see \ref{EA}). Here $\EE_A(\alpha)$ stands for 
$\bigcup \{ \EE^S_A(\alpha): S\in A\ \mbox{ pseudo-unitary}\}.$   This is definable in $\mathcal{L}_A$ since there are only finitely many $\EE^S_A$ for each $A.$

\epk
\bpk {\bf Lemma.} {\em {\rm (i)} The binary relation `` $e\in \EE_A(\alpha)$'' is definable in the language $\mathcal{L}_A$.

{\rm (ii)} The $N+1$-ary  relation `` $\{ e_0,\ldots,e_{N-1}\}$  is a canonical orthonormal $U$-basis of $\bV_A(\alpha) $'' is definable in the language $\mathcal{L}_A$.

}

{\bf Proof.} (i) is immediate by \ref{EUC}. 
(ii) is by \ref{E^u}.
$\Box$

\medskip

Note  that defining $e\in \EE_A(\alpha)$ from $e\in \EE^U_A(\alpha)$ may require formulas which depend on $A.$
\epk
\bpk {\bf Remark.} The language  $\mathcal{L}_A$ has no symbols for $\Gamma_A$ or group actions. The  group plays a role in the theory of
$\bV_A$ through  axioms describing canonical bases in $\EE_A(a)$ and the matrices of linear transformations from one base to another, see  subsection 2.1 of \cite{QZG}.
\epk

\bpk \label{almost} {\bf Definition.} Let $\mathbf{N}$ and $\M$ be structures in languages $L_N$ and $L_M,$ correspondingly.
We say that a structure {\bf $\mathbf{N}$ is almost interpretable (definable) in  $\M$} if,

(i) there is a structure $\mathbf{N}^0$ in a language $L^0_N,$ which is definable in  $\mathbf{N}$ and  interpretable in $\M;$ 

(ii) $\mathbf{N}$ is prime over $\mathbf{N}^0$ (in particular, any automorphism of $\mathbf{N}^0$
can be lifted to an automorphism of $\mathbf{N}$);

(iii) $N\subs \acl_\mathbf{N}(N^0),$ that is the universe $N$ of $\mathbf{N}$ is in the algebraic closure of $N^0$, in the sense of $\mathbf{N}.$

\medskip

{\bf Remark.} \cite{HZ}, \cite{QZG} and \cite{QHO} discuss examples of Zariski structures which are almost interpretable but not interpretable in an algebraically closed field $\F.$ 

\epk
\bpk \label{Th1} {\bf Theorem.} {\em Given  an  $A\in \A_{fin},$ 
 
1. $\bV_A$ is almost interpretable in the field $\F,$ in particular,  
the first order theory of the structure $\bV_A$ is categorical in 
uncountable cardinalities. 

2. $\bV_A$ is a Zariski geometry in the  topology given by the positive formulas with
all quantifiers restricted to the predicate $E.$ 

} \smallskip

{\bf Proof.} 1.  The field  $\F$ is definable in $\bV_A$ as  the sort $F$ by definition, see \ref{bV}. It is also clear by the construction that $\acl_{\bV_A}(F)=V_A.$
Theorem A of \cite{QZG} proves that $\bV_A$ is categorical in 
uncountable cardinalities and that $\bV_A$ is prime over $\F.$

2. This is a special case of Theorem B of  \cite{QZG}. 
$\Box$
\epk

\bpk \label{Th2-}{\bf Corollary.} 
{\em Given $\A_{fin},$  $A,B\in \A_{fin},$ 
 $\bV_B$ is almost interpretable in $\bV_A.$ .}
\epk
\bpk \label{Th2}{\bf Proposition.} {\em 
Assume $B\subset A.$ Then $\bV_B$  is definable in $\bV_A.$}

We may  assume that $A$ is the Weyl algebra generated by $\check{U}$ and $\check{V},$ $N=n_A,$ and
$B$ is generated by $\check{U}^n$ and $\check{V}^m,$ $n,m$ divide $N.$  Let $\la \check{u},\check{v}\ra\in \F^2$ define $\alpha$ as in \ref{uv}.  We deduce that   $\check{u}=u^N, \ \check{v}=v^N,$ for some
$u,v\in \F^\times$ (following notations of \ref{uuvv}) and $\beta$ is determined by
$\la u^\frac{N}{m}, v^\frac{N}{n}\ra.$
 

 The subbundle $\{ \VV_{AB}(\beta): \beta\in \spec_B\},$ of the bundle 
 $\{ \VV_A(\alpha):\alpha\in \spec_A\}$ as $ \alpha=\pi_{BA}(\beta)$
 is defined 
by the formula
$$ ``\mathbf{w}\in \VV_B(u^\frac{N}{m}, v^\frac{N}{n}) ":=  \check{U}^\frac{N}{m}\mathbf{w}=u^\frac{N}{m}\mathbf{w}\ 
\& \
 \check{V}^\frac{N}{n}\mathbf{w}=v^\frac{N}{n}\mathbf{w}.$$

Since $\VV_A(\alpha)$ are $B$-modules, the $ \VV_{AB}(\beta)$ get the structure of $B$-modules too.

Finally,  we define $$ \EE_{AB}(\beta):=\EE_A(\alpha)\cap \VV_{AB}(\beta) \mbox{ for }\alpha=\pi_{BA}(\beta).  $$

Thus we  get  the bundle of algebraic-Hilbert $B$-modules on  $\{ \VV_{AB}(\beta): \beta\in \spec_B\}.$

This is canonically (but not uniquely) isomorphic to the bundle of algebraic-Hilbert $B$-modules on  $\{ \VV_{B}(\beta): \beta\in \spec_B\}$ by fibrewise isomorphisms 
$\mathbf{p}_{BA}^\beta$ between $\VV_B(\beta)$ and the corresponding submodules 
of $\VV_{AB}(\beta)$  as defined in \ref{1.10}.
 $\Box$

\epk

\bpk {\bf Remark.} The image of a 0-definable relation on 
$\VV_B$ under ${\bf p}_{BA}$ does not depend on the choice of a representative in  ${\bf p}_{BA}$
since definable relations are invariant under isomorphisms.

Moreover,  ${\bf p}_{BA}$ sends Zariski-closed relations on $\VV_B$ to Zariski-closed ones on $\VV_A,$  that is this
 is a Zariski-continuous Zariski-closed morphism. 
\epk
\bpk \label{Iso1} { \bf Proposition.} {\em The categories $\A_{fin}(\F)$ and $\V_{fin}(\F)$ are isomorphic.
In other words, the functor
$$A\mapsto \bV_A$$
is invertible.
}

{\bf Proof.} In fact, even the weaker functor $A\mapsto \VV_A$ is invertible. We only need to show how to recover $A$ from  $\VV_A.$ Since $\VV_A$ has the structure of a bundle of $A$-modules, we are already given the action of $A$ on each module $\VV_A(\alpha).$ It remains to see that the annulator of
all the modules  $\VV_A(\alpha),$ $\alpha\in \spec_A,$ is trivial. This is immediate from the classification of irreducible representations of $A,$ see \cite{BG}. $\Box$
\epk
\section{The category $\V_{fin}$ and the sheaf on $\A_{fin}$}\label{Cat}

\bpk {\bf Category $\V_{fin}.$} The category $\V_{fin}$ consists of objects $\bV_A,$ $A\in \A_{fin},$ and morphisms ${\bf p}_{BA}: \bV_B\to \bV_A,$
 $B\subseteq A.$ 
 
 Note that ${\bf p}_{BA}$ determines both  $$\pi_{BA}:\spec_B\to \spec_A$$  and $$\pi^*_{BA}:{\rm P}\Gamma_A\to {\rm P}\Gamma_B.$$ 

It is immediate by definition that the functor  $A\to\bV_A$ from $\A_{fin}$ to $\V_{fin}$ is a  presheaf on the category $\A_{fin}$ and so a  presheaf on the category $\mathcal{C}_{fin}^{op}.$

We use the same name $\V_{fin}$ for the functor.

\epk
\bpk
{\bf Remark.} 
We have  associated with an $A\in \A_{fin}$ the (finite) set
$O(A)$ of maximal  $\mathcal{C}_{fin}$-algebras. 

It is appropriate to think about the collection $O(\A_{fin})$ of all maximal $C\in \mathcal{C}_{fin}$ as a Grothendieck site with subsets $O(A)$ 
providing ''open subsets`` of the Grothendieck topology. Any finite covering will be considered admissible. 

\epk
\bpk \label{rem2.3} {\bf Remark.} Given $A, B\in \A_{fin},$ $B\subs A$ there is a unique morphism $${\bf p}_{BA}: \bV_B\to \bV_A.$$ 
This follows from Lemma~\ref{lemmapb}.
\epk
\bpk {\bf Theorem.} {\em $\V_{fin}$ is a sheaf  over the site $O(\A_{fin}).$}

\smallskip

{\bf Proof.} Let $B_1,\ldots,B_k\in \A_{fin}$ and 
$\bV_{B_1},\ldots,\bV_{B_k},$ the corresponding objects of the presheaf. By the remark above the morphisms between the object
are uniquely determined. By \ref{ass3} and \ref{1.1old} there is $A\in \A_{fin},$ $Z(A)\in \mathcal{C}_{fin}$ such $Z(A)=Z(B_1)\cap\ldots\cap Z(B_k),$ equivalently, $O(A)=O(B_1)\cup\ldots \cup O(B_k).$

We need to prove that there is a unique object $\bV_A$ in the category with the corresponding morphisms 
${\bf p}_{B_iA}:\bV_{B_i}\to \bV_A.$ For this just take the uniquely defined $\bV_A$ and the unique morphisms  ${\bf p}_{B_iA}.$
$\Box$
\epk

\bpk {\bf Sheaf $\V_{fin}$ as a structure.}
We define the structure $\V_{fin}$ as a multi-sorted structure with sorts $\bV_A,$ for each $A\in \A_{fin},$ in the language  $\mathcal{L}_A$
expanded by unary predicates for each of the sorts $\bV_{A}.$

In this definition, each sort $\bV_A$ is a multi-sorted structure, and it is assumed that all the sorts have a common field $\F.$
\epk
\bpk \label{Th3} {\bf Theorem.} 

(i) {\em The theory of $\V_{fin}$ is categorical in uncountable cardinalities. 
Moreover, every two models over a field $\F$
are isomorphic over $\F.$}

(ii) {\em For any $A,B\in \A_{fin},$ $B\subs A,$ the morphism ${\bf p}_{BA}$ is definable. More precisely,
there is a definable sort ${\bf P}_{BA}$ and a definable relation 
$$P_{BA}\subs {\bf P}_{BA}\times \spec_B\times \VV_B\times \VV_A$$ 
 in $\V_{fin},$ such that for every $p\in {\bf P}_{BA}$ and  $\beta\in \spec_{B}$ the binary relation on $\VV_B\times \VV_A,$
 $$P_{BA}(p,\beta,v_1,v_2)$$
 is the graph of a local morphism ${\rm p}^\beta_{BA},$ and every local morphism can be obtained in this way.   }

{\bf Proof.} (i) Immediate from \ref{Th1}. 

(ii) By construction.
$\Box$
\epk
\bpk \label{glu} {\bf A symmetric pairing.}
Given $B,D\in \A_{fin}$ 
we introduce a pairing $$[\cdot\ |\ \cdot]:\ \EE_B\times \EE_D\to \R_0.$$
First, we recall that the algebra $A=\la B\cup D\ra$ is in $\A_{fin}$ and so there are morphisms ${\bf p}_{BA}:\bV_B\to \bV_A$ and  
${\bf p}_{DA}:\bV_D\to \bV_A$ as defined in \ref{dfp}.

Let $\beta\in \spec_B,\ \delta\in \spec_D.$ For  $e\in \EE_B(\beta)$ and $f\in \EE_D(\delta)$ set
$$[e|f]=\left\lbrace\begin{array}{ll}|\la {\bf p}^\beta(e)|{\bf p}^\delta(f)\ra_\alpha|^2,\mbox{ if }\pi_{BA}(\beta)=\alpha=\pi_{DA}(\delta)\\
0,\mbox{ otherwise}\end{array}\right.$$
(the absolute value of the inner product).
\epk
\bpk \label{well-glu}{\bf Lemma.} {\em The pairing $[e|f]$ is well-defined and does not depend on the choice of  ${\bf p}^\beta$ and ${\bf p}^\delta.$

The pairing is symmetric.

$$[s e|f]= |s|^2\cdot[e|f]\mbox{ and } [e|t f]=|t|^2\cdot[e|f]$$
for any $s,t\in \F_0,$ roots of unity of orders $n_B$ and $n_D$ respectively.

$$ [e|f]=s\cdot s^*\mbox{ for }s=\la {\bf p}^\beta(e)|{\bf p}^\delta(f)\ra_\alpha$$
}

{\bf Proof.} 
Note that if $\pi_{BA}(\beta)=\alpha= \pi_{DA}(\delta),$ we will have ${\bf p}^\beta(e)\in \EE_A(\alpha),$ ${\bf p}^\delta(f)\in \EE_A(\alpha)$
and so the inner product $\la {\bf p}^\beta(e)|{\bf p}^\delta(f)\ra_a$ is defined. 

It does not depend on  ${\bf p}^\beta$ and ${\bf p}^\delta$ by \ref{lemmapb}. 

Clearly, 
$$[e|f]=[f|e].$$

The last two statement are also by definition.
$\Box$

\epk
\bpk {\bf Commentaries.} (i) The physicists interpretation of the pairing $[\cdot\ |\ \cdot ]$ would be ``{\bf probability}'', whereas the inner product would correspond to ``amplitude''. We have means to define amplitude locally, on each module $\VV_A(\alpha),$ but globally we have to use probability because of non-uniqueness in the choice of embedding ${\bf p}^\beta.$ The latter forced upon us if we want uniqueness for our whole construction (Theorem~\ref{Th3}), in fact, by model-theoretic stability assumptions.

(ii) A geometric interpretation of the pairing  should be that it provides a {\bf gluing correspondence} between $\EE_B$ and $\EE_D,$ and more specifically between $\EE_B^S$ and $\EE_D^T$ for chosen pseudo-unitary $S$ and $T$ (equivalently, maximal commutative subalgebras).  Namely, the relation between $e\in \EE_B^S$ and $f\in \EE_D^T$ given by
the definable  condition $[e|f]\neq 0$ is a finite-to-finite correspondense. 

Note that the correspondence $[e|f]\neq 0$ is Zariski closed, which is easy to work out from definitions.
\epk

\bpk \label{GlCbasis}

We call {\bf an orthonormal canonical basis of $\bV_A$} a collection $$\mathbf{e}\subs \bigcup\{ \EE_A(\alpha):\alpha\in \spec_A\}$$ with the property
that $\mathbf{e}\cap \EE_A(\alpha)$ is an orthonormal canonical $C$-basis in $\bV_A(\alpha)$ for some $C\in O(A).$
  
\epk
\bpk
{\bf Lemma.} {\em Let $B,D\in \A_{fin}$ and  $\mathbf{e}$ be an orthonormal basis of $\bV_B.$
Then, for any $\delta\in \spec_D$ and $f\in \EE_D(\delta),$ 
\begin{itemize}

\item[(i)]
 there are finitely many $e\in \mathbf{e}$ such that $[e|f]\neq 0$ and

\item[(ii) ]$\ \ \  \sum_{e\in \mathbf{e}}\ [e|f]=1 .$
\end{itemize}}

{\bf Proof.} Note that $\mathbf{g}:={\bf p}_{BA}(\mathbf{e})$ is an orthonormal basis of $\bV_A,$ more precisely,
 $$\mathbf{g}\subset \bigcup\{ \VV_{AB}(\alpha): \alpha\in \spec_A\}.$$
 Now (i) follows by definition. 
 
 For (ii) note that, assuming $\alpha=\pi_{DA}(\delta)$ and using the fact that $ \mathrm{p}^\delta(f)\in \EE_A(\alpha),$ 
$$\sum_{e\in \mathbf{e}}\ [e|f]=\sum_{g\in \mathbf{g}}\ |\la g|\mathrm{p}^d(f)\ra_a|^2=\sum_{g\in \mathbf{g}\cap \EE_A(a)}\ \la \mathrm{p}^d(f)|g\ra_a\cdot \la g|\mathrm{p}^d(f)\ra_a
=\la \mathrm{p}^d(f)|\mathrm{p}^d(f)\ra_a=1.$$ $\Box$

\epk

\section{The categories $\A^*_{fin}$ and $\V^*_{fin}.$}
\bpk Define the category $\A^*_{fin}(\C)$ as consisting of algebras $A\in \A_{fin}(\C)$ (that is $\F=\C$) endowed with an involution
$X\mapsto X^*,$ which is defined on the pseudo-unitary operators by the rule
$$^*:\ U^aV^b\mapsto V^{-b}U^{-a},$$
and on $\C$ is defined as complex conjugation. 
The morphisms of  $\A^*_{fin}(\C)$ will be the same as of $\A_{fin}(\C),$ which of course preserve the involution by definition.

Given $A\in \A_{fin}(\C),$ we will write $A_*$ for the algebra in $ \A^*_{fin}(\C)$ obtained by introducing the involution. We consider the functor 
$$F_\mathcal{A}: A\mapsto A_*; \ \  \A_{fin}(\C)\to  \A^*_{fin}(\C).$$
\epk
\bpk \label{L1} {\bf Lemma.} {\em The functor $F_\mathcal{A}$ is an isomorphism of
the categories $\A_{fin}(\C)$ and $\A^*_{fin}(\C).$ }

{\bf Proof.} Immediate. $\Box$
\epk
\bpk
The objects of the category  $\V^*_{fin}(\C)$ will be defined as the {\bf real-coordinate parts} of the Zariski 
geometries $\bV_A,$ for  $A=A(a,b)\in    A^*_{fin}(\C).$  We define the {\bf real-coordinate part} 
of $\spec_A$ 
to be
$$\spec_A(\R)=\{ \la u^N,v^N\ra:   |u|=|v|=1\}, $$
that is the eigenvalues of the generators $U^a$ and $V^b$ of the algebras on the module
$\bV_A(\alpha)$ are of modulus $1$ when $\alpha\in  \spec_A(\R).$

Note that for $\alpha\in \spec_A(\R),$ pseudo-unitary operators act in the algebraic-Hilbert module $\bV_A(\alpha)$ as unitary, if one considers  $\bV_A(\alpha)$ as a Hilbert space as described in  

Now define the object  $$\bV_A(\R)=\{ \bV_A(\alpha): \ \alpha\in \spec_A(\R)\}$$ 
as the bundle of Hilbert spaces $\bV_A(\alpha)$ as defined in \ref{FC}, with  the distinguished  bases $\EE_A(\alpha).$

Clearly, $\bV_A(\R)$ is a substructure of $\bV_A.$

The morphisms in the category $\V^*_{fin}(\C)$ are given by the same maps  as in $\V_{fin}(\C)$ restricted
to the substructures  $\bV_A(\R).$

We consider the functor of taking the real co-ordinate part
$$F_\mathcal{V}: \bV_A\mapsto \bV_A(\R);\ \  \V_{fin}(\C)\to \V_{fin}(\C)$$  
\epk
\bpk {\bf Lemma.} {\em $F_\mathcal{V}$ is an isomorphism of the categories
$\V_{fin}(\C)$ and $\V^*_{fin}(\C).$}

{\bf Proof.} By \ref{L1} and \ref{Iso1} it suffices to prove that the functor $$A\mapsto \bV_A(\R)$$
is invertible. 

 Suppose not. Then there is a non-zero operator $X\in A$ which annihilates all the modules 
$\bV_A(\alpha)$ with $\alpha\in \spec_A(\R).$  

Note that the set
$$\mathrm{Null}(X)=\{ \alpha\in \spec_A: \ \forall v\in \bV_A(\alpha)\, Xv=0\}$$
is definable in the Zariski structure $\VV_A.$  Moreover, it is  Zariski closed 
since 
$$\mathrm{Null}(X)=\{ \alpha\in \spec_A: \exists u,v\, \la u^N,v^N\ra=\alpha \ \& \
\bigwedge_{k=0}^{N-1}\   \uu(uq^k, v)\in \EE^U_A\ \& \         X\uu(uq^k, v)=0 \}$$
is given by a core-formula (see \cite{QZG}). 

Our assumption implies that  $\spec_A(\R)\subs \mathrm{Null}(X).$ But  the real 2-torus $\spec_A(\R)$ is Zariski dense in the complex 2-torus  $\spec_A=\C^\times \times \C^\times,$ so 
 $$\mathrm{Null}(X)=\spec_A$$
that is $X$ annihilates all the irreducible modules. Then $X=0.$ The contradiction. $\Box$
\epk
\bpk {\bf Model-theoretic commentary.}
Note that $\spec_A(\R)$ is a real 2-torus and as such can be defined in the field $\R.$ Also
each Hilbert space $\bV_A(\alpha)$  with the distinguished  bases $\EE_A(\alpha)$
is definable in $\R$ using parameters (needed to fix the frame, the family of distinguished bases). It can be shown that the bundle
$\bV_A(\R)$ is not interpretable in $\R$.  However, one can see that 
$\bV_A(\R)$ is prime over its definable substructure $\spec_A(\R).$

From applications points of view it is  category  $\A^*_{fin}(\R)$ and objects of the form $\bV_A(\R)$ rather than $\A_{fin}(\C)$ and $\bV_A(\C)$ which are of interest. However, there is a certain gain in viewing the  real structures as the  real part of the complex ones, and conversely, to see the complex structure as  ``complexifications'' of the real ones.  The property exhibited by our real structures is that their complexifications are nice Zariski geometries.

\epk

\section{Regular unitary transformations.}
In this section we work with a (temporarily) fixed Weyl algebra $A=A(a,b)$ generated by $\hat{U}=U^a$ and $\hat{V}=V^b.$ 

We denote, for $\alpha\in \spec_A,$
 
$\EE^U_N(\alpha):=$  the set of  $\uu=\langle e_0,\ldots,e_{N-1}\rangle,$   canonical orthonormal $\hat{U}$-bases of $\bV_A(\alpha),$ $N=N_A,$

 and denote $\EE^U_N:=\{ \{ \alpha\}\times \EE^U_N(\alpha): \alpha\in \spec_A\}.$


\bpk \label{defReg} Let $B=\la S,T\ra$ and $B'=\la S',T'\ra$ be subalgebras of $A,$ with $S,T$ and $S',T'$ pseudo-unitary generators. 

Suppose that 
\begin{itemize}
\item[(i)] There is an isomorphism 
 $$\sigma: B\to B';\ \ \ S^\sigma= S',\ T^\sigma= T'. $$

\item[(ii)]  
There is 
a Zariski  
closed relation $\Lambda\subset \EE^U_N\times \EE^U_N$
defining a finite-to-finite correspondence and there is a Zariski regular map $$\lambda: \Lambda\to  \EE^S_{N_B}\times \EE^{S'}_{N_B}; \  \la \alpha, \uu, \alpha', \uu'\ra \mapsto \la \beta, \s, \beta', \s'\ra,\
\alpha=\pi_{BA}(\beta),  \alpha'=\pi_{B'A}(\beta'); $$
$\lambda$ is
surjective on the co-ordinates $\la \beta, \s\ra \in \EE^S_{N_B}.$

This data
 for each choice of  $\la \alpha, \uu, \alpha', \uu'\ra$
 determines 
  the unique  
$\F$-linear transformation  $$L^{\uu,\uu'}_{\alpha,\alpha'}: \bV_B(\beta)\to  \bV_{B'}(\beta')$$
which sends  $ \s$ to $\s'. $
 
We  assume that the transformation 
preserves 
inner product and the action, that is: for any $e\in \bV_B(\beta),$
$X\in B$ and $g,f\in \s$ 
\be\label{Lpreserves} \la g| f\ra = \la L^{\uu,\uu'}_{\alpha, \alpha'}(g)|L^{\uu,\uu'}_{\alpha, \alpha'}(f)\ra\mbox{ and }     L^{\uu,\uu'}_{\alpha, \alpha'}(Xe)=X^\sigma L^{\uu,\uu'}_{\alpha, \alpha'}( e).\ee

\item[(iii)]   There is a positive integer $M$ such that, given that
$$ \la \alpha, \uu_1, \alpha', \uu'_1\ra \mapsto \la \beta, \s_1, \beta', \s'_1\ra\mbox{ and }  \la \alpha, \uu_2, \alpha', \uu'_2\ra \mapsto \la \beta, \s_2, \beta', \s'_2\ra,$$
 there is a 
  $\zeta\in \F,$  $\zeta^M=1$ and 

$$L^{\uu_1,\uu'_1}_{\alpha, \alpha'}(\mathbf{b})=\zeta\cdot L^{{\uu_2,\uu'_2}}_{\alpha, \alpha'}(\mathbf{b})$$
for every $\mathbf{b}\in \bV_B(\beta).$

\item[(iv)] For  $\la \alpha, \uu, \alpha', \uu'\ra\in \Lambda,$ if $\alpha=\mathbf{1}$ (that is $\bV_A(\alpha)$ is a principal algebraic-Hilbert space) then $\alpha'=\alpha$.
\end{itemize}
We call the family
$$L=\{  L^{\uu,\uu'}_{\alpha, \alpha'}:\ \bV_B(\beta)\to  \bV_{B'}(\beta'),\ \langle\alpha,\uu, \alpha', \uu'\rangle\in \Lambda\} $$
a {\bf regular unitary transformation of $\bV_A$ (associated with $B$ and $\sigma$)} and write  
 $$L: \bV_A\to \bV_A.$$
We call $\bV_B$ the {\bf domain } and $\bV_{B'}$ the {\bf range } of $L.$ 
\epk
\bpk {\bf Commentary.}  Later we will apply certain meaning of a limit and assume that the index $B$ in $A$ is small relative to the dimension of  the algebraic-Hilbert space $\bV_A(\alpha).$ This will allow a meaning to the notion that  
 ''$ \bV_B(\beta)$ is dense in  $ \bV_A(\alpha)$'' and that ''$L$ can be continuously extended from  $\bV_B(\beta)$ to $ \bV_A(\alpha)$''.

\medskip

 In the second half of the paper we will only
work with principal modules $\bV_A(\mathbf{1})$ and its submodules  $\bV_B(\mathbf{1})$ and $\bV_{B'}(\mathbf{1})$.
 

\epk
\bpk {\bf Lemma.} (i) {\em (\ref{Lpreserves}) can be extended to any $g,f\in \mathrm{span}_{\F_0}(\s).$ }

(ii) {\em  $\beta'$ is determined by $\beta.$}

{\bf Proof.} (i) is obvious by definition.

(ii) (\ref{Lpreserves}) implies that for $s,t\in \F_0$ we have
$$ L^{\uu,\uu'}_{\alpha, \alpha'}:\s(s,t)\mapsto \s'(s,t'),$$
where $(t')^{N_B}= t^{N_B}$ (and the same eigenvalues $s$ on both sides). But $\la s^{N_B}, t^{N_B}\ra$ is the invariant of the $B$-module, that is determines $\beta,$ and $\la s^{N_B}, (t')^{N_B}\ra$ is the invariant of the $B'$-module, tha is determines $\beta'.$ 

It follows that the set of $\beta\in \spec_B$ for which $\Lambda$ and $\lambda$ determine
$\beta'$ uniquely is Zariski dense in $\spec_B.$ Hence this is true for all $\beta.$
$\Box$

\epk
\bpk\label{LC} {\bf Lemma} {\em Suppose $C\subset B,$ a subalgebra generated by a pair of pesudo-unitaries, $C=\la X,Y\ra.$ Then
$\sigma$ of \ref{defReg}(i) acts on $C$ and we get $C'=C^\sigma= \la X',Y'\ra.$ Let $\sigma_C$ be the restriction of $\sigma$ to $C.$

We claim that
$L$ as given above is also  a regular unitary transformation of $\bV_A$ associated with $C$ and $\sigma_C.$}

{\bf Proof.}
We start with the same $\Lambda$ and need to construct a $$\lambda_C: \Lambda\to  \EE^X_{N_C}\times \EE^{X'}_{N_C}; \  \la \alpha, \uu, \alpha', \uu'\ra \mapsto \la \gamma, \x, \gamma', \x'\ra,\ \alpha=\pi_{CA}(\gamma),  \alpha'=\pi_{C'A}(\gamma'); $$

We do it by constructing a Zariski regular map  \be\label{M} \la \beta, \s, \beta', \s'\ra \mapsto \la \gamma, \x, \gamma', \x'\ra\ee
Recall that for $\bV_B(\beta)$ splits into the direct sum of irreducible $C$-submodules. Let $\gamma$ be the invariant for one of them, $\bV_C(\gamma)\subset \bV_B(\beta).$ Let
$\x$ be  a canonical $X$-basis of   $\bV_C(\gamma).$ Then there is a $N_B\times N_C$-matrix $\mathrm{M}$ over $\F_0$ such that $\x=\s \mathrm{M}.$
Since any two  irreducible $C$-submodules are conjugated by an automorphism of $\bV_B(\beta),$ 
as we ran through all canonical  $S$-bases $\s$ of  $ \bV_B(\beta),$ the $N_C$-tuple $\x=\s \mathrm{M}$
will run through all canonical $X$-basis for all irreducible  $C$-submodules.  Moreover, by definition $L$ will take $\s \mathrm{M}$ to $\s' \mathrm{M},$ and once $\s'$ is known the module $\bV_C(\gamma')$ and 
$\gamma'$ are determined. Thus  $\mathrm{M}$ determines the right-hand side of (\ref{M}) for the given $\beta$ and $\s.$ It remains to note that the same  matrix  $\mathrm{M}$ will define $\x,$ $\x',$ $\gamma$  and $\gamma'$ for any   $ \beta$ and $\s.$ $\Box$
\epk

\bpk \label{LformulaL} {\bf Lemma.} {\em
Let $\bar{\bV}_B(\beta)$ and $\bar{\bV}_{B'}(\beta')$ stand
for the quotients of  ${\bV}_B(\beta)$ and ${\bV}_{B'}(\beta'),$ respectively, by the equivalence relation 
$$g\approx g'\mbox{ iff } \exists \zeta\  \zeta^M=1\ \& \ g'=\zeta\, g.$$ 
Then the formula 
\be \label{formulaL}  L_{\alpha, \alpha'}(\bar{\mathbf{b}})=\bar{\mathbf{b}}':\equiv\ \ 
\exists{\uu,\uu'} \ \Lambda(\alpha,\uu,\alpha', \uu') \ \& \ L^{\uu,\uu'}_{\alpha, \alpha'}(\mathbf{b})=\mathbf{b}'\ee
determines for every $\beta\in \spec_B$   a well-defined map  
$$  \bar{\bV}_B(\beta)\to  \bar{\bV}_{B'}(\beta')$$
for a corresponding unique $\beta'.$}

{\bf Proof.} Immediate from (iii) of the definition.$\Box$
\epk

\bpk {\bf Lemma.} {\em A regular unitary transformation sends a canonical $S$-basis (with respect to $T$) to a canonical $S'$-basis (with respect to $T'$) preserving eigenvalues:
\be \label{Lsr} L^{\uu,\uu'}_{\alpha, \alpha'}: \s(sq^{ck},t)\mapsto \s'(sq^{ck},t),\ k=0,\ldots, N_B-1,\ee

were $c$ is the index of $B$ in $A$ and $N_B=\frac{N}{c}.$ 
}

{\bf Proof.} Let $\s(sq^{ck},t)$ be in an $S$-basis with respect to $T,$ that is satisfying (\ref{st}). 
Let $\s'_k=L^{\uu,\uu'}_{\alpha, \alpha'}(\s(sq^{ck},t)).$ 
Then by (ii) and (i) of the definition $\s'_k$ 
$S' \s'_k= sq^{ck}\s'_k,$ that is $\s'_k=sq^{ck}\s'(sq^{ck},t)$ for the $k$-th element of a canonical $S'$-basis $\{ \s'(sq^{cm},t): m=0,1,\ldots, N_B-1\}.$ Using (ii) and (\ref{st}) we get 
$$ L\left(\s(sq^{c(k-1)},t)\right)= L\left(t\inv T\s(sq^{ck},t)\right)=  t\inv T' \s'(sq^{ck},t)= \s'(sq^{c(k-1)},t),$$  
which extents the equality to all $k$ and proves (\ref{Lsr}). $\Box$

\epk



\bpk Let $L$ be regular on $\bV_A.$ Let  $\{ \s(sq^{ck},t), 0\le k<N_B\}$
and $\{ \s'(sq^{ck},t), 0\le k<N_B\}$ be as in (\ref{Lsr}).

We have
$$\s(sq^{ck},t)=\sum_{n=0}^{N-1} \lambda_{n,k}\cdot\uu_n$$
and 
$$\s'(sq^{ck},t)=\sum_{n=0}^{N-1} \lambda'_{n,k}\cdot\uu'_n$$
for some 
coefficients $\lambda_{n,k}, \lambda'_{n,k}\in \F_0$ which are determined by
$\lambda$ of \ref{defReg}, and since the latter is Zariski regular we have that $\lambda_{n,k}, \lambda'_{n,k}\in \F_0$ and do not depend on $\alpha, \alpha', \uu,\uu'.$

Now for arbitrary $\mathbf{b}\in \bV_B(\beta)$ we have 
$$\mathbf{b}=\sum_{k=0}^{N_B-1} b_k \s(sq^{ck},t)$$
for some $b_1,\ldots,b_k\in \F,$
and hence
$$L^{\uu,\uu'}_{\alpha, \alpha'}(\mathbf{b})=\mathbf{b}'=\sum_{k=0}^{N_B-1} b_k \s'(sq^{ck},t).$$

Or
$$\mathbf{b}=\sum_{k=0}^{N_B-1}\sum_{n=0}^{N-1} \lambda_{n,k}\cdot b_k\cdot \uu_k$$
and 
$$\mathbf{b}'=\sum_{k=0}^{N_B-1}\sum_{n=0}^{N-1} \lambda'_{n,k}\cdot b_k\cdot \uu'_k$$

Thus
\be \label{Luu}\begin{array}{ll} L^{\uu,\uu'}_{\alpha, \alpha'}( \mathbf{b})=\mathbf{b}' \Leftrightarrow\ \exists\, b_1,\ldots, b_k\in \F
\\
\mathbf{b}=\sum_{k=0}^{N_B-1}\sum_{n=0}^{N-1} \lambda_{n,k}\cdot b_k\cdot \uu_k\ \& \
\mathbf{b}'=\sum_{k=0}^{N_B-1}\sum_{n=0}^{N-1} \lambda'_{n,k}\cdot b_k\cdot \uu'_k
\end{array}
\ee

Finally,  $L_{\alpha, \alpha'}(\bar{\mathbf{b}})=\bar{\mathbf{b}}' $ of
 (\ref{formulaL}) becomes 
\be \label{L}\begin{array}{ll} \exists\, \uu \in \EE^U_N(\alpha)\ \exists\, \uu' \in \EE^U_N(\alpha')\  \exists\, b_1,\ldots, b_k\in \F:
\\
\mathbf{b}=\sum_{k=0}^{N_B-1}\sum_{n=0}^{N-1} \lambda_{n,k}\cdot b_k\cdot \uu_k\ \& \
\mathbf{b}'=\sum_{k=0}^{N_B-1}\sum_{n=0}^{N-1} \lambda'_{n,k}\cdot b_k\cdot \uu'_k 
\end{array}
\ee
\epk
\bpk \label{regZar} {\bf Theorem.} {\em Given a regular unitary transformation $L,$ 

(i) the formula
 $L_{\alpha, \alpha'}(\bar{\mathbf{b}})=\bar{\mathbf{b}}' $ defines  a family of maps  $$ L_{\alpha, \alpha'}:  \bar{\bV}_B(\beta)\to  \bar{\bV}_{B'}(\beta'),\ \  \bar{\bV}_B(\beta)\subseteq \bV_A(\alpha), \bar{\bV}_B(\beta')\subseteq \bV_A(\alpha').$$

(ii) 
The family  $L$
is  Zariski closed, that is the  relation
(between $\bar{\mathbf{b}},\bar{\mathbf{b}}', \alpha, \alpha',\beta,\beta'$) given by (\ref{L}) is Zariski closed.

(iii) There is an inverse regular transformation $L\inv$ associated to $B'$ and
$\sigma\inv.$ It
defines, in notations of (i) above,  the family of maps  $$ L_{\alpha', \alpha}\inv:  \bar{\bV}_{B'}(\beta')\to  \bar{\bV}_{B}(\beta)$$
such that the  composition $$L_{\alpha', \alpha}\inv\circ 
L_{\alpha, \alpha'}:\bar{\bV}_B(\beta)\to  \bar{\bV}_{B}(\beta)$$ defines a family of isomorphisms.

(iv) Let $L^{(1)}$ and $L^{(2)}$ be both regular unitary transformations on $\bV_A$ associated with $B_1,$ $\sigma_1$ and $B_2,$ $\sigma_2$ respectively and
$C\subset B_1$  be a subalgebra generated by pseudo-unitaries such that $C^{\sigma_1}\subseteq B_2.$ Let $\sigma_C$ be  the restriction of $\sigma_2\circ \sigma_1$ to $C.$

Then the composition $L^{(2)}\circ L^{(1)}$ given by the formula
$$\exists \alpha'\ \exists \mathbf{b'}\ L^{(1)}_{\alpha, \alpha'}(\bar{\mathbf{b}})=\bar{\mathbf{b}}' \ \& \ L^{(2)}_{\alpha', \alpha''}(\bar{\mathbf{b'}})=\bar{\mathbf{b}}''$$
defines a regular unitary transformation associated to $C$ and $\sigma_C.$.
} 

{\bf Proof.} (i) is just \ref{LformulaL}.

 (ii) follows from the fact that (\ref{formulaL}) in the form (\ref{L})  
 is a canonical form of a Zariski closed relation (a {\em core formula}) as defined in \cite{QZG}, section 3.

(iii) For $L\inv$ take $\Lambda\inv$ (inverting the order of 4-tuples) in place of $\Lambda.$ Define $\lambda_{-1}$ for $L\inv$ in place of $\lambda$ accordingly: 
$$\lambda_{-1}:\la \alpha', \uu', \alpha, \uu\ra \mapsto \la \beta', \s', \beta, \s\ra.$$
By definition this satisfies the required properties.
 
(iv) Follows by definition from \ref{LC}.
$\Box$
\epk

\bpk \label{onRegTr} {\bf Lemma.} {\em  To every regular $L: \bV_A\to \bV_A$ one can associate a rational $2\times 2$-matrix $g_L=(g_{lk})$ of determinant $1$ and an automorphism
of the  rational Heisenberg groups $\HH(\Q,\Q).$
} 

{\bf Proof.} Given $L,$ by definition we are given $S,T,S',T'\in \HH(a,b)$ and the isomorphism $\sigma$ between subgroups of $\HH(a,b)$ generated by $S,T$ and $S',T'.$ 
This isomorphism corresponds to an isomorphism of Lie subalgebras of the  Lie algebra $\la P,Q\ra$
over the rationals
with generators corresponding to  $S,T$ and $S',T'.$
 
This  can be naturally extended to the automorphism of the  Lie algebra $\la P,Q\ra$ which induces
an automorphism $$\check{\sigma}: \HH(\Q,\Q)\to \HH(\Q,\Q)$$
of the rational Heisenberg groups of rank 2. We thus will
have 
$$\check{\sigma}: \begin{array}{ll} S\mapsto S' \\
T\mapsto T' 
\end{array} ;        
 \begin{array}{ll} \hat{U}\mapsto c_U \hat{U}^{g_{11}}\hat{V}^{g_{12}}\\
\hat{V}\mapsto c_V \hat{U}^{g_{21}}\hat{V}^{g_{22}}
\end{array}
$$
for some rational matrix $(g_{kl})$ and $c_S,c_T,c_U,c_V\in Z(\HH(\Q,\Q)).$ 

The fact that $\check{\sigma}$ is an automorphism implies that
$\det (g_{kl})=1$ (see a remark in \ref{Aut}). $\Box$
\epk
\bpk {\bf Proposition.} {\em  To every regular $L: \bV_A\to \bV_A$ one can associate  an automorphism $\check{\sigma}_L$ of the category
$\A^{(1)}_{fin}$ such that for any non-zero rational $r,s,$
$$\check{\sigma}_L: \la \hat{U}^r, \hat{V}^s\ra \to \la q^{-\frac{r(r-1)}{2}} c_U \hat{U}^{rg_{11}}\hat{V}^{rg_{12}}, q^{-\frac{s(s-1)}{2}}c_V \hat{U}^{sg_{21}}\hat{V}^{sg_{22}}\ra.$$

The same determines an automorphism of the algebra $$\A^{(1)}=\bigcup_{A\in \A^{(1)}_{fin}}A.$$
} 

{\bf Proof.} This follows by  \ref{onRegTr}.
 The formula for $\check{\sigma}$  is obtained by  using the Campbell - Hausdorff formula for the Lie exponentiation. $\Box$
\epk
\bpk \label{ProblemMatrix} {\bf Problem.} Prove that,
conversely to \ref{onRegTr}, given $A,$ any matrix $(g_{kl})\in \mathrm{SL}(2,\Q)$ is associated  with a regular unitary transformation
$L$ of $\bV_A,$ provided that $A$ is of the form $A=\la U^\frac{1}{\mu},   V^\frac{1}{\nu}\ra,$ $\mu, \nu \in \Z$ and both divisible by the product of denumerators of rational numbers $g_{kl}.$  

Note that by \ref{Fourmx} and \ref{gaussex} below it suffices to prove that under the above assumptions the composition of regular unitary transformation is again a  regular unitary transformation. 

 
\epk

\bpk \label{Four} Let $q, N$ and  the invariants of $A$-modules be determined  in agreement with \ref{check}. In particular, $\alpha=\la 
u^N,v^N\ra$ and, conversely, $u,v$ are defined by $\alpha$ up to roots of unity of order $N.$
Denote $\alpha'=\la v^{-N}, u^{-N}\ra.$

{\bf The Fourier transform} $\Phi_A=\Phi: \bV_A(\alpha)\to \bV_A(\alpha')$ is defined on the ${A}$-modules 
 as  
 \be \label{four}\Phi : \uu(uq^{m},v)\mapsto \frac{1}{\sqrt{N}}\sum_{k=0}^{N-1} 
q^{-mk}\uu(v\inv q^{k},u\inv),\ee  
or, according to \ref{defv},
\be\label{four1} \Phi:   \uu(uq^{m},v)\mapsto \vv(u\inv q^{m},v\inv),\ee
and the domain and range of $\Phi$ is the whole of $\bV_A(\alpha),$ that is $B=A=B',$ $\beta=\alpha,$ $\beta'=\alpha'.$
  
Applying $\Phi$ twice we   get
$$\Phi^2:   \uu(uq^{m},v)\mapsto \uu(uq^{-m},v).$$ 

Note that $\Phi$ restricted to  the principal $A$-module is a transformation of the module $$\Phi: \bV_A(\mathbf{1})\to \bV_A(\mathbf{1}).$$
\epk
\bpk
{\bf Proposition.} \label{Fourmx} {\em $\Phi$ is a regular unitary transformation associated with the matrix
$$\left(\begin{array}{ll} 0\ \ \ 1\\ -1\  \ 0\end{array}\right)$$
and corresponds to the automorphism of the category

$$\hat{\sigma}:\ \ \begin{array}{ll} U^r\mapsto\ {V}^{r} \\ V^s \mapsto \ U^{-s}\end{array}.$$
}

{\bf Proof.} (i),(ii) and (iv)  of the definition \ref{defReg} are obvious. In order to prove (iii) we consider the effect of replacing $u$ by $uq$ and $v$ by $vq$ in the formula (\ref{four}).  Applying these operations several times one gets an arbitrary  case of (iii).

Substituting $u:=uq$ in (\ref{four}) and
using \ref{twobases} we get $ \uu(uq^{m+1},v)$ on the left of $\mapsto,$
and
$$ \frac{1}{\sqrt{N}}\sum_{k=0}^{N-1} 
q^{-mk}\uu(v\inv q^{k},u\inv q\inv)=  \frac{1}{\sqrt{N}}\sum_{k=0}^{N-1} 
q^{-mk}q^{-k}\xi \uu(v\inv q^{k},u\inv)=\xi\vv(u\inv q^{m+1},v\inv) $$
on the right, for some scalar $\xi.$ According to \ref{plustwobases} $\xi^N=1.$ This satisfies (iii).

Substituting $v:=vq$ we get $\uu(uq^m,vq)=\xi q^m\uu(uq^m,v)$ on the left, and 
$$\frac{1}{\sqrt{N}}\sum_{k=0}^{N-1} 
q^{-mk}\uu(v\inv q^{k-1},u\inv)=q^{-m} \vv(u\inv q^{m},v\inv) $$
on the right, which again agrees with the requirement (iv). $\Box$ 
\epk
\bpk \label{Gauss0} {\bf Gaussian transformation.}
We introduce a special transformation $\gau_A=\gau$ of the  bundle $\bV_{A}.$ We  set $q^\frac{1}{2}$ to be a root of order $2$ of $q.$  We assume $\alpha=\la 
u^N,v^N\ra$ and  $\alpha'=\la 
(uv)^N,v^N\ra$.

Set
\be \label{fe1g0}\gau:\,\uu(uq^{m},v)\mapsto \frac{\cc  }{\sqrt{N}}\sum_{\l=0}^{N-1} 
q^{\frac{(\l-m)^2}{2}}\uu(uv q^{\l},v), \ \ \ \bV_A(\alpha)\to \bV_A(\alpha')\ee
where $\cc  \in \F_0$ is a constant of modulus 1 to be determined later and $\uu(uq^{m},v)$ and $\uu(uv q^{\l},v)$ are assumed  to be from $\EE_A^U.$  By definition $\beta=\alpha$ and $\beta'=\alpha'.$

\epk
\bpk {\bf Lemma.} {\em $\gau$ is unitary. Moreover, $$\{ \gau\,\uu(uq^m,v): m=0,1,\ldots,N-1\}$$
is a canonical $S$-basis, for 
$$S:=q^{-\frac{1}{2}}\check U\check{V}\inv.$$
More precisely, for
$\s(uq^m,v):=\gau\,\uu(uq^m,v)$
we have 
\be \label{SV}\begin{array}{ll}
S: \s(uq^m,v)\mapsto uq^m \s(uq^m,v)\\
V: \s(uq^m,v)\mapsto v \s(uq^{m-1},v)
\end{array}\ee
and
\be \label{Sv0}S \gau \uu(uq^m,v)=uq^m\gau \uu(uq^m,v)=\gau\check{U}\uu(uq^m,v).\ee 
}

{\bf Proof.} (\ref{SV}) is by direct application of $S$ to (\ref{fe1g0}). (\ref{Sv0}) is a corollary of (\ref{SV}).

 Since $S$ is unitary by construction,
the basis is orthogonal, so Lemma follows. $\Box$

\epk
\bpk \label{P^20} We can calculate the action of $\gau$ on the basis $\{ \vv(vq^{n},u):\ n=0,1,\ldots,N-1\}.$

Substituting (\ref{fe1g0}) in  formula (\ref{defv}) for $\vv(vq^n,u)$ 
we get
$$\gau \vv(vq^n,u)=\frac{1}{\sqrt{N}}\sum_{m=0}^{N-1} 
q^{nm}\gau\uu(u q^{m},v)=\frac{\cc  }{N}\sum_{m=0}^{N-1} 
q^{nm}\sum_{\l=0}^{N-1}q^{\frac{(\l-m)^2}{2}}\uu(uv q^{\l},v)=$$ 
$$=\frac{\cc  }{N}\sum_{\l=0}^{N-1}\{\sum_{m=0}^{N-1} q^{\frac{(\l-m)^2}{2}+nm}\}\uu(uv q^{\l},v)$$
Note that $(\l-m)^2+2nm=[-n^2+2n\l]+(m+n-\l)^2.$ Hence we may continue the above
$$=\frac{\cc  }{N}\sum_{\l=0}^{N-1}q^{-\frac{n^2}{2}+n\l}\{\sum_{m=0}^{N-1} q^{\frac{(m+n-\l)^2}{2}}\}\uu(uvq^{\l},v)=
\frac{\cc  }{N}G(N)q^{-\frac{n^2}{2}}\sum_{\l=0}^{N-1}q^{n\l}\uu(uvq^{\l},v)$$
where $$G(N)=\sum_{m=0}^{N-1} q^{\frac{(m+n-\l)^2}{2}}=\sum_{m=0}^{N-1} q^{\frac{m^2}{2}}$$
is a {\em quadratic Gauss sum}, see \cite{IK}.
The last expression by (\ref{defv}) can be rewritten as 
$$=\frac{\cc  }{\sqrt{N}}G(N)q^{-\frac{n^2}{2}}\vv(vq^n,uv).$$
We have proven that  $$\gau: \vv(vq^n,u)\mapsto \frac{\cc  }{\sqrt{N}}G(N)q^{-\frac{n^2}{2}} \vv(vq^n,uv),$$ 
in particular,
 for $v=1$  the $ \vv(vq^n,u)$
 are eigenvectors of $\gau$ with eigenvalues $\frac{\cc  }{\sqrt{N}}G(N)q^{-\frac{n^2}{2}}.$
\epk 
\bpk \label{c0N0} We set $$\cc  =\frac{\sqrt{N}}{G(N)}.$$
By \ref{P^20} this is equivalent to the assumption that
\be \label{ga V0} \gau \vv(vq^n,u)=q^{-\frac{n^2}{2}}\vv(vq^n,uv)\ee
Moreover, reversing the calculation in \ref{P^20} we see that (\ref{ga V0}) is equivalent to (\ref{fe1g0}).
\epk
\bpk \label{4.6+} {\bf Proposition.} {\em Assume that $N$ is even. Then $$\cc  =\e^{-i\frac{\pi}{4}}.$$}

{\bf Proof.} Recall the Gauss formula, for integers $a$ and $L,$
\be \label{Gauss} \sum_{n=0}^{L-1}\e^{\pi i \frac{a}{L}n^2}=\sqrt{|\frac{L}{a}|}\cdot\e^{\pi i \frac{|aL|}{4aL}}\cdot\sum _{n=0}^{a-1}\e^{\pi i \frac{L}{a}n^2}, \ee
 see \cite{IK}. The statement follows, if we take $a=1$ and $L=N$ even.
 $\Box$
\epk
\bpk \label{gaussex} A general form of Gaussian transformation more useful  in applications is given by the associated matrix defined in \ref{onRegTr}, of the form
$$\left(\begin{array}{ll} 1\  -\frac{b}{d}\\ 0\ \ \ 1\end{array}\right)$$
for some positive integers $b$ and $d.$

This corresponds to 

$$B= \la \hat{U}^d, \hat{V}^b\ra=B'= \la \hat{U}^d\hat{V}^{-b}, \hat{V}^b\ra.$$

$$\hat{\sigma}:\ \ \begin{array}{ll} \hat{U}^d\to \hat{U}^d\hat{V}^{-b} \\ \hat{V}^b \to \ \ \hat{V}^b\end{array}$$

\epk
\bpk \label{Reg} {\bf Proposition.} {\em $\gau_B$ is a regular unitary
transformation.}

{\bf Proof.}
For simplicity (but without loss of generality) we consider the example with $b=1.$

We assume $A=\la \hat{U}, \hat{V}\ra$ and correspondingly 
$$B= \la \hat{U}^d, \hat{V}\ra=B'= \la \hat{U}^d\hat{V}\inv, \hat{V}\ra.$$

Let $\beta\in \spec_B$ and 
$\alpha$ be an element of $\spec_A$ such that $\bV_B(\beta)\subs \bV_A(\alpha).$ We may identify $\alpha$ with a pair 
$\la u^N, v^N\ra\in \F^\times \times \F^\times,$ and
according to \ref{1.7}-\ref{3.45}, $\beta=\la u^N,v^\frac{N}{d}\ra,$
so we have exactly $d$ distinct $B$-submodules in the $A$-module.

A canonical $\hat{U}^d$ basis of  $\bV_B(\beta)$ can be represented in the form
$$\{ \uu^{d,1}(u^d q^{dm},v): m=0,\ldots, \frac{N}{d}-1\}$$
defined by the formula (\ref{uu1}).  In this notation the definition (\ref{fe1g0}) becomes
 \be \label{fe1g2}\gau_B\uu^{d,1}(u^d q^{dm},v)=\cc \sqrt{\frac{d}{N}}\sum_{\l=0}^{\frac{N}{d}-1} 
q^{d\frac{(\l-m)^2}{2}}\uu^{d,1}(u^dv q^{d\l},v)=:\s(u^dq^{dm},v)\ee
 where $\s(u^dq^{dm},v)$ is an element of a canonical $\hat{U^d}\hat{V}\inv$-basis as described in (\ref{SV}).
 
 So, the formula (\ref{fe1g2}) defines, for the given choice of parameters $u$ and $v,$ the map
 $$\bV_B(\beta)\to \bV_{B'}(\beta'),\ \
 \beta'=\beta$$
 between the submodules of $\bV_A(\alpha).$ 
This verifies conditions (i),(ii) and (iii) of the definition \ref{defReg}. 

To check (iv) note that  $\beta(u,v)=\beta(u',v')$ if and only if $u'=uq^k$ and $v'=vq^{dn}$ for some $k,n\in \Z.$ 

Suppose $u=u'$ and consider the substitution $v'=vq^{dn}$ for $v$ in (\ref{fe1g2}).  
By
\ref{twobases}
$\uu^{d,1}(u^d q^{dm},vq^{dn})=c_1 q^{dnm} \uu^{d,1}(u^d q^{dm},v)$ and\\
$\uu^{d,1}(u^d vq^{dl},vq^{dn})=c_2 q^{dnl} \uu^{d,1}(u^dv q^{dl},v)$ with $c_1,c_2$ roots of unity of order $N$ (since the choice of the bases is within $\EE_B$). 
Write $\gau'_B$ for the transformation defined by
$\Lambda(\mathbf{b},\mathbf{b}', u',v').$

We will have
$$\gau'_B\uu^{d,1}(u^d q^{dm},vq^{dn})
:=
\cc \sqrt{\frac{d}{N}}\sum_{\l=0}^{\frac{N}{d}-1} 
q^{d\frac{(\l-m)^2}{2}}\uu^{d,1}(u^dv q^{d(l+n)},vq^{dn})$$
$$=c_2\cc \sqrt{\frac{d}{N}}\sum_{\l=0}^{\frac{N}{d}-1} 
q^{d\frac{(\l-m)^2}{2}}q^{dn(l+n)} \uu^{d,1}(u^dv q^{d\l},v)$$
and 
$$\gau'_B\uu^{d,1}(u^d q^{dm},vq^{dn})=
c_1 q^{dnm}\gau'_B  \uu^{d,1}(u^d q^{dm},v).$$
Hence
$$\gau'_B  \uu^{d,1}(u^d q^{dm},v)=c_1\inv c_2\cc \sqrt{\frac{d}{N}}\sum_{\l=0}^{\frac{N}{d}-1} 
q^{d\frac{(\l-m)^2}{2}}q^{dn(l+n-m)} \uu^{d,1}(u^dv q^{d(l+n)},v)=$$
$$=c_1\inv c_2q^{\frac{n^2}{2}} \cc \sqrt{\frac{d}{N}}\sum_{\l=0}^{\frac{N}{d}-1} 
q^{d\frac{(\l+n-m)^2}{2}}\uu^{d,1}(u^dv q^{d(l+n)},v)=c_1\inv c_2q^{\frac{n^2}{2}}=c_1\inv c_2q^{\frac{n^2}{2}}\s(u^d q^{dm},v).$$
Set $\zeta=c_1\inv c_2q^{\frac{n^2}{2}}.$ Clearly, $\zeta^{2N}=1.$ $\Box$ 
\epk

\bpk {\bf Remark.} If we only require the coefficient $\zeta$ to be an element of $\F_0$ of modulus $1,$ this can be shown by a simpler argument:

Note that the substitution $u:=u'$ and $v:=v'$ by \ref{twobases}
transforms the basis $\{ \uu^{d,1}(u^d q^{dl},v): 0\le l <\frac{N}{d}\}$ by the application of a transformation of the form $c\gamma,$ where $\gamma$ is a $\F_0$-unitary matrix and $c$ a non-zero constant. It follows that
the basis  $\{ \gau_B\uu^{d,1}(u^d q^{dl},v): 0\le l <\frac{N}{d}\}$ under the substitution transforms by the application of  a $\F_0$-unitary matrix $\gamma'$ to a basis 
$\{ \s'(u^d q^{dl},v): 0\le l <\frac{N}{d}\}$ satisfying 
(\ref{SV}). Now \ref{twobases} imply that $ \s'(u^d q^{dl},v)=\zeta \cdot \s(u^d q^{dl},v)$ for some $\zeta\in \F_0$ of modulus $1.$ $\Box$

\epk
\bpk \label{diag} Finally, we consider  {\bf diagonal transformations} $D_m$ on $\bV_A$ which correspond to the matrices of the form $$\left(\begin{array}{ll} m\  \ 0\\  0\ \ \frac{1}{m}\end{array}\right).$$

We assume that $N$ is divisible by $m.$

We construct a respective regular unitary transformation by considering the subalgebras $B=\la \hat{U},\hat{V}^m\ra$ and $B'=\la \hat{U}^m,\hat{V}\ra$ and the isomorphism 
$$\sigma: \hat{U}\mapsto \hat{U}^m,\ \ \hat{V}^m\mapsto \hat{V}.$$

A canonical $U$-base for a $B$-module is of the form
$$\{ \uu(uq^{mk},v): k=0,1\ldots, \frac{N}{m}-1\}$$
and that of $B'$
$$\{ \uu^m(u^mq^{mk},v): k=0,1\ldots, \frac{N}{m}-1\}$$
 
where by the definition  (\ref{uu1})
$$\uu^m(u^mq^{mk},v^m)=\frac{1}{\sqrt{m}}\sum_{l=0}^{m-1}  \uu(uq^{k}\zeta^l,v^m),$$
for $\zeta$ a root of $1$ of order $m,$ and respectively,
$$\begin{array}{ll} U^m:\uu^m(u^mq^{mk},v^m)\mapsto u^mq^{mk}\uu^m(uq^{mk},v^m),\\ V\ :\uu^m(u^mq^{mk},v^m)\mapsto v^m\uu^m(u^mq^{mk-m},v)\end{array}.$$

In terms of bases the transformation is
$$L^{u^m,u}_{\alpha,\alpha'}: \uu(u^mq^{mk},v)\to \uu^m(u^mq^{mk},v^m)$$
where for $\alpha=\la u^{mN},v^{N}\ra$ the correspondence $\Lambda$ determines $\alpha'=\la u^{N},v^{mN}\ra.$

\epk
\section{The limit coordinate system}

\bpk Consider  $\EE_A(\alpha)$ defined in \ref{bV} 
as the union of $\EE^S_A(\alpha),$
canonical  $\la S,R\ra$-bases (see
\ref{E^u}).

 We note first that each  $\EE^S_A(\alpha)$  is in a  natural way coordinatised by  the residue ring $\Z_{\slash N}.$  

The group $\Sigma_A(\alpha)=\langle \mu\ra$ of definable permutations of a basis $\{ \s(sq^n,r): n\le N-1\}$ can be realised as a group of 
$0$-defined operators 
  $R^n,$ or equivalently as pairs    $\s(sq^k,r)\mapsto \s(sq^{k+n},r)$ of elements of   $\EE^S_A(\alpha),$ up to the obvious equivalence.
 

Using parameters $s$ and $r$ 
the inner product formula
$$\langle \s(sq^{n_1},r)| \rr(rq^{m_1},s)\rangle=q^{n_1m_1}\cdot\langle \s(s,r)| \rr(r,s)\rangle=\langle \s(sq^{n_2},r)| \rr(rq^{m_2},s)\rangle
$$ 
allows to recognise the 4-ary relation
$$n_1m_1\equiv n_2m_2\mbox{ mod}\, N .$$ 
Since we can choose and fix $n_2=1$ we obtain 
the relation $n_1m_1=m_2.$ 
\epk
\bpk \label{mu}
Here we start the analysis of how the above coordinatisation behaves under {\em passing to an infinite limit,} which we formalise through considering a nonstandard model of integers and an infinite value of $N$ in it.

Let $\ZZ$ be the non-standard model of the ring of integers and
$\mu\in \ZZ$ divisible by all standard integers.
Let $N=\mu^2h\inv,$ for some $h\in \Q.$  Let 
$\bar{\R}=\R\cup \{ -\infty, +\infty\}$ and $\mathbb{S}=\R/\Z.$     
We construct  a surjective map
$$\mathrm{st}_\mu: \ZZ_{\slash N}\to \bar{\R}.$$

We identify an element $\bar{m}$ of  the residue ring  $\ZZ_{\slash N}$ with the 
unique (non-standard) integer $m\in [-\frac{N}{2}, \frac{N}{2})\cap \ZZ.$ Note that $\frac{m}{N}$ and $\frac{m}{\mu}$ are elements of $\QQ.$

Set $$\mathrm{st}_\mu: \bar{m}\mapsto \mathrm{st}(\frac{m}{\mu})$$
where $\mathrm{st}$ is the standard part map  $\QQ\to \bar{\R}.$ 

\pk
{\bf Lemma.} {\em
\begin{itemize}
\item[(i)] $\mathrm{st}_\mu$ is an additive (semi)group homomorphism;
\item[(ii)] $\mathrm{st}_\mu$ preserves the linear pre-order on $[-\frac{N}{2}, \frac{N}{2})\cap \ZZ,$ that is
$$m_1\le m_2 \Rightarrow  \mathrm{st}_\mu(m_1)\le \mathrm{st}_\mu(m_2);$$
\item[(iii)] if $m_1n_1=m_2n_2$ does hold in $\ZZ_{\slash N}$ 
and the images of $m_1,\ n_1,\ m_2$ and $n_2$ are finite
then
$$\mathrm{st}_\mu m_1\cdot\mathrm{st}_\mu n_1\equiv \mathrm{st}_\mu m_2\cdot \mathrm{st}_\mu n_2\,\mathrm{mod}\, \Z$$
\end{itemize}}

{\bf Proof.}
(i) and (ii) are obvious. 

(iii). We have the unique representation $m_1=a_1\mu +c_1,$ 
$m_2=a_2\mu +c_2,$
$n_1=b_1\mu +d_1,$
$n_1=b_2\mu +d_2,$ and $$|a_1|,|a_2|,|b_1|,|b_2|,c_1,c_2,d_1,d_2\in \ZZ\cap [0,\mu).$$

The assumption of finiteness implies that $a_1,a_2,b_1,b_2\in \Z$ (standard).
Now $$m_1n_1-m_2n_2=(a_1b_1-a_2b_2)\mu^2+(a_1d_1+b_1c_1-a_2b_2-b_2c_2)\mu +c_1d_1-c_2d_2  $$
Note that $a_1b_1-a_2b_2\in \Z,$
$|{a_1d_1+b_1c_1-a_2b_2-b_2c_2}|<k\mu$ for some positive $k\in \Z$ and $|c_1d_1-c_2d_2|< \mu^2.$
It follows that
 $$m_1n_1-m_2n_2\equiv 0\,\mbox{mod}\, \mu^2\ \mbox{ iff }\
 m_1n_1-m_2n_2= (a_1b_1-a_2b_2)\mu^2.$$
 Hence the condition in (ii) implies that 
 $$\frac{m_1}{\mu}\frac{n_1}{\mu}-\frac{m_2}{\mu}\frac{n_2}{\mu}=a_1b_1-a_2b_2\in \Z.$$
 It is now immediate from the definition  
$$\mathrm{st}_\mu m_1\cdot\mathrm{st}_\mu n_1- \mathrm{st}_\mu m_2\cdot \mathrm{st}_\mu n_2=a_1b_1-a_2b_2$$
thus the statement follows. $\Box$
\medskip

For $\alpha,\beta\in \ZZ$ let $|\alpha|<<|\beta|$ stand for ``$\frac{\alpha}{\beta}$ is an infinitesimal''.
And let  $\alpha<<\beta$ mean $\alpha<0\ \&\ |\beta|<<|\alpha|\ \vee\ \beta>0\ \& \ |\alpha|<< |\beta|.$

\pk \label{wring} {\bf Corollary.} {\em $\mathrm{st}_\mu$ is a homomorphism of the additive ordered  group $\{ m\in \ZZ: -\frac{\mu^2}{2}<< m<<\frac{\mu^2}{2}\}$ onto the additive ordered  group $\R$
  and $$m_1n_1=m_2n_2\Rightarrow \e^{2\pi i x_1y_1}= \e^{2\pi ix_2y_2}$$
for $x_1:=\mathrm{st}_\mu m_1,\ y_1:=\mathrm{st}_\mu n_1,$
$x_2:=\mathrm{st}_\mu m_2$ and $ y_2:=\mathrm{st}_\mu n_.$ 

In particular, the image of the partial ring structure on $\{ m\in \ZZ: 0\le m<\mu\}$ induces via $\mathrm{st}_\mu$ 
the partial addition and multiplication on the real interval $[0,1).$}
\epk
\bpk Definition. The {\bf weak ring} structure $\R_w$ on the reals $\R$
is given by the addition + and the ternary relation
$$P^4(x_1, y_1, x_2, y_2) :\equiv\ \e^{2\pi i x_1y_1} = \e^{2\pi i x_2y_2}.$$

The compactified weak ring structure $\bar\R_w$ on the compactified
reals $\bar{\R} = \R ∪ \{ -\infty, +\infty\}$ is given by the extension of + to $\bar{\R}$ in
the natural way as the ternary relation 
$S^3(x, y, z)$ (in particular,
$S^3(-\infty, +\infty, z)$ holds for any $z ∈ \bar{\R}$) and the extension of $P^4$ to $\bar{\R}$
defined by the condition
$$\bar{\R}_w\vDash P^4(a_1, b_1, a_2, b_2)\mbox{ if } \pm\infty\in \{a_1, b_1, a_2, b_2\}.$$
Clearly, one can always consider the 4-ary relation  
 $a_1b_1 - a_2b_2 = 0$
on an arbitrary ring $R.$
For a ring with 1 this relation is 0-definably equivalent to
the multiplication. In this context any ring $R$ is a weak
ring.
\epk

\bpk {\bf Corollary to \ref{wring}}. {\em $\mathrm{st}_\mu$ is a surjective homomorphism of the weak ring
$^*\Z/N$ onto the compactified weak ring structure $\bar{\R}_w.$}
\epk

\bpk \label{rescale0} {\bf Dirac rescaling.} By assuming the (practically) infinite dimensionality of the algebraic-Hilbert spaces we deal with, we encounter an immediate problem of vanishing values of inner product
since the absolute value of $\la e|f\ra$ in $\bV_A(\alpha)$ is proportional to  
$\frac{1}{\sqrt{N}},$ for $N=n_A.$

In order to remedy this problem we rescale the definition of the inner product and introduce another pairing,    {\bf Dirac inner product} $\la e|f\ra_{Dir}$.  The rescaling depends on the number $N$ as well as on a ``parametrisation'' by a variable $x$ based on the following data.

Let $B,D\subseteq A$ be two Weyl subalgebras of $A$ with pseudo-unitary generators $B=\la R, R_\dagger\ra$ and $D=\la S,S_\dagger\ra.$

Suppose also that there is 
a regular unitary transformation  $L$ of the bundle $\bV_A(\alpha)$ such that for each $\alpha\in \spec_A$ the domain of $L_\alpha$ is a $B$-submodule $\bV_B(\alpha)$ and the range of
$L_\alpha$ is a $D$-submodule $\bV_D(\alpha)$ of $\bV_A(\alpha),$ both of dimension $n_B,$
the domain is spanned by
the $R$-eigenvectors 
$$\{ \rr(rq^{dk}, r_\dagger): k=0,1,\ldots, n_B-1\}$$
and the range  by $S$-eigenvectors 
$$\{ \s(sq^{dm}, s_\dagger): m=0,1,\ldots, n_B-1\}$$
 canonical $R$ and $S$ bases of $\bV_B(\alpha)$ and $\bV_D(\alpha)$ respectively, such that
\be \label{BD1} L_\alpha: \rr(rq^{dk}, r_\dagger)\mapsto \s(sq^{dk},\ s_\dagger),\ k=0,1,\ldots, n_B-1.\ee



Our next step is to define a rescaling parameter called $\Delta k.$

If $R$ and $S$ commute, then $\rr=\s$ and 
we define
$$\Delta k=\dd\sqrt{\frac{d}{N}},$$
where $\dd$ is a fixed parameter to be defined later, independent of $A$ and $L.$

Otherwise we assume that $RS\neq SR.$ 

Let
$b<N$ be the minimal positive integer with the property 
$$RS=q^bSR\mbox{ or }SR=q^{b}RS.$$

 Let $a_R, a_S$ be   the maximal positive integers with the property that
 $$R^\frac{1}{a_R} \in A\mbox{ and }S^\frac{1}{a_S} \in A.$$

  Set $$\Delta k=\frac{b\dd}{a_Ra_S\sqrt{N}}.$$

 We can now define the {\bf Dirac delta-function}, which depends on the parametrisation, that is on  $A$ and $L.$ 
   $$ \delta(p)=
  \left\lbrace \begin{array}{ll} \frac{1}{\Delta k}, \mbox{ if } p=0\\
  0, \mbox{ otherwise.}
  \end{array}\right. $$

   Further set the rescaled product, dependent on the chosen parametrisation:
   
\be \label{dir_s?} \la \rr(rq^{dk}, r_\dagger)|\s(sq^{dm}, s_\dagger)\ra_{Dir}=  \frac{1}{\Delta k}\la \rr(rq^{dk}, r_\dagger)|\s(sq^{dm}, s_\dagger)\ra\ee
\epk

\bpk \label{exm_dir}
{\bf An example}. Let $A=\la \check{U}, \check{V}\ra.$ 

$L$ be the Fourier transform $\Phi:\bV_A\to \bV_A,$
which  by definition is a bijection of $\bV_A$-modules. That is $B=D=A.$ 

We respectively will have $R=\check{U}=S_\dagger,$  $S=\check{V}=R_\dagger,$  
and $\la B,D\ra=A.$

Hence $\Delta k=\frac{1}{\dd\sqrt{N}}$
$$\la \uu(uq^k,v)|\vv(vq^m,u)\ra_{Dir}=\dd\cdot \sqrt{N}\la \uu(uq^k,v)|\vv(vq^m,u)\ra=\dd   q^{km}$$
 and does not depend on $N.$
 
\epk
\bpk {\bf A non-example.} Consider the  transform
$$\Phi_{B(e,f)}: \uu^{e,f}(u^e q^{ek}, v^f)\mapsto \vv^{f,e}(v^f q^{fk}, u^e).$$
This is not a restriction of $\Phi_A,$ so \ref{BD+} is not applicable. Moreover, condition (\ref{BD1}) for parametrisation is not satisfied.
\epk
\bpk \label{BD+} {\bf Proposition.} {\em Let $e,f$ be positive integers and
$$B'=\la R^e, R_\dagger^f\ra,\ \ D'=\la S^e, S_\dagger^f\ra,$$
subalgebras of $B$ and $D$ of \ref{rescale0} respectively. Consider the restriction $L'$ of $L$ sending $B'$-submodules onto $D'$-submodules
$$L'_\alpha: \rr^{e,f}(r^eq^{dek}, r^f_\dagger)\mapsto \s^{e,f}(s^eq^{dek},\ s^f_\dagger),\ k=0,f,2f,\ldots, \frac{N}{def}-f.$$
Then
$$\la \rr^{e,f}(rq^{dek}, r^f_\dagger)|\s^{e,f}(s^eq^{dem}, s^f_\dagger)\ra_{Dir}=\la \rr(rq^{dk}, r_\dagger)|\s(sq^{dm}, s_\dagger)\ra_{Dir},$$
where the left-hand side is rescaled with respect to $L',$
and in the case $RS=SR$ both sides of the equality take value equal to $\delta(k-m),$ with respective  interpretations of this
 symbol on the left- and on the right-hand side.
}

{\bf Proof.} It is enough to prove the statement in the two cases: 

Case 1. $e=1.$

and

Case 2. $f=1.$

The relation between the $\rr^{e,f}$ and $\rr$ (and between $\s^{e,f}$ and $\s$) has been analysed in \ref{3.45}.

In the first case, the respective bases of the submodules are just subsets of the initial bases, so nothing changes and we have the  required equality.

In the second case, the new bases are different, however calculating with formula (\ref{uu1}) we get
$$\la \rr^{e,f}(rq^{dek}, r^f_\dagger)|\s^{e,f}(s^eq^{dem}, s^f_\dagger)\ra=\la \rr(rq^{dk}, r_\dagger)|\s(sq^{dm}, s_\dagger)\ra.$$
Now we look at the new $\Delta k.$ Clearly, $b'=e^2b,$  $a'_R=ea_R$ and $a'_S=ea_S.$ Hence $\Delta k$ remains the same and so the required equality holds. $\Box$
\epk
\bpk \label{prob_measure} We can now introduce the {\bf Dirac rescaled pairing} (probability measure) just replacing the inner product in \ref{glu} by the Dirac rescaled inner product.

$$[e|f]_{Dir}:=(\frac{1}{\Delta k})^2\cdot [e|f].$$

\epk

\bpk \label{OAB1} {\bf Lemma -- definition.} {\em
Let $B\subs A,$ $A,B\in \A_{fin}$ and $C\in O(A).$ Then the commutative subalgebra $C_B=\la C\cap B, Z(B)\ra$ (generated by  $C\cap B$ and  $Z(B)$) is in $O(B).$ Moreover, the map
$$\pi^*_{BA}: O(A)\to O(B), \ \ C\mapsto C_B,$$
is surjective. }

\epk

\bpk \label{OAB2} {\bf Lemma -- definition.} {\em For a pseudo-unitary $S$ define $$C(S):=\la S^a: a\in \Q\ra, $$  
that is the subalgebra of $\Aa^{(1)}$ generated by all elements $S^\frac{1}{k},$ $k\in \Z.$ 

$C(S)$ is a maximal commutative subalgebra of $\Aa^{(1)}.$

This has natural generalisation to maximal commutative subalgebra of $\Aa^{(n)}.$
}

{\bf Proof.} Immediate. $\Box$
\epk

\section{Limit objects} \label{limitobjects}
Our aim here is to extend the categories $\A^*_{fin}(\C)$ and $\V^*_{fin}(\C)$ by adding some limit objects.
This follows the idea that objects in $\A^*_{fin}(\C)$ and $\V^*_{fin}(\C)$ {\em approximate} objects from a larger categories $\A^*(\C)$ and  $\V^*(\C)$
in a functorial
correspondence to the way the rational Weyl algebras approximate
(in the sence of normed algebras) subalgebras of the whole Weyl algebra. 

The approximation 
will be carried out in terms of {\em structural approximation} described in \cite{Appr}, sections 3 and 4.

One particular limit object of $\V^*_{fin}(\C)$
will be called the {\bf space of states} and will take the role of the rigged Hilbert space of mathematical physics.

We assume $\F_0\subset \C$ when $\F_0$ is mentioned and  will mostly drop referring to $\C$ below.

\bpk   
We fix an ultrapower $\ZZ$ (equivalently, a saturated enough model) of the ring of integers. 

Fix a non-standard integer $\mu \in \ZZ$ which in addition to assumptions in \ref{mu} we assume to have
the {\bf high divisibility property}: \be \label{div} \mu \mbox{ is divisible by any standard } m\in \Z\ee 

We also fix for the rest of the paper a (standard) positive rational number $\hh.$ \footnote{The theory which follows is also applicable to a non-standard  rational $\hh\in {^*\Q}_{fin}$ provided the numerator of $\hh$ is ``much less'' than $\mu$ and divides $\mu.$ }

The definitions and statements of previous sections are applicable in the new context, with 'rational' and 'integer' replaced by more general  'non-standard rational' and 'non-standard integer'. In particular, in the definitions  we consider 'non-standard' Heisenberg group $\tilde{\HH}(\frac{1}{\mu},\frac{\hh}{\mu})$ consisting of non-standard pseudo-unitary operators generated (in the non-standard sense) by $U^\frac{1}{\mu}$ and $V^\frac{\hh}{\mu}.$ More precisely,  $\tilde{\HH}(\frac{1}{\mu},\frac{\hh}{\mu})$ can be identified with the ultraproduct of
the groups  ${\HH}(\frac{1}{m},\frac{\hh}{m}),$ $m\in \N,$ along an ultrafilter $\mathcal{D}_{div}$ which for every positive integer $n$ contains the set
$$\{ m\in \N:  n|m\}.$$

We use, for a positive $d\in \ZZ$ the notation ``much less'' 
$$d\prec \mu$$ with the meaning  that $\frac{d}{\mu}$ is an infinitesimal and $d$ divides $\mu.$
\epk
\bpk\label{*A} {\bf A universal nonstandard-rational Weyl algebra.} 
Define the algebra ${\tilde{\mathrm{A}}}  $ to be the  ultraproduct of
the algebras  $A(\frac{1}{m}, \frac{\hh}{m}),$ $m\in \N,$ along the ultrafilter $\mathcal{D}_{div}.$ One may think of  ${\tilde{\mathrm{A}}}  $
as being generated, in a pseudo-finite sense, 
 by $U^{\frac{1}{\mu}}$ and $V^{\frac{\hh}{\mu}}.$ A pseudo-unitary operator $W$ in  ${\tilde{\mathrm{A}}}  $ is by definition any element
 of $\tilde{\HH}(\frac{1}{\mu},\frac{\hh}{\mu}).$

 \epk
 \bpk
 We call a pseudo-unitary $W\in {\tilde{\mathrm{A}}}  $ {\bf finite} if it can be presented in the form 
   \be \label{W} W=\e^\frac{2\pi i\hh\nu}{\mu^2}U^{\frac{\rho}{\mu}}V^{\frac{\hh \tau}{\mu}}\ee
for nonstandard integers $\nu,\rho, \tau$ such that $|\nu|< \mu^2$ and $\frac{\rho}{\mu}, \frac{\tau}{\mu}\in \QQ_{fin}.$

Define $\tilde{\mathrm{A}}_{fin}$ to be the subalgebra of $ \tilde{\mathrm{A}}$ generated by all
the finite  pseudo-unitary $W.$

Note that  
 \be \label{maj} \mathrm{A}\subset {\tilde{\mathrm{A}}}  \ee
 and so  ${\tilde{\mathrm{A}}}  $ is an upper bound for $\A_{fin}.$

\epk
\bpk
We  set \be\label{hbar} \hbar:=2\pi \hh\ee
and the constant introduced in \ref{rescale0} 
\be\label{dd} \dd:=\sqrt{2\pi \hbar}.\ee

\epk
\bpk
We are interested in the physically meaningful ${\tilde{\mathrm{A}}}  $-module corresponding to $U^{\frac{1}{\mu}}$- and $V^{\frac{\hh}{\mu}}$-eigenvalues $u=1$ and $v=1,$ which is the principal module $\bV_{\tilde{\mathrm{A}}}  (\mathbf{1})$ (see also \ref{principal}).  We abbreviate the latter to 
$\bV_*(\mathbf{1}).$ 

It follows from definitions that  $\bV_*(\mathbf{1})$ is the ultraproduct of $\bV_A(\mathbf{1}),$
$A=\la U^\frac{1}{m}, V^{\frac{\hh}{m}}\ra$ along the ultrafilter $\mathcal{D}_{div}.$
 In order to understand $\bV_*(\mathbf{1})$ we study some construction on the $\bV_A(\mathbf{1}).$
 
\epk

\bpk \label{7.4} 
Let $\Nn$ be the denominator of the reduced fraction $\frac{\hh}{\mu^2}.$ By definitions

$$\Nn=\dim 
\bV_*(\mathbf{1}).$$

By  divisibility of $\mu$ we have \be \label{N} \Nn=\frac{\mu^2}{\hh}=\frac{2\pi \mu^2}{\hbar}.\ee 

If not stated otherwise, we denote
$$q=\e^\frac{2\pi i\hh}{\mu^2}=\e^\frac{i\hbar}{\mu^2}.$$
\epk
\bpk\label{gl_lan} {\bf The global language $L^{Glob}_{\bV}.$}

We are interested in switching to a language which is, on each
$\bV_{A(\frac{1}{m},\frac{\hh}{m})},$  interdefinable with our original language but has the advantage not to refer to the parameter $m,$ (equivalently $\Nn$), or similar parameters which change with the ``size'' of $\bV_{A}$. 
 
For each $C\in O(\Aa)$ the language $L^{Glob}_{\bV}$ contains the unary 
predicate $\EE^C,$ which we also may denote $\EE^S$ if $C=C(S).$ In each
$\bV_A(\alpha)$ the interpretation of $e\in \EE^C$ is $e\in \EE^{C_A}_A(\alpha).$

The unitary predicate $\EE$ is interpreted in each $\bV_A$ as 
$$\EE\equiv  \bigvee_{C\in O(A)}  \EE^C.$$

We are going to ``forget'' the names of pseudo-unitary predicates
such as $U^{\frac{1}{m}}$ and $V^{\frac{\hh}{m}}$ and instead
introduce to 
$L^{Glob}_{\bV}$  names for pseudo-selfadjoint operators
$R=R_W,$ for each $W\in \tilde{\mathrm{A}}_{fin},$ or more generally
of the form $$W=\e^\frac{2\pi i\hh n}{m^2}U^{\frac{r}{m}}V^{\frac{\hh t}{m}}$$
where $n,m,r$ and $t$ can be standard or non-standard integers.   

Interpret $R_W$ 
  in $\bV_{A}$ as a linear operator
\be \label{Rw}
R_W=\frac{W-W^{-1}}{2i\{ |\frac{r}{m}|+|\frac{t}{m}| \}}.\ee

\medskip

In particular, we use special names for operators

$$\Qq=\frac{U^{a}-U^{-a}}{2ia}\mbox{ \ \ \ and \ \ \ }\Pp=\frac{V^{a\hh}-V^{-a\hh}}{2ia},\ \ a=\frac{1}{m}.$$

 Note $R_W$ are in ${\tilde{\mathrm{A}}}_{fin}  .$

Clearly, for each $A,$ the $\Qq$ is interdefinable with $U^{a}$ and $\Pp$    with $V^{a\hh}.$  However the interpretation is given by different formulas depending on $A.$

A concrete  algebraic Hilbert space $\bV_A(\alpha)$ is given in terms of
global language $L^{Glob}_{\bV}$ 
as the two sorted structure $(\EE,\F)$ with 
\begin{itemize}
\item unary predicates $\EE^C,$ for each $C\in O(\Aa);$
\item pseudo-selfadjoint operators $R_C$ for each $C\in \mathcal{C}_{fin};$
\item regular unitary transformations $L$ interpreted for each $A$ by the formula in (\ref{Luu}) and a fixed associated matrix $g_L$ (see \ref{onRegTr}).
\end{itemize}
As before $\F$ will stand in $L^{Glob}_{\bV}$ for the field sort with addition and multiplication defined on it.

The formal linear equalities $\mathrm{Eq}(x_1,\ldots,x_n,e_1,\ldots,e_n):$
$$x_1e_1+\ldots+ x_ne_n=0,\ x_1,\ldots,x_n\in \F,\ e_1,\ldots,e_n\in \EE $$
are expressible in $L^{Glob}_{\bV},$ and are interpreted in each
$\bV_A(\alpha)$ according to the values taken by the linear combination in the  $\F$-vector space. Note that one can interpret arbitrary elements of the vector space $\VV_A(\alpha)$ in terms of $\mathrm{Eq}.$

Later we will add to the global language predicates encoding the Dirac inner product (see \ref{rescale0} and \ref{prob_measure}).

  We call these globally defined operators and predicates. 
\epk

\bpk \label{notat}
Below we standardise our notation for the generators of ${\tilde{\mathrm{A}}}  ,$
$$\hat{U}:=U^\frac{1}{\mu},\ \hat{V}=V^\frac{\hh}{\mu}.$$
More generally,  an important role will be played by pairs of pseudo-unitary generators of ${\tilde{\mathrm{A}}}  ,$
$\hat{S}$ and $\hat{T}$ conjugated to $\hat{U}$ and $\hat{V}$ by
a $\gamma\in \mathrm{Aut}(\HH(\frac{1}{\mu},\frac{\hh}{\mu}):$ 
$$\hat{S}=\hat{U}^\gamma \mbox{ and }\hat{T}=\hat{V}^\gamma$$ (see \ref{w}).  Canonical $\hat{U},$ $\hat{V}$ and generally $\hat{S}$-bases for
$\bV_*(\mathbf{1})$ will be denoted
$\uu(q^k),$ $\vv(q^k)$ and $\s(q^k),$ $0\le k<N,$ respectively. (See the definition (\ref{st}) for $\s$).
    
We will also often work in subalgebras
of ${\tilde{\mathrm{A}}}  $ of the form $B(a,b)=\la \hat{U}^a, \hat{V}^b\ra,$
 with $a,b,c,d\in \ZZ,$ 
$0< a,b<\mu.$ 
 The corresponding canonical bases for the principal module $\bV_{B(a,b)}(\mathbf{1})$ will be referred to as
$\uu^{a,b}(q^m),$ $\vv^{b,a}(q^m),$ and $\s^{a,b}(q^m),$ respectively.  

\epk
\bpk 
We define a pseudo-metric (that is the distance between distinct point can be $0$)
on the group $\tilde{\HH}(\frac{1}{\mu},\frac{\hh}{\mu})_{fin}$  of  all the finite pseudo-unitary elements of  $\tilde{\mathrm{A}}_{fin}.$

For $W$ and $W'$ in the form (\ref{W}) define the pseudo-distance, 
$$\mathrm{dist}(W,W'):=\Lim\{ |\frac{\rho}{\mu}-\frac{\rho'}{\mu}|+  |\frac{\tau}{\mu}-\frac{\tau'}{\mu}|+ |\frac{\nu}{\mu^2}-\frac{\nu'}{\mu^2}|\},$$
where $\nu', \rho'$ and $\tau'$ are the respective parameters of $W'$ in the representation (\ref{W})
and $\Lim$ is just the standard part map $\CC\to \C.$ 

Note that \be \mathrm{dist}(\hat{U},1)=0=\mathrm{dist}(\hat{V},1).\ee

Let  $$\tilde{\mathrm{A}}_{fin}^0=\{ R_W\in  \tilde{\mathrm{A}}_{fin}:  \nu=0\ \&\ \frac{\rho}{\mu}, \frac{\tau}{\mu}\in \QQ_{fin}\} $$ which we call the {\bf basic} pseudo-selfadjoint operators.
 
The pseudo-metric on     $\tilde{\mathrm{A}}_{fin}^0$ is given by 
$$\mathrm{dist}(R_W,R_{W'}):=\Lim\{  |\frac{\rho}{|\rho|+|\tau|}-\frac{\rho'}{|\rho'|+|\tau'|}|+
 |\frac{\tau}{|\rho|+|\tau|}-\frac{\tau'}{|\rho'|+|\tau'|}|\}+\mathrm{dist}(W,{W'}).$$

\epk

\bpk Define $$W\approx W' \mbox{ iff } \mathrm{dist}(W,W')=0,$$
an equivalence relation on the group $ \tilde{\HH}(\frac{1}{\mu},\frac{\hh}{\mu})_{fin}.$ It is immediate from the definition that $\approx$ is invariant under the group operation.
Set $$\HH_\R=\tilde{\HH}(\frac{1}{\mu},\frac{\hh}{\mu})_{fin}/_\approx.$$
This is a real Heisenberg group.

\medskip
Define on  $\tilde{\mathrm{A}}_{fin}^0$
 $$R\approx R' \mbox{ iff } \mathrm{dist}(R,R')=0.$$
The elements of the quotient $$\tilde{\mathrm{A}}_{fin}^0/\approx$$
will be called  the {\bf basic selfadjoint operators}. The pseudo-metric becomes a metric on the set of basic selfadjoint operators.
\footnote{The basic self-adjoint operators can be interpreted as the elements of the real Lie algebra generated by
the classical $\Pp$ and $\Qq$ (the Heisenberg algebra)

It is possible in principle to consider in the similar way the associative algebra of operators generated by these operators (the Weyl algebra). This, however, in our setting becomes rather messy business.  
Instead, we will study in later section  operators which correspond in mathematical terms to automorphisms of the Lie algebra  and in physical terms to time evolution operators, without trying to identify these operators with elements of the Weyl algebra.}

\epk

\bpk \label{Oeq} {\bf Test-equivalence on $\EE.$} 
Let $a,b,c,d\in \ZZ,$ 
$0< a,b,c,d\prec\mu.$

For
$e\in {\EEE}_{B(a,b)}(\mathbf{1})$ we define $W_e$ to be  the unique element  of   $\tilde{\HH}(\frac{1}{\mu},\frac{\hh}{\mu})_{fin}$  with the properties:
\begin{itemize}
\item  $\nu=0$ in the form (\ref{W}) for $W_e,$
\item $e$ is an eigenvector of $W_e,$
\item $\rho>0$ or $\rho=0$ and $\tau>0,$
\item $|\rho|+|\tau|$ is minimal. 
\end{itemize}

Given 
$e,e'\in {\EE}_{B(c,d)}(\mathbf{1})$ let 
$W=W_e,$ and $W'=W_{e'}$ and $R=R_W,$ $R'= R_{W'}$
be the  corresponding pseudo-selfadjoints. 

Define the two elements to be {\bf test-equivalent},
$$e\approx e'$$ if
$$R\approx R' \mbox{ and }\la e|R e\ra \sim \la e'|R' e'\ra.$$
\epk
\bpk \label{t-equiv}
 Note that  
$$e\approx re,$$ for $r$ a root of unity. In other words, $\approx$ identifies elements representing the same  physical state.

Also, crucially with the use of regular unitary transformations, $\approx$ agrees with the relation introduced in \ref{LformulaL}.
\epk

\bpk \label{EH}{\bf Lemma.} {\em 
 $f_1\approx f_2$ is an equivalence relation.

}

{\bf Proof.} (i) Immediate by definition.
 $\Box$
\epk
\bpk\label{def_appr} {\bf Extending the global language by $\approx_\epsilon.$}
We introduce for each  positive $\epsilon\in \Q$ a new
 binary predicate $e_1\approx_\epsilon e_2,$   interpreted in each $\EE_A$ as a statement:
  
  ``$e_1,e_2\in \EE$ and there are pseudo-selfadjoint operators $R_1,R_2\in A$ such that for some $w_1,w_2\in \F,$ 
  $$R_1 e_1=w_1e_1\ \& \   R_2 e_2=w_2e_2\ \& \ |w_1-w_2|<\epsilon\mbox{ ''}.$$

This is definable by a positive quantifier-free formula since for every choice of $A$ the existential quantifier runs through finitely many choices. 

\medskip

Note that the relation $\approx$ on  $\EEE$ is definable by a type consisting of positive formulas (positively type-definable):
$$e_1\approx e_2 \ \equiv\ \bigwedge_{n\in \N} e_1\approx_{\frac{1}{n}} e_2.$$
\epk

\bpk \label{Rwlemma} {\bf Lemma.} {\em Let $R=R_W\in   {\tilde{\mathrm{A}}}_{fin}  $ be a  selfadjoint operator and $e\in \EEE$ its eigenvector. 
Then 
$$e\approx \hat{U}e \mbox{ and }  e\approx \hat{V}e.$$}
{\bf Proof.} Clearly, $e$ is a $W$-eigenvector too. That is
$We=we$ for some $w\in {^*\F}_0^\times.$ 

Let $W':= \hat{U} W\hat{U}\inv.$ Then,   using the commutation identity (\ref{ccr1}),
$$W'= \e^{2\pi i\hh \frac{\tau}{\mu^2}} W$$
and 
$$W'e'=\hat{U} W\hat{U}\inv \hat{U}e=we'.$$

It follows that $$R_{W'}e'=\frac{1}{i\sigma}\left(\e^{2\pi i\hh \frac{\tau}{\mu^2}} w - \e^{-2\pi i\hh \frac{\tau}{\mu^2}}w\inv\right)e',$$
for $\sigma=|\frac{\rho}{\mu}|+ |\frac{\tau}{\mu}|$ as defined in (\ref{W}). 

We claim that 
 $$\frac{1}{\sigma}\left(\e^{2\pi i\hh \frac{\tau}{\mu^2}} w - \e^{-2\pi i\hh \frac{\tau}{\mu^2}}w\inv\right)\sim 
\frac{1}{\sigma}(w-w\inv) $$
that is the $R_{W'}$-eigenvalue of $e'$ is infinitesimally close to the $R_{W}$-eigenvalue of $e.$ Indeed, this is immediate from the fact that 
$$\frac{1}{\sigma}\cdot\frac{\tau}{\mu^2}$$
is infinitesimal. This proves the statement for $e'=\hat{U}e.$

The proof for $e'=\hat{V}e$ is similar. $\Box$

\epk

\bpk \label{appr1} {\bf Lemma.} Let $k,l\in \ZZ,$ $\frac{k}{\mu}, \frac{l}{\mu}\in \QQ_{fin}.$ Then
\be \label{dir_u} \uu^{a,b}(q^{ak})\approx \uu(q^{k}),\ee
and 
\be \label{dir_v}\vv^{b,a}(q^{bl})\approx \vv(q^{l}).\ee

{\bf Proof.} Note that for finite values of $\frac{k}{\mu}$ calculating in the $B(a,b)$-module
$$\la \uu^{a,b}(q^{ak})|\, \Pp\, \uu^{a,b}(q^{ak},1)\ra =\frac{q^{ak}-q^{-ak}}{2ia/\mu}\sim \frac{q^{k}-q^{-k}}{2i/\mu}\sim \frac{k\hbar}{\mu}.$$

And in the ${\tilde{\mathrm{A}}}  $-module
$$\la \uu(q^{k})| \Pp 
\uu(q^{k})\ra=
\frac{q^{k}-q^{-k}}{2i/\mu}\sim \frac{k\hbar}{\mu}.$$
This proves the first equivalence. The second and third equivalences follows by the same argument. $\Box$
\epk

\bpk {\bf Remark.} Under the above assumptions we have 
 $a|l$ and $b|k$ by definition of a canonical basis. We then have 
 by calculation in \ref{exm_dir} in the ambient module $B(a,b):$

\be \label{prod_ab}\la \uu^{a,b}(q^{ak})|\vv^{b,a}(q^{bl})\ra_{Dir}= \dd\,q^{kl} \ee
\epk
\bpk {\bf Lemma.} {\em Let $k$ be as in \ref{appr1} and $d\in \ZZ,$ $0\le d\prec\mu.$ Then \be \label{ap_u}\uu(q^{k+d})\approx \uu(q^{k})\ee
and \be \label{ap_v}\vv(q^{k+d},1)\approx \vv(q^{k})\ee}

{\bf Proof.} Using the same same calculations as in \ref{appr1} we would need to prove 
$$\frac{q^{k}-q^{-k}}{2i/\mu}\sim \frac{q^{k+d}-q^{-(k+d)}}{2i/\mu}.$$
Recall that $q^k=\exp \frac{i\hbar k}{\mu^2},$ $q^{k+d}=\exp \frac{i\hbar (k+d)}{\mu^2}$ and so
$$\frac{q^{k+d}- q^{k}}{2i/\mu}= q^k \mu\frac{\e^{i \alpha}-1}{2i}$$
where $\alpha\mu$ is infinitesimal. The equivalence follows. $\Box$    

\epk
\bpk \label{appr2} {\bf Corollary.} The statements of \ref{appr1} hold for any pair of pseudo-unitary operators $S$ and $T$ and their eigenvectors 
in place of $U^a$ and $V^b.$
\epk
\bpk\label{space} {\bf The space of states and the bra-ket notations.}

We define the space of states as the quotient space
$$\mathbb{S}:= \EEE/\approx$$ 
and the canonical quotient map as 

\be \label{limE}\Lim: {^*\EE}(\mathbf{1})\to \mathbb{S}.\ee
Note that by definition ${^*\EE}(\mathbf{1})$ is a pseudo-finite union of definable subsets of the form $\EE^C,$ $C=C(S),$ $S\in {\tilde{\mathrm{A}}},$ $S=U^aV^b,$ $a,b\in\QQ_{fin}.$ Thus 
$$\Lim S=:\bar{S}= U^\alpha V^\beta, \mbox{ where } \alpha=\Lim a,\ \beta=\Lim b,$$
According to \ref{appr1} and \ref{appr2}
$$\Lim {\EE^C}=\{ \Lim \s(\e^\frac{is}{\mu}): s\in\frac{1}{\mu}\ZZ\cap \QQ_{fin}\}= \{ |s\ra: s\in \R\},$$
the image of the $S$-basis.

For the given $\bar{S}=U^\alpha V^\beta$ we denote the image of the basis in $\mathbb{S},$
$$\mathbb{S}_{\alpha/\beta}:= \{ |s\ra: s\in \R\},$$
and call it a {\bf Lagrangian subspace of $\mathbb{S}.$}\footnote{Note that $U^\alpha V^\beta$ and $U^{m\alpha} V^{m\beta}$ commute and so generate the same Lagrangian subspace, hence the subspace is determined by the fraction $\alpha/\beta.$} 
Lagrangian subspaces with
$\alpha, \beta$ rational will be referred to as {\bf rational  
Lagrangian subspaces}.

Thus 
$$\mathbb{S}=\bigcup_{\alpha,\beta\in \R^\times}\mathbb{S}_{\alpha/\beta},$$
and we also will use the rational part of the space of states,
$$\mathbb{S}_\Q:=\bigcup_{\alpha,\beta\in \Q^\times}\mathbb{S}_{\alpha/\beta}.$$
\epk
\bpk
Keeping with physics notation we set
$$|x\ra =\Lim \uu(\e^\frac{ix'}{\mu})$$
$$|p\ra =\Lim \vv(\e^\frac{ip'}{\mu})$$
for  \be \label{xp_condition} x',p'\in \frac{\hbar}{\mu}\cdot\ZZ,\ x',p'\in \QQ_{fin},\ee 
Note that these assumptions imply that  $x',p'\in [-\frac{\mu}{2\hbar}, \frac{\mu}{2\hbar})$ 
and $\Lim x'\in \R,$ $\Lim p'\in \R,$ taking all possible values in $\R.$

These notation rely on distinguishing between the so called {\bf position eigenstates} $|x\ra$ and the {\bf momentum eigenstates} $|p\ra$ just by the letters $x$ or $p$ used to denote possibly the same number, the eigenvalue of the corresponding operator (to be explained below).

Often we simply right $$|x\ra =\Lim \uu(\e^\frac{ix}{\mu})$$
and
$$|p\ra =\Lim \vv(\e^\frac{ip}{\mu})$$
assuming that the passage from the right-hand side to the left is clear.

\epk
\bpk We extend the domain of $\Lim$ to $\Ee_{B(a,b)}(\mathbf{1}),$ the eigenvector-bases of submodules, 
and define 

\be \label{def_x}\Lim \uu^{a,b}(\e^\frac{iax}{\mu})=|x\ra\ee
\be \label{def_p}\Lim \vv^{b,a}(\e^\frac{ibp}{\mu})=|p\ra\ee
and
\be \label{def_s}\Lim \s^{a,b}(\e^\frac{ias}{\mu})=|s\ra\ee

This agrees with $\approx$ by (\ref{dir_u}) and (\ref{dir_v}).

\epk
\bpk \label{dense0}
We say that a subset $E\subset \EEE$ is {\bf dense in } $\mathbb{S}_{\alpha/\beta}$ if $\Lim E=\mathbb{S}_{\alpha/\beta}.$

Note that $\Ee_{B(a,b)}(\mathbf{1})\subset \bV_*(\mathbf{1})$ (see \ref{1.7}) and 
$\Ee_{B(a,b)}(\mathbf{1})$ can be embedded into $\EEE.$

Assuming that $B(a,b)$ is a subalgebra of $\tilde{\mathrm{A}}$ of finite index, $a,b\in \QQ,$   $S\in B(a,b),$  it follows from \ref{appr1} and \ref{appr2} that $\Ee^S_{B(a,b)}(\mathbf{1})$ is dense in   $\mathbb{S}_{\alpha/\beta}.$

\epk

{\bf We study $\mathbb{S}$ along with the structure $\bV_*$ and the map
$$\Lim: \EEE\to \mathbb{S}.$$}

\bpk  \label{defpairing}{\bf Dirac product on $\mathbb{S}$.} This is defined in accordance with \ref{rescale0} with usually fixed parametrisation in variables $x$ or $p.$

We will work in the situation when we are given canonical bases
$\uu(\e^\frac{ix}{\mu}),$
 $\rr(\e^\frac{ix}{\mu}),$ and $\s(\e^\frac{ix}{\mu})$ 
such that both  $\rr(\e^\frac{ix}{\mu})$ and $\s(\e^\frac{ix}{\mu})$ are parametrised by regular transformations
$$L_r:  \uu(\e^\frac{ix}{\mu})\mapsto \rr(\e^\frac{ix}{\mu})\mbox{ and }L_s:\uu(\e^\frac{ix}{\mu})\mapsto \s(\e^\frac{ix}{\mu}).$$ 
 This will define $\Delta u$, which we following above notations
 will write as $\Delta x.$
 
We will also use the notation 
$$\Lim \rr(\e^\frac{ix}{\mu})= |r(x)\ra \mbox{ and }  \Lim \s(\e^\frac{ix}{\mu})= |s(x)\ra.$$

Now set \be \label{dir_s}
\la r(x_1)|s(x_2)\ra_{Dir}:=
\Lim  \frac{1}{\Delta x} \left\lbrace\max_{x'_1\sim x_1,\ x'_2\sim x_2}\la \rr(\e^\frac{ix'_1}{\mu})|  \s(\e^\frac{ix'_2}{\mu})\ra\right\rbrace\ee
Here the quantifier {\bf max} chooses $x'_1$ and $x'_2$ so that the modulus of $\la \rr(\e^\frac{ix'_1}{\mu})|  \s(\e^\frac{ix'_2}{\mu})\ra$ reaches maximum. The argument of this expression depends continuously on $x'_1$ and $x'_2$ (see the definition and \ref{exm_uv}). Hence the right-hand side of (\ref{dir_s}) is well-defined.

On the left-hand side of (\ref{dir_s}), the use of bra-ket notation assumes that we rescale the product. 

\medskip

{\bf Warning.} (\ref{dir_s}) only makes sense when we have chosen the corresponding canonical bases, that is the two consecutive vectors of each basis. Otherwise, we can only determine the {\bf absolute value of  $\la r(x_1)|s(x_2)\ra_{Dir}$}.

Nevertheless, (\ref{dir_s}) is a function of $x_1,x_2$ defined ``up to the phase'', that is the scalar of modulus 1.  

\medskip

Similarly one defines the parametrisation on variable $p,$ just replacing $ \uu(\e^\frac{ix}{\mu})$ by  $\vv(\e^\frac{ip}{\mu}).$

We will have to use respectively $\Delta p$ in the definition of
the Dirac product. But note, that the application of the Fourier
transform will produce the equality
$$\Delta p= \Delta x.$$
\epk
\bpk {\bf Example.} Reinterpreting the calculation in \ref{exm_dir} and taking into account (\ref{dd})
we have the classical
$$\la x|p\ra_{Dir} =\frac{1}{\sqrt{2\pi \hbar}}\e^{ixp}.$$
\epk

\bpk \label{rescale2} We also use the Dirac delta 
in $\bV_*(\mathbf{1})$ as defined in \ref{rescale0}. Since the dimension of the module is pseudo-finite
the $\Delta x$ and
$\delta$ are nonstandard integers-valued. In the limit structure on $\mathbb{S}$ 
we  introduce the Dirac $\delta(x)$  as an 
   ``infinite constant''-valued function, or symbol, along with symbol $\mathrm{d}x$ 
   and set $$\Lim \Delta x:= \mathrm{d}x,$$
   
$$ \Lim \delta(x):=\delta(\Lim x).$$


\epk

\bpk \label{dx} {\bf Using integration notation.} It is helpful to re-write the definition (\ref{dir_s}) in the following suggestive terms.


From the notations agreement above we get in particular,
\be \label{Dir3x} d x\, \la r(x_1)|s(x_2)\ra=\la \rr(\e^\frac{ix_1}{\mu})|\s(\e^\frac{ix_2}{\mu})\ra.\ee

In functional Hilbert spaces elements  $|r(x)\ra$ and $|s(x)\ra$
are understood as a family of functions $\rho(x,p)$ and $\sigma(x,p),$ with variable, say $p.$
Now 
a summation formula
$$\rr(\e^\frac{ix}{\mu})= \sum_y \la \rr(\e^\frac{ix}{\mu})| \s(\e^\frac{iy}{\mu})\ra \s(\e^\frac{iy}{\mu})$$
corresponds to the integral 
$$\rho(x,p)=\int_{y\in \R} d y\, \la r(x)|s(y)\ra\cdot \sigma(y,p).$$

E.g. the summation formula for the Fourier transform (\ref{four}) 
becomes, for $\frac{1}{\sqrt{2\pi \hbar}}\e^\frac{ixp}{\hbar},$ an eigenvector of the operator $f\mapsto -i\hbar\frac{\mathrm{d}f}{\mathrm{d}x}$, and $\delta(x-p),$
an eigenvector of the operator $f\mapsto x f,$ 

$$\delta(x-p)=\int_{y\in \R} \mathrm{d}y \frac{1}{\sqrt{2\pi \hbar}}\e^{-\frac{iyx}{\hbar}}\frac{1}{\sqrt{2\pi \hbar}}\e^\frac{iyp}{\hbar},$$
i.e.
$$\delta(0)= \frac{1}{{2\pi \hbar}}\int_{y\in \R} \mathrm{d}y $$

Conversely, 
$$\frac{1}{\sqrt{2\pi \hbar}}\e^{ixp}=\int_{y\in \R} \mathrm{d}y \frac{1}{{2\pi \hbar}}\e^{iyx}\delta(p-y),$$
\epk

\bpk Note also that by definition the integral reinterpretation of definitions \ref{rescale0} yield
$$\int_\R \delta(x)  \mathrm{d}x=1.$$
\epk

\bpk {\bf Topology on the  space of  states.}
Note that the construction of the space of  states $(\mathbb{S},\C)$ as the limit (\ref{limE}) of the pseudo-finitary space $\bV_*(\mathbf{1})=({^*\EE}(\mathbf{1}),{^*\F_0})$ introduces a topology on  the pseudo-Hibert space. Namely, $\C$ canonically acquires the  metric topology from  the standard part map $\Lim: {^*\F_0}\to \C.$  The predicates $e_1\approx_\frac{1}{n}e_2$ define topology on  $\mathbb{S}.$ More generally,
 on Cartesian products of $\mathbb{S}$ and $\C$ the topology is given by declaring closed subsets those which are definable by positive quantifier free formulas in the language $L_{\bV}^{Glob}$ extended by metric relation $|x|\le 1$ on $\C.$ All such relations are preserved by map $\Lim.$
\epk
\bpk {\bf Theorem.} {\em The object $\bV(\R)$ is universally attracting for the category $\V_{fin}.$
In other words, for every $\bV_A\in \V^*_{fin}$ there is a unique morphism 
$$\mathbf{p}_A: \bV_A\to  \bV(\R).$$}

{\bf Proof. } First we consider the trivial morphism $$\pi_A: \spec_A(\R)\to \mathbf{1},$$
the right-hand side being the spectrum of $\bV(\R),$ by definition.

Now we need to define the embeddings  $$\mathbf{p}_A^\beta: \bV_A(\beta)\to \bV(\R)$$
for every $\beta\in  \spec_A(\R).$

Since $A\subseteq A(\frac{1}{m}, \frac{\hh}{m})$ for $m\in \Z$ large enough (in terms of divisibility), such that $\frac{m}{\hh}$ is an integer, we may assume  $A= A(\frac{1}{m}, \frac{\hh}{m}).$
Recall that  $N=n_A=\frac{m^2}{\hh}$ is the dimension of  $\bV_A(\beta).$

By definition, $\beta=\la u^N, v^N\ra$ 
for some $u,v\in \C$ of modulus
$1,$ that is $$u=\e^{i\frac{x}{m}}, \ v=\e^{i\frac{p}{m}}, \mbox{ for some unique real } x,p\in [0, 2m\pi).$$ 

 For $b=\frac{\mu}{m}$
 consider the subalgebra $B(1, b)=\la \hat{U}, V^\frac{\hh}{m}\ra$ and a
  canonical $\hat{U}$-basis 
$\{ \uu^{1,b}(q^k): k=0,\ldots, \frac{m\mu}{\hh}-1\}$ of $\bV_{B(1,b)}(\mathbf{1}),$ 
$q=\e^\frac{2\pi i \hh}{m\mu}.$

Since $\mu$ is an infinite integer, for $k=0,\ldots, m\mu-1,$
$\Lim  q^k$ takes all the complex values of modulus $1.$ Let $k_0$ be such that 
$$\Lim  q^{k_0}=u$$ and denote
$$|uq^{bn}\ra =\Lim  \uu^{1,b}(q^{k_0+bn}):\ \ \  n=0,1,\ldots \frac{m^2}{\hh}-1.$$ 

By definition
$$U^\frac{1}{m}: |uq^{bn}\ra\mapsto uq^{bn}|uq^{bn}\ra$$
and 
$$V^\frac{\hh}{m}: |uq^{bn}\ra\mapsto |uq^{b(n-1)}\ra.$$

Write  the $U^\frac{1}{m}$-basis of $ \bV_A(\beta)$ as $\{  \uu(uq^{bn}, v): n=0,1,\ldots, \frac{m^2}{\hh}-1\}$ and 
 define  $\mathbf{p}_A^\beta$ on the basis as
 $$\mathbf{p}_A^\beta: \uu(uq^{bn}, v)\mapsto v^{-n}|uq^{bn}\ra, \ \ 
 n=0,1,\ldots, \frac{m^2}{\hh}-1.$$
It is clear that  $\mathbf{p}_A^\beta$ commutes with $U^\frac{1}{m}.$
 One also checks that  $$V^\frac{\hh}{m}: \mathbf{p}_A^\beta \uu(uq^{bn}, v)=
 v^{-n} |uq^{b(n-1)}\ra=v \mathbf{p}_A^\beta \uu(uq^{b(n-1)}, v), $$
which proves that  $\mathbf{p}_A^\beta$ commutes with $V^\frac{\hh}{m}.$
Hence it is an isomorphism between the $A$-modules. 

$\Box$
\epk

\section{ Some basic calculations.}
\bpk \label{some}
We calculate with operators $\Qq$ and $\Pp$ defined in \ref{gl_lan}.

\be \label{Qx} \begin{array}{ll}\Qq |x\ra=\Lim  \frac{\hat{U}-\hat{U}\inv}{\frac{2i}{\mu}} \uu(\e^\frac{i x}{\mu})=\\
=
\Lim \frac{\mu}{2i} (\e^{\frac{ix}{\mu}}-\e^{-\frac{ix}{\mu}}) \uu(\e^\frac{i x}{\mu})=x|x\ra.\end{array} \ee

It is clear, that this calculation is invariant under replacing $x$ by $x',$ with $x\sim x'.$

 Hence
$$\Qq |x\ra =x|x\ra.$$

On the other hand
\be \label{Px}  \begin{array}{ll}\Pp |x\ra=\Lim  \frac{\hat{V}-\hat{V}\inv} \uu(\e^\frac{i x}{\mu})=\\
=
\Lim  \frac{\mu}{2i} \left\lbrace\uu(\e^{(\frac{i x}{\mu}-\frac{i\hbar}{\mu^2})})-\uu(\e^{(\frac{i x}{\mu}+\frac{i\hbar}{\mu^2})})\right\rbrace \end{array}\ee

is ill-defined since the vector in the bracket is an infinitesimal and 
 $\mu$ is infinite.

Similarly, 
$$ \Pp |p\ra=p|p\ra$$ but 
$$ \Qq |p\ra\mbox{ is undefined}.$$
\epk
\bpk {\bf Comments.} The real reason for $\Pp$ and $\Qq$ to be only partially defined operators on $\EE$ is that
these are  {\em unbounded} operators, that is 
$$||\Pp||=\infty=||\Qq||.$$
For $\Qq$ this follows from the calculation (\ref{Qx}) and similarly can be checked for $\Pp.$

In mathematical physics this is the main difficulty for treating
the Heisenberg algebra as a $C^*$-algebra. 
\epk
\bpk \label{QP}
Let us calculate the application of $\Qq\Pp-\Pp\Qq$ to $|x\ra$ and $|p\ra.$ To simplify the computation, we rewrite
$$\Qq=\frac{1}{2}(\Qq'+\Qq''),\ \ \Pp=\frac{1}{2}(\Pp'+\Pp'')$$
where
$$\Qq'=\frac{\hat{U}-1}{i/\mu},\ \Qq''=\frac{1-\hat{U}\inv}{i/\mu},$$
$$\Pp'=\frac{\hat{V}-1}{i/\mu},\ \Pp''=\frac{1-\hat{V}\inv}{i/\mu}.$$

First we calculate
$$\begin{array}{ll}(\Qq'\Pp'-\Pp'\Qq')\uu(\e^\frac{i x}{\mu})=-4\mu^2\left( \hat{U}\hat{V}- \hat{V}\hat{U}\right)\uu(\e^\frac{i x}{\mu})=\\
-\mu^2\left(1-\e^\frac{i\hbar}{\mu^2}\right)\hat{U}\hat{V}\uu(\e^\frac{i x}{\mu})=
-\mu^2\left(1-\e^\frac{i\hbar}{\mu^2}\right)\e^{\frac{i x}{\mu}-\frac{i\hbar}{\mu^2}}\cdot \uu(\e^{\frac{i x}{\mu}-\frac{i\hbar}{\mu^2}})
\end{array}$$ 
Now note that by definition and (\ref{ap_u}) $$\Lim \, \uu(\e^\frac{i x}{\mu})=|x\ra=\Lim \, \uu(\e^{\frac{i x}{\mu}-\frac{i\hbar}{\mu^2}}),$$
so, applying $\Lim $ to the initial and final term of the equality 
and using that $$\Lim \ \mu^2(\e^\frac{i\hbar}{\mu^2}-1)\cdot\e^{\frac{i x}{\mu}-\frac{i\hbar}{\mu^2}}=i\hbar$$
we get 
$$ (\Qq'\Pp'-\Pp'\Qq')|x\ra =i\hbar|x\ra.$$
Similarly we get the same result for
$ (\Qq'\Pp''-\Pp''\Qq'),$ $(\Qq''\Pp'-\Pp'\Qq'')$ and $(\Qq''\Pp''-\Pp''\Qq'').$ It follows
$$ (\Qq\Pp-\Pp\Qq)|x\ra =i\hbar|x\ra.$$

Analogously one gets $$ (\Qq\Pp-\Pp\Qq)|p\ra =i\hbar|p\ra.$$
In fact, these are the instances of the more general fact. 
\epk
\bpk \label{StoneNeumann} {\bf Theorem.} {\em The Canonical Commutation Relation {\rm (\ref{ccr})} holds in $ \mathbb{S}:$
$$\Qq\Pp-\Pp\Qq=i\hbar I.$$
In other words, given $|s\ra \in \mathbb{S}$
$$(\Qq\Pp-\Pp\Qq)|s\ra=i\hbar |s\ra.$$ }

{\bf Proof.} By definition  $|\s\ra =\Lim \s(\e^{i\frac{s}{\mu}}),$ 
for some 
$\s(\e^{i\frac{s}{\mu}})\in \EEE,$ an element of a canonical $S$-basis. 

Now we apply $(\Qq'\Pp'-\Pp'\Qq')$ to $\s(\e^{i\frac{s}{\mu}})$ and get by 
the calculation in \ref{QP}
$$(\Qq'\Pp'-\Pp'\Qq')\s(\e^{i\frac{s}{\mu}})=\mu^2\left(\e^\frac{i\hbar}{\mu^2}-1\right)\hat{U}\hat{V}\s(\e^{i\frac{s}{\mu}}).$$

By \ref{Rwlemma} $\hat{U}\hat{V}\s(\e^{i\frac{s}{\mu}})\approx\s(\e^{i\frac{s}{\mu}}).$ Also  $\mu^2\left(\e^\frac{i\hbar}{\mu^2}-1\right)\sim i\hbar.$
Hence 
$$(\Qq'\Pp'-\Pp'\Qq')\s(\e^{i\frac{s}{\mu}})\approx i\hbar\, \s(\e^{i\frac{s}{\mu}})$$
and hence, $$(\Qq\Pp-\Pp\Qq)\s(\e^{i\frac{s}{\mu}})\approx i\hbar\, \s(\e^{i\frac{s}{\mu}}).$$ $\Box$
\epk
\bpk By the argument in \ref{QP} we also have for $\mathbb{S}$
$$\Qq'=\Qq\mbox{ and }\Pp'=\Pp.$$
\epk
\bpk \label{Four1}
{\bf The Fourier transform} is defined on $\mathbb{S}$ by applying $\Lim$ to (\ref{four}):

$$\Phi |x\ra= |p\ra,\ \mbox{ for } p=x.$$ 

Applying $\Phi$ twice we get
 
$$\Phi^2: |x\ra\mapsto |-x\ra.$$
\epk

\bpk \label{commentE}
The structure on $\EEE$ in the global language  plays a central role in the theory below. 
This is a pseudo-finite structure which can be seen as a model of ``huge finite universe'' of quantum mechanics.
Its
specialisation (image under $\Lim$ ), the ``standard'' structure $\mathbb{S},$  the space of states, is a close analogue of the standard Hilbert space where operators of quantum mechanics are being represented. The calculations in the next sections demonstrate that we may carry out rigorous quantitative  analysis of processes in the    ``huge finite universe'' and then, specialising these to the standard space of states, obtain well-defined formulas in complete agreement with traditional methods of quantum mechanics.

\epk
\section{Regular unitary transformations on $\bV_*$ and $\mathbb{S}.$}
\bpk A {\bf regular unitary transformation $L$ of finite type on $\bV_*$} is defined as in \ref{defReg}(i)-(iv) for the algebra $\tilde{\mathrm{A}}$ with the extra assumption:

\medskip 

(v) The index $\nu_B:=(\Nn:n_B)$ of the subalgebra $B$ in  $\tilde{\mathrm{A}}$ is finite.
 
Then we have the following refinement of Theorem \ref{regZar}.
\epk
\bpk \label{Group1} {\bf Theorem.} (i) {\em A  regular unitary transformation $L$ of finite type on $\bV_*$ associated with the subalgebra $B=\la \hat{U}^a\hat{V}^b,\hat{U}^c\hat{V}^d\ra,$ $a,b,c,d\in \Z,$ and the isomorphism $\sigma: \la \hat{U}^a\hat{V}^b,\hat{U}^c\hat{V}^d\ra\to \la \hat{U}^{a'}\hat{V}^{b'},\hat{U}^{c'}\hat{V}^{d'}\ra$
induces a bijection
$$\hat{L}:\mathbb{S}_{a/b}\to \mathbb{S}_{a'/b'}.$$}

(ii) {\em
If $C=\la \hat{U}^{a_C}\hat{V}^{b_C},\hat{U}^{c_C}\hat{V}^{d_C}\ra\subset B$ is another subalgebra of finite index, $\sigma_C$ is the restriction of $\sigma$ to $C$ and  $L_C$ the transformation associated to $C$ and $\sigma_C,$ then $\hat{L}_C=\hat{L}.$ In particular, 
$$\hat{L}: \mathbb{S}_{a_C/b_C}\to \mathbb{S}_{a'_C/b'_C}  ,$$  
and so $\hat{L}$ is defined on all rational Lagrangian subspaces $$\hat{L}:\mathbb{S}_\Q\to \mathbb{S}_\Q.$$
}

(iii) {\em The set $\mathcal{G}_{fin}$ of all transformations $\hat{L},$ for 
  regular unitary transformation $L$ of finite type on $\bV_*$, 
  forms a group acting on $\mathbb{S}_\Q.$ The group is isomorphic to $\mathrm{SL}(2,\Q).$} 
  
  \medskip
  {\bf Proof.} By definition and \ref{appr2} we have
 $$\hat{L}: \{ |s\ra : s\in \R\}\to   \{ |s'\ra : s'\in \R\}$$
 where $S=\hat{U}^{a}\hat{V}^{b}$ and $S'= \hat{U}^{a'}\hat{V}^{b'}.$ This is exactly the statement of (i).
 
 (ii) follows by \ref{LC}.

(iii). It follows from \ref{regZar}(iii)-(iv) that $\mathcal{G}$ is a group. Moreover, a regular transformations
acts on elements of the rational Heisenberg group $$\hat{U}^{a}\hat{V}^{b}\mapsto \hat{U}^{a'}\hat{V}^{b'}$$
as an automorphism (and so symplectomorphism). 
So $\mathcal{G}$ is the subgroup of all (rational) symplectomorphisms on $\mathbb{S}_\Q,$ that is $\mathrm{SL}(2,\Q).$ On the other hand \ref{Fourmx}, \ref{gaussex} and \ref{diag} prove that the generators of $\mathrm{SL}(2,\Q)$ are in $\mathcal{G}.$ So $\mathcal{G}=\mathrm{SL}(2,\Q).$
$\Box$

\epk
\bpk A {\bf tame regular unitary transformation $L$  on $\bV_*$} is defined as in \ref{defReg}(i)-(iv) for the algebra $\tilde{\mathrm{A}}$ with the extra assumption:

\medskip 

(v) The index $\nu_B:=(\Nn:n_B)$ of the subalgebra $B$ in  $\tilde{\mathrm{A}}$ is ``small compared to $\mu,$ that is
$$\nu_B\prec \mu.$$

This condition is equivalent to $B=\la \hat{U}^a\hat{V}^b,\hat{U}^c\hat{V}^d\ra,$ $a,b,c,d\in \ZZ,$ $$a,b,c,d\prec \mu.$$
\epk
\bpk\label{Group2} {\bf Theorem.} (i) {\em A tame regular unitary transformation $L$  on $\bV_*$ associated with the subalgebra $B=\la \hat{U}^a\hat{V}^b,\hat{U}^c\hat{V}^d\ra,$ $a,b,c,d\in \ZZ,$ and the isomorphism $\sigma: \la \hat{U}^a\hat{V}^b,\hat{U}^c\hat{V}^d\ra\to \la \hat{U}^{a'}\hat{V}^{b'},\hat{U}^{c'}\hat{V}^{d'}\ra$
induces a bijection
$$\hat{L}:\mathbb{S}_{\alpha/\beta}\to \mathbb{S}_{\alpha'/\beta'}$$
where $\alpha/\beta=\Lim a/b,$ $\alpha'/\beta'=\Lim a'/b',$
}

(ii) {\em The set $\mathcal{G}$ of all transformations $\hat{L},$ for tame 
  regular unitary transformation $L$  on $\bV_*$, 
  forms a group acting on $\mathbb{S}.$ The group is isomorphic to $\mathrm{SL}(2,\R).$} 

{\bf Proof.} The same arguments as in \ref{Group1}. $\Box$
\epk
\section{Time evolution operators. The case of the free particle}
\bpk \label{def_free}
Let $t$ be a rational number. We introduce a regular unitary transformation $\ga^t$ on the space of states, which corresponds to the so called {\em time evolution operator for the free particle}. In terms of the Heisenberg-Weyl algebra this operator could be defined as
$$\e^{-it\frac{\Pp^2}{2\hbar}}.$$
Equivalently, this is the operator acting on the position eigenstates as
\be \label{free}\e^{-it\frac{\Pp^2}{2\hbar}}|p\ra = \e^{-it\frac{p^2}{2\hbar}}|p\ra.\ee

Our $\ga^t$ satisfies the same property.

\medskip

As above $\ga^t$ will first be defined on the pseudo-finitary  space $\bV_*(\mathbf{1})$ and then transferred to $\mathbb{S}$ by $\Lim.$
\epk
\bpk In fact, we can only fully define $\ga^t$ on a submodule of $\bV_*(\mathbf{1})$ the construction of which depends on $t.$

We assume that $t=\frac{b}{d},$ where 
 $b,d\in \ZZ,$ $|b|,|d|\prec\mu.$

 We calculate over the subalgebra
${B}(b,d)$ of ${\tilde{\mathrm{A}}}  .$ 

By \ref{1.7} the ${\tilde{\mathrm{A}}}  $-module $\bV_*(\mathbf{1})$ splits into the direct sum of $d$ of its ${B}(b,d)$-submodules. We define $\ga^t$ on the submodule
$\bV_{B(b,d)}(\mathbf{1}).$ 

 Set $$\check{q}=\e^\frac{i bd\hbar}{\mu^2}=q^{bd}
  \mbox{ and } \check{N}=\frac{\mu^2}{bd\hh}=\frac{\Nn}{bd}.$$
Clearly,
$$\check{N}=\dim  \bV_{B(b,d)}(\mathbf{1}).$$
Choose a canonical $\hat{U}^d$-basis of $\bV_{B(b,d)}(\mathbf{1})$ in the form
$$\{ \check{\uu}(\check{q}^{m}):\ m=0,\ldots, N-1\}$$
satisfying
$$ \begin{array}{ll} \hat{U}^d: \check{\uu}(\check{q}^{m})\mapsto \check{q}^{m}\check{\uu}(\check{q}^{m})\\
\hat{V}^b: \check{\uu}(\check{q}^{m})\mapsto \check{\uu}(\check{q}^{m-1})
\end{array}
$$

Set $\ga^t$ on the $U$-basis by:
\be \label{fe1g}\ga^t\check{\uu}(\check{q}^{m})=\frac{\cc  }{\sqrt{\check{N}}}\sum_{l=0}^{\check{N}-1} 
\check{q}^{\frac{(l-m)^2}{2}}\check{\uu}(\check{q}^{l})=:\s(\check{q}^m) \ee
where $\cc  $ is defined in \ref{4.6+}.

\medskip

One sees immediately that $\ga^t$ is the Gaussian transformation with the domain of definition and range $\bV_{B(b,d)}(\mathbf{1}).$ So we may use \ref{Gauss0} - \ref{Reg}.

\epk

\bpk \label{mot1} {\bf An alternative definition of $\ga^t.$} According to quantum mechanics the time evolution operator for the free particle 
is $\ga^t:=\e^{-it\frac{\Pp^2}{2\hbar}}$
and the expression on the right makes sense in the  operator algebra
 Then
the canonical commutation relation (\ref{ccr}) gives 
$$[\Qq,\ga^t]= t\Pp \ga^t.$$  
Hence, 
$\Qq \ga^t -\ga^t\Qq=t\Pp \ga^t$ and $$\ga^t i\Qq \ga^{-t}= i(\Qq-t\Pp).$$


Note that by the Baker-Campbell-Hausdorff formula, for any $r\in \R$

$$U^r\cdot V^{-{r\hh t}}=\exp ir\Qq\,\cdot\, \exp\{ -i r t\Pp\}=
\exp (i r\Qq-irt\Pp+\frac{1}{2}[ir\Qq,-i{r t}\Pp])=$$ 
$$=\exp (ir\Qq- ir t\Pp+i{\pi r^2t \hh})=\e^{\pi i\hh r^2 t} K^tU^r K^{-t}.$$

Letting $t=\frac{b}{d}$ and $r=\frac{b}{\mu}$ as above, we get 
\be \label{Sv2} \ga^t \hat{U}^d \ga^{-t}=S,\ \ S=\check{q}^{-\frac{1}{2}}\hat{U}^d\hat{V}^{-b}\ee
which is  (\ref{Sv0}) in corresponding notation.

Another property that follows from the form of the time evolution operator, is its  commutation with $\Pp$ and so
with $\hat{V}^b:$ 
\be \label{w2} K^t \hat{V}^b K^{-t} = \hat{V}^b\ee

(\ref{Sv2}) and (\ref{w2}) can be taken as the defining properties of 
 $\ga^t.$ Note that these two algebraic properties alone do not allow to determine  the scalar coefficient $\cc  .$

\epk

\bpk Now we  define $\ga^t$ on $\mathbb{S}$  {\em by continuity}, that is set
 $$\Lim\ga^t \s(\check{q}^m)=\ga^t \Lim \s(\check{q}^m),$$ 
for all $\s\in \EEE_{B(d,b)}.$
 
 By (\ref{def_p}), remembering that $$\check{\vv}(\check{q}^n)= \vv^{\frac{b}{\mu},\frac{d}{\mu}}(\e^{\frac{i\hbar nbd}{\mu^2}})=
\vv^{b,a}(\e^{\frac{ibp}{\mu}}),$$
where $$p=\frac{nd\hbar}{\mu}$$
satisfies (\ref{xp_condition}). We have by definition
$$|p\ra=\Lim \hat\vv^{\frac{b}{\mu},\frac{d}{\mu}}(\e^{\frac{ibp}{\mu}}).$$

We also note that $$\check{q}^{-\frac{n^2}{2}}=\e^{-t\frac{p^2}{2}},\ \ t=\frac{b}{d}.$$

By 
 (\ref{ga V0}) as
\be \label{ga V3} \ga^t |p\ra= e^{-i  t \frac{p^2}{2\hbar}}|p\ra. \ee
Respectively, in
 (\ref{fe1g}) we set
$$x_1=\frac{lb\hbar}{\mu},\ \ x_2=\frac{mb\hbar}{\mu},$$ 
we get by definition and recalling that $$\check{\uu}(\check{q}^l)=\uu^{b,d}(\e^\frac{lbd\hbar}{\mu^2})=\uu^{b,d}(\e^\frac{dx_1}{\mu}),$$
$$\check{\uu}(\check{q}^m)=\uu^{b,d}(\e^\frac{mbd\hbar}{\mu^2})=\uu^{b,d}(\e^\frac{dx_2}{\mu}),$$
$$|x_1\ra=\Lim \check{\uu}(\e^\frac{dx_1}{\mu}) $$
and $$\ga^t |x_2\ra =\Lim \s(\e^\frac{dx_2}{\mu}). $$ 
\epk
\bpk Our aim now is to determine the value of 
$$\la \check{\uu}(\e^\frac{dx_1}{\mu})| \s(\e^\frac{dx_2}{\mu})\ra_{Dir},$$
for which we need to determine  ${\Delta x}.$
This was determined in \ref{rescale0} taking into account (\ref{dd})  and gives for $B=D=B(d,b)$  $$\Delta x=\frac{b\hbar}{\mu}.$$

By (\ref{Dir3x}) and (\ref{fe1g})
$$ \la \check{\uu}(\e^\frac{dx_1}{\mu})| \s(\e^\frac{dx_x}{\mu})\ra_{Dir}=\frac{1}{\Delta x}\la \check{\uu}(\check{q}^l)| \s(\check{q}^m)\ra=\frac{\sqrt{bd\hh}}{\mu \Delta x}\e^{\frac{-\pi i}{4}}\check q^{\frac{(\l-m)^2}{2}}$$

which gives us by application of $\Lim:$

\be \label{free_prop}\la x_1| \ga^t x_2\ra_{Dir}= \e^{-\frac{\pi i}{4}}\sqrt{\frac{d\hh}{ b}}\cdot\frac{1}{\hbar} \e^\frac{i(x_1-x_2)^2}{2}=\frac{1}{\sqrt{2\pi i\hbar t}}\e^\frac{i(x_1-x_2)^2}{2t\hbar}\ee

 The left-hand side of (\ref{free_prop}) is called {\bf the kernel of a time evolution operator}  (Feynman propagator for the corresponding particle). 
\epk
\section{Harmonic oscillator.}

\bpk In the Heisenberg Weyl algebra consider the operator $$H=\frac{\Pp^2+\Qq^2}{2},$$
the {\em Hamiltonian for the Harmonic oscillator}, where  we suppressed some parameters of physical
significance). 

We are interested in the time evolution operator for the Harmonic oscillator, i.e. the operator
$$K^t:=K^t_{HO}:=\e^{-it\frac{H}{\hbar}}.$$

In order to motivate our technical definition we
assuming again as in \ref{mot1} that $\Pp$ and $\Qq$ belong to  a Banach algebra. Then
well-known calculations, which we omit here, produce the following  
formulae
\be \label{Ucos}K^t U^aK^{-t}=\e^{-a^2\pi i \hh \sin   t\cos   t} 
U^{a\cos  t}V^{-a\hh\sin  t}.\ee
and
\be \label{Vsin}K^t V^{a\hh} K^{-t}=\e^{a^2\pi i \hh\sin   t\cos   t} 
U^{a\sin  t}V^{a\hh \cos  t}.\ee
\epk
\bpk
\label{US} We will assume that $  t\in \R$ is such that
$\sin   t$ and $\cos   t$ are rational (there is a dense subset of $t\in (0,2\pi]$ satisfying this condition), and even allow these to be non-standard rational.

For the rest of this section we fix 
$$\sin   t=\frac{e}{c},\ \ \cos   t=\frac{f}{c},\ \ \mbox{ for some 
nonzero }
e,f,c\in \ZZ,\ 0\le |e|,|f|,  c\prec\mu. $$

We work in the principal ${\tilde{\mathrm{A}}}  $-module $\bV_*(\mathbf{1})$ and
in the space of states $\mathbb{S}$ and 
 want to construct, as suggested by (\ref{Ucos}) and (\ref{Vsin}), an operator $K^t$  so that
\begin{align}K^t\hat{U}^c(K^t)\inv=q^{-\frac{1}{2}ef} \hat{U}^f
\hat{V}^{-e}\label{KU} \end{align}

\begin{align}(K^{t})\hat{V}^{ce} (K^{t})\inv=q^{\frac{1}{2}e^3f}
\hat{U}^{e^2}\hat{V}^{fe}\label{-KU} \end{align}
where $$q^\frac{1}{2}=\e^{\frac{\pi i \hh}{\mu^2}}.$$

\epk
\bpk Denote $$S_t:=q^{-\frac{1}{2}ef}\hat{U}^f
\hat{V}^{-e},\  R_t:=q^{\frac{1}{2}ef}\hat{U}^e
\hat{V}^{f} \mbox{ and  check that }  R_t^m:=q^{\frac{1}{2}efm^2}
\hat{U}^{em}\hat{V}^{fm}.$$

Note that the algebra $\la S_t, R_t^e\ra$ is isomorphic to 
$\la \hat{U}^c,\hat{V}^{ce}\ra$ by the isomorphism given by conjugation by $K^t,$ so its irreducible modules are  of
the same dimension $\frac{\Nn}{c^2e}.$
  
Clearly, $S_t$ and $R_t$ are pseudo-unitary.  Recall that with our notation $$\dim \bV_{B(1,e)}(\mathbf{1})=\frac{\Nn}{e}.$$ Note also that
the $\hat{U}$ basis $\uu^{1,e}(q^{el})$ can be identified with the subset $\{ \uu(q^{k}): e|k\}$ of the 
$\hat{U}$-basis of $\bV_*(\mathbf{1}).$
\epk
\bpk \label{defK}
Define the operator\footnote{We state without proof that $K^t$ can be defined as a regular unitary operator on $\bV_{\tilde{A}}.$} 
$$K^t: \bV_{B(c,ce)}(\mathbf{1})\to \bV_{*}(\mathbf{1})$$
by the action on an $U$-basis, 

\be \label{Kt} K^t: \uu^{c,ce}(q^{c^2em})\mapsto \s(q^{c^2em}):=C_0\sqrt{\frac{e}{\Nn}}\sum_{\l=0}^{\frac{\Nn}{e}-1} q^{ef\frac{l^2-e^2m^2}{2}-e^3ml}
\uu(q^{e(l+mf)}),\ee
where $C_0\in \F_0$ is a constant of modulus $1,$ the exact value of which we determine at the end of this section.

\epk
\bpk {\bf Lemma.} {\em For any integer $m,$
$$\begin{array}{ll} S_t: \s(q^{c^2em})\mapsto q^{c^2em} \s(q^{c^2em})\\
R_t^e: \s(q^{c^2em})\mapsto  \s(q^{c^2e(m-1)}).\end{array}$$}

{\bf Proof.} One checks that 
$$S_t: \uu(q^{e(l+mf)})\mapsto q^{ef(l+\frac{1}{2})+emf^2} \uu(q^{e(l+1+mf)})$$
and thus, using $f^2+e^2=c^2,$ 
$$S_t: q^{ef\frac{l^2-e^2m^2}{2}-e^3ml}\uu(q^{e(l+mf)})\mapsto q^{c^2em}\cdot q^{ef\frac{(l+1)^2-e^2m^2}{2}-e^3m(l+1)} \uu(q^{e(l+1+mf)}).$$ This implies the first statement.


For the second statement first note
that
$$\s(1):=C_0\sqrt{\frac{e}{\Nn}}\sum_{\l=1}^{\frac{\Nn}{e}-1} q^{ef\frac{l^2}{2}}
\uu(q^{el})$$
and that 
$$R_t^{-em}: \uu(q^{el})\mapsto q^{k(e,f,m,l)}\uu(q^{el+emf})$$
for $$k(e,f,m,l)={ -e^3f\frac{m^2}{2}-e^3ml}.$$
It follows
$$ \label{defR} R_t^{-em}: \s(1)\mapsto C_0\sqrt{\frac{e}{\Nn}}\sum_{\l=1}^{\frac{\Nn}{e}-1} q^{ef\frac{l^2-e^2m^2}{2}-e^3ml}
\uu(q^{e(l+mf)})=\s(q^{c^2em}).$$
$\Box$

\epk

\bpk For $n,m\in \ZZ$ we have, expressing $\uu^{c,ce}(q^{c^2en})$
in terms of $\uu$ (see \ref{3.45}), 
$$\la \uu^{c,ce}(q^{c^2en})|\s( q^{c^2em})\ra=$$
$$= \frac{C_0}{\sqrt{c}}\cdot \sqrt{\frac{e}{\Nn}}\la
\sum_{k=0}^{c-1}\uu(q^{cen+k\frac{\Nn}{c}})|\sum_{l=0}^{\frac{\Nn}{e}-1}q^{ef\frac{l^2-e^2m^2}{2}-e^3ml}
\uu(q^{e(l+mf)})\ra         $$
Set $M:=\frac{\Nn}{c}$ and $l_k=kM+l_0,$ for $k=0,\ldots,c-1.$

Note that the necessary condition for the product $\la \uu^{c,ce}(q^{c^2en})|\s( q^{c^2em})\ra$  to be non-zero is that
\be \label{cond_nl}  cen+kM=el+emf\mbox{ for some } l=0,\ldots, M-1,\ k  =0,\ldots, c-1.\ee

Let $l_0=cn-mf$. 

Claim. $$fl_0-e^2m\equiv 0 \,\mathrm{mod}\,c.$$
Note that
$fl_0-e^2m=fcn- mf^2 -me^2=fcn-mc^2.$ Claim proved.

 So
$$cn+kM=l_k+mf.$$
Recall that $0<|c|, |e|\prec\Nn$ which implies that
$$M^2\equiv 0\,\mathrm{mod}\, \Nn.$$
 
Now we can continue
$$\la \uu^{c,ce}(q^{c^2en})|\s( q^{c^2em})\ra=$$
$$=C_0\sqrt{\frac{e}{\Nn c}}\sum_{k=0}^{c-1}\la \uu(q^{cen+k\frac{\Nn}{c}})|q^{ef\frac{l_k^2-e^2m^2}{2}-e^3ml_k}
\uu(q^{e(l_k+mf)})\ra=     $$
$$=C_0\sqrt{\frac{e}{\Nn c}}\sum_{k=0}^{c-1}q^{ef\frac{l_k^2-e^2m^2}{2}-e^3ml_k}= C_0\sqrt{\frac{e}{\Nn c}}q^{ef\frac{l_0^2-e^2m^2}{2}-e^3ml_0}\sum_{k=0}^{c-1}q^{efl_0kM-e^3mkM}$$
Now note that $efl_0kM-e^3mkM=ekM(fl_0-e^2m) $ is divisible by
$\Nn$ by Claim above. Hence the last sum is equal to $c$ and we have finally
\be \label{eq9.6} \la \uu^{c,ce}(q^{c^2en})|\s( q^{c^2em})\ra=C_0\sqrt{\frac{ec}{\Nn }}q^{ef\frac{l_0^2-e^2m^2}{2}-e^3ml_0}. \ee

\epk
\bpk Note by direct substitution and using $f^2+e^2=c^2$ that
\be \label{9.7} ef\frac{l_0^2-e^2m^2}{2}-e^3ml_0=\frac{1}{2}ec^2\{ (n^2+m^2)f-2cnm\}.\ee

 Now rename $$x_1:=\frac{cen\hbar}{\mu},\ x_2:=\frac{cem\hbar}{\mu}.$$
 
Then, using \ref{*A}, \ref{7.4} and (\ref{9.7}), the equality (\ref{eq9.6}) becomes 
\be \label{eq9.8} \la \uu^{c,ce}(q^{c^2en})|\s( q^{c^2em})\ra=C_0 \frac{\sqrt{ec\hh}}{\mu }\exp  i\frac{(x_1^2+x_2^2)f-2x_1x_2c}{2e\hbar}=\ee
$$=C_0 \frac{\sqrt{ec\hh}}{\mu }\exp i\frac{(x_1^2+x_2^2)\cos   t-2x_1x_2}{2\hbar \sin   t}.$$
\epk
\bpk \label{final}  We rescale the last product according to \ref{rescale0}  and
(\ref{dir_s?}). We have $\la \hat{U}^c, \hat{V}^{ce}\ra$ in place of $\la R,R_\dagger\ra,$ and $\la S_t, R_t^e\ra$ in place of $\la S, S_\dagger\ra.$ 
 $N=\Nn=\frac{\mu^2}{\hh}.$

We have $$SR=q^{ce}RS$$ and $a_R=c,$ the maximal with the property
$$R^{\frac{1}{c}}\in {\tilde{\mathrm{A}}}  ,$$ while for $S$ the corresponding number $a_S=1.$

Hence, by \ref{rescale0} and (\ref{dd})
$$\Delta x = \dd\frac{e\sqrt{\hh}}{\mu}=\frac{e\hbar}{\mu}.$$
Thus
$$\la \uu^{c,ce}(q^{c^2en})|\s( q^{c^2em})\ra_{Dir}=\frac{\mu}{e\hbar}\la \uu^{c,ce}(q^{c^2en})|\s( q^{c^2em})\ra=$$ $$=\frac{\mu}{e\hbar}C_0 
\frac{\sqrt{ec\hh}}{\mu }\exp i\frac{(x_1^2+x_2^2)\cos   t-2x_1x_2}{\hbar\sin   t}=\frac{C_0  }{\sqrt{2\pi \hbar \sin   t}}
\exp i\frac{(x_1^2+x_2^2)\cos   t-2x_1x_2}{2\hbar \sin   t}.$$
\epk

\bpk 
For the Feynman propagator we need to apply the limit
$$\la x_1|K^t|x_2\ra=\Lim \la \uu^{c,ce}(q^{c^2en})|\s( q^{c^2em})\ra_{Dir}$$

which by the final formula of \ref{final} above gives us
$$\la x_2|K^t_{HO}|x_1\ra=C_0 \sqrt{\frac{1}{2\pi \hbar\sin   t}}\,
\exp i\frac{(x_1^2+x_2^2)\cos   t-2x_1x_2}{2\hbar\sin   t}.$$

This is in accordance with the well-known formula \cite{Zeidler II}, p552, if we set
\be\label{c_0} C_0=\cc=\e^{-\frac{\pi i}{4}} .\ee

Our definition can only determine $K^t$ up to the coefficient $C_0$ of modulus $1.$

\epk
\bpk In this subsection we will motivate a stronger definition for $K^t_{HO}$ which will have (\ref{c_0}) as a corollary. 

First we need to formulate and prove several claims. Note, that in the definition (\ref{Kt}) $C_0$ may depend on  $t.$ We need to have reasons to claim that it does not. 

Claim 1.  Given  $ t_1$ and $t_2$ as in \ref{US} we have the following operator identity
$$K^{t_1}K^{t_2}=K^{t_1+t_2}$$
where $K^{t_1},$ $K^{t_2}$ and $K^{t_1+t_2}$ are defined by  (\ref{Kt}) with the same $C_0$ (but with different values of $e$ and $c,$ which depend on $t$).

Proof of the Claim. TODO.

Corollary.  For all $t_1\in \Q\cdot t,$  $$C_0(t_1)=C_0(t).$$

Now we may use the fact used by physicists and motivated by simple algebraic (the Trotter product formula) and continuity arguments which imply that the physicists' values of the operator must satisfy
$$\lim_{t\to 0}\frac{\la x_1|K^t_{HO}x_2\ra}{\la x_1|K^t_{\mathrm{Fr}}x_2\ra}=1.$$

Postulating that this identity holds for the operators $K^t_{HO}$ and $K^t_{\mathrm{Fr}}$ as defined above and
comparing the corresponding formulas we conclude that (\ref{c_0}) holds. 
\epk
\section{Trace of the time evolution  operator for the Harmonic oscillator.}

\bpk \label{na2}  First note that by (\ref{cond_nl})
$$\la \uu^{c,ce}(q^{c^2en})|\s( q^{c^2en})\ra\neq 0 \Leftrightarrow
 en(c-f)+kM=el\mbox{ for some } l=0,\ldots, \frac{\Nn}{e}-1,\ k  =0,\ldots,c-1.$$
 Hence, the $n$ on the right-hand side can only take values
 $$n=0, 1,\ldots, \frac{\Nn}{ec(c-f)}-1$$
\epk
\bpk 
 Note  for $n=m$ (\ref{9.7}) is equal to
 $$-ec^2(c-f)n^2.$$
 
 Hence, by (\ref{eq9.6}) and \ref{na2}
$$\mathrm{Tr}(K^t):=  \cc  \sqrt{\frac{ec}{\Nn }}\sum_{n=0}^{\frac{\Nn}{ec(c-f)}-1}
q^{-ec^2(c-f)n^2}.$$

Denote $L=\frac{\Nn}{ec^2(c-f)}.$ We can rewrite the above as
$$\mathrm{Tr}(K^t)=  \cc \sqrt{\frac{1}{c(c-f)L }}\sum_{n=0}^{cL-1}
\e^{-\frac{2\pi i}{L}n^2}.$$

By the Gauss quadratic sums formula (\ref{Gauss})

$$ \sum_{n=0}^{L-1}
\e^{-\frac{2\pi i}{L}n^2}=2\e^{-\frac{\pi i }{4}}\sqrt{\frac{L}{2}}=\e^{-\frac{\pi i }{4}}\sqrt{2L}; \ \ \sum_{n=0}^{cL-1}
\e^{-\frac{2\pi i}{L}n^2}=c\sqrt{2L}.$$
Hence
$$\mathrm{Tr}(K^t)=c\, \cc ^2\sqrt{2L} \sqrt{\frac{1}{c(c-f)L }}= \cc^2\,\sqrt{\frac{2}{(1-\frac{f}{c}) }}.$$
Returning to the original notation \ref{US} we have 
$$ \sqrt{\frac{2}{(1-\frac{f}{c}) }}= \sqrt{\frac{2}{(1-\cos t) }}= \frac{1}{ |\sin \frac{t}{2}|}.$$
Since $\cc^2=-i,$ we finally get,
$$\mathrm{Tr}(K^t_{HO})=\frac{1}{ i|\sin \frac{t}{2}|}.$$
\epk



\thebibliography{mphys}
\bibitem{BG} K.Brown and K.Goodearl, {\bf Lectures on algebraic quantum groups}, Birkhauser, 2002

\bibitem{BY} I.Ben-Yaacov, {\em Positive model theory and compact abstract theories}, J. Math. Log. 3, 85-118 (2003)
\bibitem{withA} A. Cruz Morales and B.Zilber, {\em The geometric semantics of algebraic quantum mechanic} Phil. Trans. R. Soc. A, 2015, v 373, issue 2047
\bibitem{Efem} U.Efem, {\em Specialisations on algebraically closed fields}, arXiv  1308.2487
\bibitem{FH} R.Feynman and A. Hibbs, {\bf Quantum Mechanics and Path Integrals}, Dover Books on Physics
\bibitem{DG} J. Derezi\'nski and  C.G\'erard, {\bf  Mathematics of Quantization and Quantum Fields}, CUP, 2013
 \bibitem{ID0} A.D\"oring and C.Isham, Chris, {\em What is a Thing?: Topos Theory in the Foundations of Physics}, {\bf  New Structures in Physics}, Chapter 13, pp. 753--940, Lecture Notes in Physics, 813, Springer, 2011,
\bibitem{ID} A.Doering and C.Isham,{\em A Topos Foundation for Theories of Physics:
I. Formal Languages for Physics}, J. Math. Phys. 49, 053515 (2008)
\bibitem{GN} S.Gudder and V.Naroditsky,{ \em
Finite-dimensional quantum mechanics}, International Journal of Theoretical Physics; v. 20(8) p. 619-643;  1981
\bibitem{HZ} E.Hrushovski and B.Zilber, {\em Zarsiki geometries.}
 J. Amer. Math. Soc. 9 (1996), no. 1, 1-56.
\bibitem{IK} H.Iwaniec, E. Kowalski {\bf Analytic number theory} American Mathematical Society,  2004

\bibitem{LV} G.Lion,  and M.Vergne. {\bf The Weil Representation, Maslov Index, and
Theta Series} Birkhauser, 1980 ̈

\bibitem{QHO} V.Solanki, D.Sustretov and B.Zilber,
{\em The quantum harmonic oscillator as a Zariski geometry}, Annals of Pure and Applied Logic, v.165 (2014), no. 6,  1149 -- 1168

\bibitem{Zeidler I} E.Zeidler, {\bf Quantum Field Theory} v.1, 2009

\bibitem{Zeidler II} E.Zeidler, {\bf Quantum Field Theory} v.2, 
2009

\bibitem{QZG} B. Zilber, \textit{A Class of Quantum Zariski Geometries}, in Model Theory with Applications to Algebra and Analysis, I, Volume 349 of LMS Lecture Notes Series, Cambridge University Press, 2008.

\bibitem{Appr} B. Zilber, \textit{Perfect infinities and finite approximation} In: {\bf Infinity and Truth}. IMS Lecture Notes Series, V.25, 2014


\bibitem{Zbook} B. Zilber, \textit{Zariski Geometries}, CUP, 2010
\end{document}